# CONSEQUENCES OF THE FERMAT'S ANISOTROPIC UNIAXIAL PRINCIPLE ON THE REFLEXION AND TRANSMISSION FACTORS FOR ONE-DIMENSIONAL UNIAXIAL CRYSTAL SLABS

Vital LE DEZ and Hamou SADAT

Institut PPRIME, Université de Poitiers, 40 Avenue du Recteur Pineau, 86022 Poitiers, France

**Abstract** – A coherent definition of the reflection and transmission factors at a plane interface separating two uniaxial crystals is proposed, from the photons impulsion-energy 4-vectors conservation. This definition, different from the classical electromagnetic one, is compatible with the completely resolved extended Fermat's principle of the geometric optics for extraordinary luminous rays inside uniaxial media, and allows the exact calculation of the transmission factors at the plane interface for any practical configuration, combining all possible optical axes and anisotropy factors variations. Furthermore, this particular technique points out the existence of quasi-particles strongly associated to the photons, whose behaviour is highly correlated to the photons transmission/reflection possibilities.

## I - INTRODUCTION

In a recent work dealing with the derivation of the radiative transfer equation inside homogeneous moving semi-transparent media [1], we pointed out the fact that a moving isotropic medium could be formally interpreted, from a radiative point of view, as an uniaxial anisotropic crystal of optical characteristics depending on the refractive index of the isotropic medium so as the speed and direction of motion, from which it should exist a consistent formulation of the phenomenological radiative transfer inside uniaxial crystals, opening the way to quantitative studies on radiative transfer in such media. Up to now however, the phenomenological theory of radiative transfer assumes that the radiative energy is transported along trajectories which are the minimal paths of the luminous rays, loci of the travelling photons seen as zero mass particles, and should solely deal with coherent physical quantities related to photons understood as particles, since a consistent electromagnetic description of radiative transfer, in infra-red spectrum for instance, has not still been yet achieved: it is so true that the fundamental variable used in the phenomenological radiative transfer equation, the intensity defined from the radiative flux, cannot be simply related to the electromagnetic energy transported by the electromagnetic waves, so as the electromagnetic flux of the associated Poynting vector cannot be replaced by the radiation flux defined from the radiative intensities without any precaution. Then the directional radiative intensity appearing in the radiative transfer equation has to be understood as a physical quantity evolving in a given medium along particular trajectories which are the luminous rays, governed by the geometric optics laws: an interesting comment of this situation is found in [2], and if an electromagnetic description of some short scale radiative effects, in isotropic so as in anisotropic media, has retained a special attention, an average description at larger scales of the radiative phenomena based on light geometric optics still suffers from a lack of satisfactory investigation in anisotropic media; in this context, this point of view, commonly used for isotropic media, should also be extended to any uniaxial anisotropic medium, since there exists for any uniaxial homogeneous medium a mathematical application formally transforming a given crystal into an homogeneous moving isotropic medium of given refractive index, direction of motion and speed, where the radiative transfer equation can be derived as for non moving media; hence it is possible to find an equivalent physical signification for the radiative intensity in uniaxial media, this intensity being, as for isotropic media, understood as a physical quantity evolving on the luminous rays governed by the extended anisotropic explicit geometric optics laws. Furthermore, if an internal description of the anisotropic geometric optics has to be correctly achieved, a complete description of the luminous rays at a boundary between two anisotropic media has also to be understood, since from a practical point of view, radiative transfer is examined inside media of finite dimensions surrounded by an external environment: this introduces the problem of definition and calculation of the reflection/transmission factors, explicitly appearing as boundary conditions for the radiative intensities: indeed, is it judicious to use an electromagnetic description in the determination of the reflection/transmission factors when the radiative intensities bound to luminous rays cannot always been related to the wave energy densities, so that the boundary intensities let appear both quantities obtained from electromagnetic theory and geometric optics?

The problem dealing with the determination of the optical rays inside an anisotropic medium has been examined from different approaches for several years [3-6]: Carinena and Nasarre for instance generalise their successful presymplectic geometry tool applied in isotropic media, to anisotropic media for which the refractive index is a function of the ray direction, and Newcomb applies a principle of least time in geometrical optics referred to as the Fermat's principle for a general medium of arbitrary anisotropy; in [6] the authors give a particular form to the Hamilton equation of the rays and obtain a so-called Fermat's principle by minimising the corresponding Lagrange function along a luminous ray parallel to the group velocity; it is important to remember here that this problem can equivalently be solved in isotropic media either by minimising the electromagnetic eikonal function or by determining the light geodesics for a given metric tensor; this latter point of view developed in [7], i.e. the determination of the geodesics associated to a metric tensor of a Finsler space $\overline{\overline{g}}\left(x_k, \dot{x}_k\right)$, depending both on space and ray direction parallel to the group velocity, shows that the travel time along the Fermat's functional, that is along a luminous ray, is directly related to the group velocity $U$ and that the luminous ray trajectories are the ones for which the travel time is the minimal quantity, has been detailed and applied in [8], leading to a compact form of the so-call anisotropic Fermat's principle for geometric optics: in this present work, we shall give in section II an explicit solution of this principle in the case of a parallel slab of uniaxial crystal, leading to the generalization of the Descartes' law, which is, in the frame of the anisotropic geometric optics, the basic equation in the comprehension of the reflection and transmission phenomena of luminous rays at a plane interface separating two anisotropic crystals, fully described in section III. At this stage, we shall notice that using an electromagnetic description in the scope of calculating the reflection/transmission factors at a separating interface is not compatible with the generalized Descartes' law of the anisotropic geometric optics.

On the other hand, the calculation of the wave amplitude reflection/transmission factors, directly related to the wave energy factors, understood in the frame of the electromagnetic theory, has been investigated by many authors for many years, up to a very recent point [9, 10]: Lekner for instance found explicit formulae for the reflection/transmission factors for incident polarized waves of any direction at an interface between an uniaxial absorbing crystal and external environment, applied in [11] for calculation of a temperature field inside an anisotropic crystal at radiative equilibrium, and more recently, Sluijter and al. proposed a polarized ray tracing technique useful to describe the luminous ray propagation inside an heterogeneous uniaxial crystal, with insights to calculate the ray reflections and energy transfers in the bulk and at a separating curved interface; in all these approaches however, the electromagnetic theory basis, namely continuity of the fields at the interface, conservation of the electromagnetic flux and invariance of the field phase are the fundamental keys in obtaining the transmission factors, which obviously are electromagnetic quantities: consistent quantities for approaches discussed in [2], they may be irrelevantly used in descriptions based on the geometric optics laws. In this present work, we shall produce a consistent definition of the reflection and transmission factors at a plane interface separating two uniaxial crystals, based on the photons impulsion-energy 4-vectors conservation, and compatible with the geometric optics laws: first established in the case of isotropic media where the notion of photons associated normal quasi-particles is introduced, and compared with the electromagnetic definition, it shall be extended in the most general case to the anisotropic media and will let appear another specific class of associated quasi-particles. Numerical examples and a short conclusion will finally end this study.

## II – THE ANISOTROPIC UNIAXIAL FERMAT'S PRINCIPLE RESOLUTION IN A PLANE PARALLEL SLAB OF UNIAXIAL CRYSTAL

The system to be studied consists in a plane parallel slab, filled in with a non scattering grey semi-transparent uniaxial material characterised by its real dielectric diagonal tensor $\overline{\overline{\varepsilon}} = diag\left[\varepsilon'_{||}, \varepsilon'_{\perp}, \varepsilon'_{\perp}\right]$ in a convenient basis $\left(\vec{e}_{||}, \vec{e}_{\perp_1}, \vec{e}_{\perp_2}\right)$, depending on the location inside the medium, hereafter called the optical basis.

Because of the internal dielectric tensor gradient, the trajectories on which the radiative energy is transported are not straight lines but curved ones determined by the Fermat's principle which shall be developed now; let us first remind some classical but important results for uniaxial media [12]: from the Fresnel equation, namely $\left(p^2 \overline{\overline{I}} - \overline{\overline{\varepsilon'}} - \overline{\overline{p}}\right)\vec{E} = \vec{0}$, where $\vec{p} = p\,\vec{\Omega}$ is the wave vector and $p_{ik} = \left(\vec{p} \otimes \vec{p}\right)_{ik} = p_i\, p_k$, with $p^2 = \sum_{i=1}^{3} p_i^2 = n^2$, $n$ being the refractive index; in the optical basis, there exist two angles $\theta_D$ and $\varphi_D$ such that the unit wave vector can be expressed as $\vec{\Omega}=\begin{pmatrix} \cos\theta_D \\ \sin\theta_D\cos\varphi_D \\ \sin\theta_D\sin\varphi_D \end{pmatrix}$; to have a non zero solution to the Fresnel equation, the determinant of the linear system must be 0, so that:

$$\Delta=\det\left(p^2\bar{\bar{I}}-\bar{\bar{\varepsilon}}'-\vec{p}\otimes\vec{p}\right)=0=\begin{vmatrix} p_{2D}^2+p_{3D}^2-\varepsilon'_{||} & -p_{1D}p_{2D} & -p_{1D}p_{3D} \\ -p_{1D}p_{2D} & p_{1D}^2+p_{3D}^2-\varepsilon'_{\perp} & -p_{2D}p_{3D} \\ -p_{1D}p_{3D} & -p_{2D}p_{3D} & p_{1D}^2+p_{2D}^2-\varepsilon'_{\perp} \end{vmatrix}, \qquad (1)$$

from which one easily obtains $\Delta=\left(\varepsilon'_{\perp}-p^2\right)\left[\varepsilon'_{||}p_{1D}^2+\varepsilon'_{\perp}\left(p_{2D}^2+p_{3D}^2\right)-\varepsilon'_{\perp}\varepsilon'_{||}\right]=0$; hence it exists an ordinary isotropic refractive index $n_o=\sqrt{\varepsilon'_{\perp}}$ and an extraordinary wave refractive index depending on the unit wave vector direction $n_e^2=\frac{\varepsilon'_{\perp}\varepsilon'_{||}}{\varepsilon'_{||}\cos^2\theta_D+\varepsilon'_{\perp}\sin^2\theta_D}$; it is important to notice that if the Fresnel equation has two distinct solutions, it is however unable to decide if the two solutions are both possible in all situations; we call $f(n_e)=\varepsilon'_{||}p_{1D}^2+\varepsilon'_{\perp}\left(p_{2D}^2+p_{3D}^2\right)-\varepsilon'_{\perp}\varepsilon'_{||}$ the index function; defining the ray vector $\vec{s}$ such that $\vec{s}=\alpha\frac{\partial f(n_e)}{\partial\vec{p}}$ and $\vec{p}\,\vec{s}=1$, where $\alpha$ is a constant, one easily deduces that $\vec{s}=-\frac{1}{n_e^2}\frac{\partial n_e}{\partial\vec{\Omega}}$, or equivalently $\vec{s}=\frac{\vec{p}}{\varepsilon'_{||}}+\left(\vec{p}\,\vec{e}_{||}\right)\left(\frac{1}{\varepsilon'_{\perp}}-\frac{1}{\varepsilon'_{||}}\right)\vec{e}_{||}$ where $\vec{e}_{||}$ is the optical axis direction inside the medium; note that $\vec{s}$ is in the direction of the Poynting's vector transporting the radiative energy, and this allows to define the extraordinary energy refractive index as the inverse of the norm of the ray vector as $N_e^2=\frac{\varepsilon'_{\perp}\varepsilon'_{||}\left(\varepsilon'_{||}\cos^2\theta_D+\varepsilon'_{\perp}\sin^2\theta_D\right)}{\varepsilon'^2_{||}\cos^2\theta_D+\varepsilon'^2_{\perp}\sin^2\theta_D}$; from what precedes, the unit tangent vector to the trajectory of energy is obtained from $\vec{t}=N_e\vec{s}$, and there exist two angles $\Theta_D$ and $\Phi_D$ such that the unit energy vector can be expressed as $\vec{t}=\begin{pmatrix} \cos\Theta_D \\ \sin\Theta_D\cos\Phi_D \\ \sin\Theta_D\sin\Phi_D \end{pmatrix}$ in the optical basis, from which one deduces that $tg\,\Theta_D=\frac{\varepsilon'_{\perp}}{\varepsilon'_{||}}tg\,\theta_D$ and $\Phi_D=\varphi_D$; hence it comes $N_e^2=\varepsilon'_{||}+\left(\varepsilon'_{\perp}-\varepsilon'_{||}\right)\cos^2\Theta_D$, or expressed in a more general form:

$$N_e^2 = \varepsilon'_{||} + \left(\varepsilon'_{\perp} - \varepsilon'_{||}\right)\left(\vec{t}\,\vec{e}_{||}\right)^2 , \tag{2}$$

On the other hand, the wave refractive index can be expressed under the general following form:

$$\frac{1}{n_e^2} = \frac{1}{\varepsilon'_{||}} + \left(\frac{1}{\varepsilon'_{\perp}} - \frac{1}{\varepsilon'_{||}}\right)\left(\vec{\Omega}\,\vec{e}_{||}\right)^2 , \tag{3}$$

combination of the last two expressions easily leads to $N_e n_e \vec{\Omega}\,\vec{e}_{||} = \varepsilon'_{\perp}\,\vec{t}\,\vec{e}_{||}$ and for the unit ray and wave vectors:

$$\begin{aligned} \vec{t} &= N_e n_e \left[ \frac{\vec{\Omega}}{\varepsilon'_{||}} + \left(\frac{1}{\varepsilon'_{\perp}} - \frac{1}{\varepsilon'_{||}}\right)\left(\vec{\Omega}\,\vec{e}_{||}\right)\vec{e}_{||} \right] \\ \vec{\Omega} &= \frac{1}{N_e n_e}\left[ \varepsilon'_{||}\,\vec{t} + \left(\varepsilon'_{\perp} - \varepsilon'_{||}\right)\left(\vec{t}\,\vec{e}_{||}\right)\vec{e}_{||} \right] \end{aligned} , \tag{4}$$

One considers now that the speed of energy along the unit energy vector parallel to the unit ray vector is $\frac{c_0}{N_e}$, that is $ds = \frac{c_0}{N_e} dt$ where $c_0$ is the light celerity in vacuum. Hence the determination of the extraordinary optical trajectories on which the radiative energy is transported, is equivalent to the search of the light geodesics inside a curved space with a Finsler metric tensor [13], that is the following minimisation problem:

$$\delta\left[ \int_s N_e\left(\vec{q}, \dot{\vec{q}}\right) ds \right] = 0 , \tag{5}$$

The minimization of Eq. (5) has been completely achieved in [8], leading to the anisotropic uniaxial Fermat's principle expressed as:

$$\frac{d}{ds}\left\{ \frac{1}{N_e}\left[ \varepsilon'_{||}\,\vec{t} + \left(\varepsilon'_{\perp} - \varepsilon'_{||}\right)\left(\vec{t}\,\vec{e}_{||}\right)\vec{e}_{||} \right] \right\} = \overrightarrow{grad}_M N_e , \tag{6}$$

or equivalently with the help of Eq. (4) one has the elegant and compact form of the energetic uniaxial anisotropic extended Fermat's principle:

$$\frac{d}{ds}\left(n_e \vec{\Omega}\right)=\overrightarrow{grad}_M N_e, \tag{7}$$

where $s$ is the curvilinear abscissa along the energy trajectory, $N_e$ the ray refractive index and $\vec{t}$ is the unit tangent vector to the radiative energy trajectory parallel to the group celerity, while $\vec{\Omega}$ is the unit associated wave vector and $n_e$ the wave refractive index; let us notice that for isotropic media, $\varepsilon'_\perp=\varepsilon'_{||}=\varepsilon'$ so that $N_e=n_e=\sqrt{\varepsilon'}$, and $\vec{\Omega}=\vec{t}$, from which one has $\frac{d}{ds}\left(N_e \vec{t}\right)=\overrightarrow{grad}_M N_e$, which is the classical Fermat's principle for isotropic media.

*II-1 – Expression of the uniaxial Fermat's principal in its normal form:*

From Eq. (7) one immediately obtains $\vec{t}\,\frac{d}{ds}\left(n_e \vec{\Omega}\right)=\frac{d\,n_e}{ds}\vec{t}\,\vec{\Omega}+n_e\,\vec{t}\,\frac{d\,\vec{\Omega}}{ds}$, and since from the definition of the ray refractive index and Eq. (4) it easily comes $\vec{t}\,\vec{\Omega}=\frac{N_e}{n_e}$, one deduces that $\frac{d}{ds}\left(\vec{t}\,\vec{\Omega}\right)=\frac{\vec{n}\,\vec{\Omega}}{R}+\vec{t}\,\frac{d\,\vec{\Omega}}{ds}=\frac{1}{n_e}\frac{d\,N_e}{ds}-\frac{N_e}{n_e^2}\frac{d\,n_e}{ds}$, where $R$ is the curvature radius of the trajectory and $\vec{n}$ the unit normal vector to the trajectory, from which $n_e\,\vec{t}\,\frac{d\,\vec{\Omega}}{ds}=\frac{d\,N_e}{ds}-\frac{N_e}{n_e}\frac{d\,n_e}{ds}-\frac{n_e}{R}\vec{n}\,\vec{\Omega}$; then one has

$$\vec{t}\,\frac{d}{ds}\left(n_e \vec{\Omega}\right)=\frac{d\,N_e}{ds}-\frac{n_e}{R}\vec{n}\,\vec{\Omega}, \tag{8}$$

so that replacing $n_e\vec{\Omega}$ by its value in terms of $\frac{\vec{t}}{N_e}$ leads to the following expression:

$$\vec{t}\,\frac{d}{ds}\left(n_e \vec{\Omega}\right)=\frac{d\,N_e}{ds}-\frac{\varepsilon'_\perp-\varepsilon'_{||}}{R\,N_e}\left(\vec{t}\,\vec{e}_{||}\right)\left(\vec{n}\,\vec{e}_{||}\right), \tag{9}$$

Since $N_e$ is a function depending on both space and direction, its curvilinear derivative is defined by $\frac{d\,N_e}{ds}=\vec{t}\,\overrightarrow{grad}_M N_e+\frac{\vec{n}}{R}\frac{\partial N_e}{\partial \vec{t}}$; but from Eq. (2) one has $\frac{\partial N_e}{\partial \vec{t}}=\frac{\varepsilon'_\perp-\varepsilon'_{||}}{N_e}\left(\vec{t}\,\vec{e}_{||}\right)\vec{e}_{||}$, from which it comes the tangential projection of the Fermat's principle:

$$\vec{t}\,\frac{d}{ds}\left(n_e\vec{\Omega}\right)=\vec{t}\,\overrightarrow{grad}_M N_e\,, \tag{10}$$

Hence the uniaxial Fermat's equation multiplied by the unit tangent to the trajectory is identically verified like in isotropic media.

By definition of the unit wave vector $\vec{\Omega}=\frac{1}{N_e n_e}\left[\varepsilon'_{||}\,\vec{t}+\left(\varepsilon'_{\perp}-\varepsilon'_{||}\right)\left(\vec{t}\,\vec{e}_{||}\right)\vec{e}_{||}\right]$, it easily comes

$$\frac{d}{ds}\left(n_e\vec{\Omega}\,\vec{n}\right)=n_e\vec{\Omega}\left(\tau\,\vec{b}-\frac{\vec{t}}{R}\right)+\vec{n}\,\overrightarrow{grad}\,N_e\,, \tag{11}$$

where $\tau$ is the torsion of the trajectory and $\vec{b}$ the associated torsion unit vector, from which one has

$$\begin{aligned}\vec{n}\,\overrightarrow{grad}\,N_e=&\frac{d}{ds}\left[\frac{\varepsilon'_{\perp}-\varepsilon'_{||}}{N_e}\left(\vec{t}\,\vec{e}_{||}\right)\left(\vec{n}\,\vec{e}_{||}\right)\right]\\ &-\frac{1}{N_e}\left\{\tau\left(\varepsilon'_{\perp}-\varepsilon'_{||}\right)\left(\vec{t}\,\vec{e}_{||}\right)\left(\vec{b}\,\vec{e}_{||}\right)-\frac{1}{R}\left[\varepsilon'_{||}+\left(\varepsilon'_{\perp}-\varepsilon'_{||}\right)\left(\vec{t}\,\vec{e}_{||}\right)^2\right]\right\},\end{aligned} \tag{12}$$

Hence, by definition of the ray refractive index, and since $\frac{\varepsilon'_{\perp}-\varepsilon'_{||}}{N_e}\left(\vec{t}\,\vec{e}_{||}\right)\vec{e}_{||}=\frac{\partial N_e}{\partial\vec{t}}$, the previous equation reduces to

$$\frac{N_e}{R}=\vec{n}\,\overrightarrow{grad}\,N_e+\frac{\vec{t}}{R}\frac{\partial N_e}{\partial\vec{t}}-\vec{n}\,\frac{d}{ds}\left(\frac{\partial N_e}{\partial\vec{t}}\right), \tag{13}$$

from which it easily comes $\frac{1}{R}\left(N_e-\vec{t}\,\frac{\partial N_e}{\partial\vec{t}}\right)=\frac{\varepsilon'_{||}}{R\,N_e}=\vec{n}\left[\overrightarrow{grad}\,N_e-\frac{d}{ds}\left(\frac{\partial N_e}{\partial\vec{t}}\right)\right]$; then, using the curve derivative along a path and the definition of the ray refractive index leads to:

$$\frac{d}{ds}\left(\frac{\partial N_e}{\partial\vec{t}}\right)=\overrightarrow{grad}\left(\vec{t}\,\frac{\partial N_e}{\partial\vec{t}}\right)+\frac{\varepsilon'_{\perp}-\varepsilon'_{||}}{R}\left[\frac{\partial}{\partial\vec{t}}\left(\frac{\vec{t}\,\vec{e}_{||}}{N_e}\right)\otimes\vec{e}_{||}\right]\vec{n}\,, \tag{14}$$

with $\frac{\partial}{\partial \vec{t}}\left(\frac{\vec{t}\,\vec{e}_{||}}{N_e}\right)=\frac{\varepsilon'_{||}}{N_e^3}\vec{e}_{||}$, so that $\frac{d}{ds}\left(\frac{\partial N_e}{\partial \vec{t}}\right)=\vec{grad}\left(\vec{t}\frac{\partial N_e}{\partial \vec{t}}\right)+\frac{\varepsilon'_{||}\left(\varepsilon'_{\perp}-\varepsilon'_{||}\right)}{R\,N_e^3}\left(\vec{e}_{||}\otimes\vec{e}_{||}\right)\vec{n}$. Hence one deduces from what precedes that:

$$\frac{\varepsilon'_{||}}{R\,N_e}\left[1+\frac{\varepsilon'_{\perp}-\varepsilon'_{||}}{N_e^2}\vec{n}\left(\vec{e}_{||}\otimes\vec{e}_{||}\right)\vec{n}\right]=\vec{n}\,\vec{grad}\left(N_e-\vec{t}\frac{\partial N_e}{\partial \vec{t}}\right)=\vec{n}\,\vec{grad}\left(\frac{\varepsilon'_{||}}{N_e}\right), \qquad (15)$$

with $\vec{n}\left(\vec{e}_{||}\otimes\vec{e}_{||}\right)\vec{n}=\left(\vec{n}\,\vec{e}_{||}\right)^2$ : for an isotropic medium, Eq. (15) reduces to $\frac{1}{R}=\vec{n}\,\vec{grad}\left[Log\left(\sqrt{\varepsilon'}\right)\right]$ which is the classical normal form of the isotropic Fermat's principle. Doing so for the torsion projection gives, after calculating the curvilinear derivative of $n_e\vec{\Omega}\,\vec{n}$, to:

$$\vec{b}\,\vec{grad}\,N_e=\frac{d}{ds}\left[\frac{\varepsilon'_{\perp}-\varepsilon'_{||}}{N_e}\left(\vec{t}\,\vec{e}_{||}\right)\left(\vec{b}\,\vec{e}_{||}\right)\right]-\frac{\tau\left(\varepsilon'_{\perp}-\varepsilon'_{||}\right)}{N_e}\left(\vec{t}\,\vec{e}_{||}\right)\left(\vec{n}\,\vec{e}_{||}\right), \qquad (16)$$

that is, from what precedes:

$$0=\vec{b}\,\vec{grad}\,N_e+\tau\frac{\partial N_e}{\partial \vec{t}}\vec{n}-\frac{d}{ds}\left(\frac{\partial N_e}{\partial \vec{t}}\vec{b}\right), \qquad (17)$$

But $\tau\frac{\partial N_e}{\partial \vec{t}}\vec{n}-\frac{d}{ds}\left(\frac{\partial N_e}{\partial \vec{t}}\vec{b}\right)=-\vec{b}\frac{d}{ds}\left(\frac{\partial N_e}{\partial \vec{t}}\right)$, from which one obtains:

$$\frac{\varepsilon'_{||}\left(\varepsilon'_{\perp}-\varepsilon'_{||}\right)}{R\,N_e^3}\vec{b}\left(\vec{e}_{||}\otimes\vec{e}_{||}\right)\vec{n}=\vec{b}\left(\vec{grad}\,N_e-\vec{t}\frac{\partial N_e}{\partial \vec{t}}\right)=\vec{b}\,\vec{grad}\left(\frac{\varepsilon'_{||}}{N_e}\right), \qquad (18)$$

with $\vec{b}\left(\vec{e}_{||}\otimes\vec{e}_{||}\right)\vec{n}=\left(\vec{b}\,\vec{e}_{||}\right)\left(\vec{n}\,\vec{e}_{||}\right)$: for an isotropic medium, Eq. (15) reduces to $0=\vec{b}\,\vec{grad}\left[Log\left(\sqrt{\varepsilon'}\right)\right]$ which is the classical equation for torsion projection of the isotropic Fermat's principle. Finally the two developed normal equations of the uniaxial Fermat's principle are rewritten under the fundamental form:

$$\frac{1}{R}\frac{1-(1-\eta)\left(\vec{b}\,\vec{e_{||}}\right)^2}{1+(1-\eta)\left(\vec{t}\,\vec{e_{||}}\right)^2}=\vec{n}\,\vec{grad}\left\{Log\left[\frac{\eta\sqrt{\varepsilon'_\perp}}{\sqrt{\eta+(1-\eta)\left(\vec{t}\,\vec{e_{||}}\right)^2}}\right]\right\}$$
$$\frac{1-\eta}{R}\frac{\left(\vec{b}\,\vec{e_{||}}\right)\left(\vec{n}\,\vec{e_{||}}\right)}{1+(1-\eta)\left(\vec{t}\,\vec{e_{||}}\right)^2}=\vec{b}\,\vec{grad}\left\{Log\left[\frac{\eta\sqrt{\varepsilon'_\perp}}{\sqrt{\eta+(1-\eta)\left(\vec{t}\,\vec{e_{||}}\right)^2}}\right]\right\}, \quad (19)$$

where $\eta$ is the anisotropy factor, such that if $\eta>1$ then the crystal is positive, and if $\eta<1$ the crystal is negative; for homogeneous non isotropic media, $\vec{grad}\left\{Log\left[\frac{\eta\sqrt{\varepsilon'_\perp}}{\sqrt{\eta+(1-\eta)\left(\vec{t}\,\vec{e_{||}}\right)^2}}\right]\right\}=\vec{0}$ , so that assuming $\frac{1}{R}\neq 0$ leads to $\left(\vec{b}\,\vec{e_{||}}\right)^2=\frac{1}{1-\eta}\Rightarrow\eta<1$ and $\vec{n}\,\vec{e_{||}}=0$, from which one obtains $\left(\vec{t}\,\vec{e_{||}}\right)^2=1-\left(\vec{b}\,\vec{e_{||}}\right)^2=-\frac{\eta}{1-\eta}$ which is impossible with $\eta<1$; then for homogeneous anisotropic media, the ray trajectory is such that $\frac{1}{R}=0$ and is then a straight line like for isotropic media

*II-2 – Determination of the plane ray paths inside an uniaxial parallel plane slab of depth d and infinite extension*

One defines the *z* coordinate along the geometrical axis $\vec{e_z}$ of the slab with $z\in[0,d]$ and the *x*, *y* coordinates such that the $\vec{e_y}$ axis is orthogonal to the plane $\left(\vec{e_x},\vec{e_z}\right)$ which contains both the geometrical axis $\vec{e_z}$ and the optical axis $\vec{e_{||}}$; because of the infinite extension of the slab in directions $\vec{e_x}$ and $\vec{e_y}$, all the physical quantities are depending only on *z*, so that $\vec{grad}=\frac{\partial}{\partial z}\vec{e_z}$; let us introduce the basis $\left(\vec{e_X},\vec{e_Y},\vec{e_Z}=\vec{e_{||}}\right)$ associated to the coordinates *X*, *Y* and *Z*, with:

$$\vec{e}_X = \cos\alpha\,\vec{e}_x - \sin\alpha\,\vec{e}_z \qquad X = x\cos\alpha - z\sin\alpha \qquad x = X\cos\alpha + Z\sin\alpha$$
$$\vec{e}_Y = \vec{e}_y \qquad Y = y \quad \Leftrightarrow \quad y = Y$$
$$\vec{e}_Z = \sin\alpha\,\vec{e}_x + \cos\alpha\,\vec{e}_z \qquad Z = x\sin\alpha + z\cos\alpha \qquad z = -X\sin\alpha + Z\cos\alpha$$

where $\alpha$ is the angle between the optical and geometrical axis of the slab, hereafter supposed constant for convenient calculations; let us consider now a ray path characterized by $X = X(Z)$ and $Y = Y(Z)$, with $ds^2 = dZ^2\left[1+\left(\frac{dX}{dZ}\right)^2+\left(\frac{dY}{dZ}\right)^2\right]$ the curvilinear abscissa element; noting $U = \dot{X} = \frac{dX}{dZ}$ and $V = \dot{Y} = \frac{dY}{dZ}$, it obviously comes for the unit tangent vector to the path in the optical basis $\left(\vec{e}_X, \vec{e}_Y, \vec{e}_Z\right)$,

$$\vec{t} = \frac{d\vec{M}}{ds} = \frac{1}{\sqrt{1+U^2+V^2}}\begin{pmatrix} U \\ V \\ 1 \end{pmatrix} = \begin{pmatrix} \sin\xi\cos\varphi \\ \sin\xi\sin\varphi \\ \cos\xi \end{pmatrix}$$

, where the two angles $\xi$ and $\varphi$ have obvious signification; then the local curvature radius of the ray path is $R = \frac{1}{\cos\xi\sqrt{\dot{\xi}^2+\dot{\varphi}^2\sin^2\xi}}$, while the normal and torsion unit vectors of the direct Frenet's trihedron of the path are defined by

$$\vec{n} = \frac{1}{\sqrt{\dot{\xi}^2+\dot{\varphi}^2\sin^2\xi}}\left[\dot{\xi}\begin{pmatrix} \cos\xi\cos\varphi \\ \cos\xi\sin\varphi \\ -\sin\xi \end{pmatrix} + \dot{\varphi}\sin\xi\begin{pmatrix} -\sin\varphi \\ \cos\varphi \\ 0 \end{pmatrix}\right],$$

and $\vec{b} = \frac{1}{\sqrt{\dot{\xi}^2+\dot{\varphi}^2\sin^2\xi}}\left[\dot{\xi}\begin{pmatrix} -\sin\varphi \\ \cos\varphi \\ 0 \end{pmatrix} - \dot{\varphi}\sin\xi\begin{pmatrix} \cos\xi\cos\varphi \\ \cos\xi\sin\varphi \\ -\sin\xi \end{pmatrix}\right]$; hence the two equations (19) can be rewritten under the equivalent form:

$$\left(\frac{\dot{\xi}^2}{\cos^2\xi+\eta\sin^2\xi}+\dot{\varphi}^2\sin^2\xi\right)\cos\xi = \left(\dot{\xi}\cos\xi\cos\varphi - \dot{\varphi}\sin\xi\sin\varphi\right)\frac{\partial F}{\partial X} - \dot{\xi}\sin\xi\frac{\partial F}{\partial Z},$$
$$\frac{(1-\eta)\dot{\xi}\dot{\varphi}\cos\xi\sin^3\xi}{\cos^2\xi+\eta\sin^2\xi} = \left(\dot{\xi}\sin\varphi + \dot{\varphi}\sin\xi\cos\xi\cos\varphi\right)\frac{\partial F}{\partial X} - \dot{\varphi}\sin^2\xi\frac{\partial F}{\partial Z} \qquad (20)$$

with $F=Log\left(\frac{\eta\sqrt{\varepsilon'_\perp}}{\sqrt{cos^2\xi+\eta\, sin^2\xi}}\right)$, for $\frac{\partial}{\partial Y}=0$; in the case where $\dot{\xi}=0$, that is $\xi$ = *constant*, it comes $\dot{\varphi} sin\xi\left(\dot{\varphi} sin\xi\, cos\xi+sin\varphi\frac{\partial F}{\partial X}\right)=0$ and $\dot{\varphi} sin\xi\left(cos\xi\, cos\varphi\frac{\partial F}{\partial X}-sin\xi\frac{\partial F}{\partial Z}\right)=0$; if $\xi$ = *0*, the ray is a straight line parallel to the optical axis, and if $\dot{\varphi}=0$ the ray is a straight line of constant direction $(\varphi,\xi)$; one supposes $\dot{\varphi} sin\xi\neq 0$, then $\dot{\varphi} sin\xi\, cos\xi=sin\varphi\, sin\alpha\frac{\partial F}{\partial z}$ and $-cos\xi\, cos\varphi\, sin\alpha=sin\xi\, cos\alpha$ since $\frac{\partial}{\partial X}=-sin\alpha\frac{\partial}{\partial z}$ and $\frac{\partial}{\partial Z}=cos\alpha\frac{\partial}{\partial z}$; the 2nd equation leads to $cos\varphi=-\frac{tg\,\xi}{tg\,\alpha}=constant$ so that $\dot{\varphi}=0$, from which one deduces that the only possible paths for which $\dot{\xi}=0$ are straight lines; ones studies now curves with $\dot{\xi}\neq 0$ and $\dot{\varphi}=0$, so that it comes for the two normal Fermat's equations $\frac{\dot{\xi}\, cos\xi}{cos^2\xi+\eta\, sin^2\xi}=-(sin\alpha\, cos\xi\, cos\varphi+cos\alpha\, sin\xi)\frac{\partial F}{\partial z}$ and $sin\varphi\, sin\alpha\frac{\partial F}{\partial z}=0$; from the 2nd relation one must have $\alpha$ = *0* and/or $\varphi$ = *0*: if $\alpha$ = *0*, the optical axis is the geometric axis so that the system is non varying by rotation around the axis and all the ray paths are plane curves; on the other hand, if $\alpha\neq 0$, the only plane curves are the ones for which $\varphi$ = *0*, that is paths in the plane $\left(\vec{e_x},\vec{e_z}\right)$.

II-2a: case $\alpha$ = *0.*

When the optical axis is the geometrical axis, all the ray paths are plane curves so that one may choose $\varphi$ = *0*, and since $\alpha$ = *0*, one has $\frac{\partial}{\partial X}=0$ and $\frac{\partial}{\partial Z}=\frac{\partial}{\partial z}$, from which the trajectory equation reduces to:

$$\frac{\xi'\, cos\xi}{cos^2\xi+\eta\, sin^2\xi}=-sin\xi\frac{\partial}{\partial z}\left[Log\left(\frac{\eta\sqrt{\varepsilon'_\perp}}{\sqrt{cos^2\xi+\eta\, sin^2\xi}}\right)\right], \tag{21}$$

with $\xi'=\dot{\xi}=\frac{d\xi}{dz}$, or expressed with the previously introduced variable $u=\frac{dx}{dz}=tg\,\xi$:

$$\frac{u'}{\eta u^2+1}=-u\frac{\partial}{\partial z}\left[Log\left(\frac{\eta\sqrt{\varepsilon'_\perp}}{\sqrt{\eta u^2+1}}\right)\right], \tag{22}$$

For non zero $u$ solutions, since $u$ is not an explicit function of $z$ and both $\eta$ and $\varepsilon'_{\perp}$ are not explicit functions of $u$, the previous equation can equivalently be rewritten as:

$$u'\frac{\partial}{\partial u}\left[Log\left(\frac{\eta u\sqrt{\varepsilon'_{\perp}}}{\sqrt{\eta u^2+1}}\right)\right]+\frac{\partial}{\partial z}\left[Log\left(\frac{\eta u\sqrt{\varepsilon'_{\perp}}}{\sqrt{\eta u^2+1}}\right)\right]=\frac{d}{dz}\left[Log\left(\frac{\eta u\sqrt{\varepsilon'_{\perp}}}{\sqrt{\eta u^2+1}}\right)\right]=0\,, \tag{23}$$

for which the immediate solution is $\frac{\varepsilon'_{||}u}{\sqrt{\varepsilon'_{||}u^2+\varepsilon'_{\perp}}}=\frac{\varepsilon'^{(0)}_{||}u_0}{\sqrt{\varepsilon'^{(0)}_{||}u_0^2+\varepsilon'^{(0)}_{\perp}}}$, where the (0) subscript indicates initial conditions on the ray path, given by:

$$\frac{1}{\sqrt{1+x'^2(z_0)}}\begin{bmatrix}x'(z_0)\\1\end{bmatrix}=\frac{1}{\sqrt{1+u_0^2}}\begin{pmatrix}u_0\\1\end{pmatrix}=\begin{pmatrix}sin\ \xi_0\\cos\ \xi_0\end{pmatrix}, \tag{24}$$

where $\xi_0$ is the angle between the geometrical (or here optical) axis of the slab and the unit initial tangent to the ray path, that is $u_0=tg\,\xi_0$, so that the 1st order differential equation of the path is $\frac{\varepsilon'_{||}(z)x'(z)}{\sqrt{\varepsilon'_{||}(z)x'^2(z)+\varepsilon'_{\perp}(z)}}=\frac{\varepsilon'_{||}(z_0)sin\xi_0}{N_e(z_0,\xi_0)}$; at each point located on the path, there is an angle $\xi(z)$ between the slab's axis and the unit tangent vector, so that $u(z)=x'(z)=tg\,\xi(z)$ : hence, the previous relation leads to:

$$\frac{\varepsilon'_{||}(z)sin[\xi(z)]}{N_e[z,\xi(z)]}=\frac{\varepsilon'_{||}(z_0)sin\xi_0}{N_e(z_0,\xi_0)}\,, \tag{25}$$

which is a generalised form of the Descartes' law with the directional effective refractive index $\hat{n}[z,\xi(z)]=\frac{\varepsilon'_{||}(z)}{N_e[z,\xi(z)]}$; then an infinite extended slab of finite depth filled in with uniaxial crystal has an isotropic-like behaviour when its optical axis coincides with its geometrical axis; it is easy then to obtain the 1st integral of the ray path, given by:

$$x(z)-x(z_0)=\hat{n}_0\ sin\xi_0\int_{u=z_0}^{z}\sqrt{\frac{\varepsilon'_{\perp}(u)}{\varepsilon'_{||}(u)}}\frac{du}{\sqrt{\varepsilon'_{||}(u)-\hat{n}_0^2 sin^2\xi_0}}\,, \tag{26}$$

with $\hat{n}_0=\frac{\varepsilon'_{||}(z_0)}{N_e(z_0,\xi_0)}$, the curvilinear abscissa element being $\frac{ds}{dz}=\sqrt{\frac{\varepsilon'^2_{||}(z)+\left[\varepsilon'_\perp(z)-\varepsilon'_{||}(z)\right]\hat{n}_0^2\sin^2\xi_0}{\varepsilon'_{||}(z)\left[\varepsilon'_{||}(z)-\hat{n}_0^2\sin^2\xi_0\right]}}$ to compare with the isotropic case $\frac{ds}{dz}=\frac{n(z)}{\sqrt{n^2(z)-n_0^2\sin^2\xi_0}}$; it is to observe that there is a solution if and only if $\sqrt{\varepsilon'_{||}(z)}>\hat{n}_0\sin\xi_0$, which is the generalisation of the isotropic condition $n(z)>n_0\sin\xi_0$: hence, like for isotropic media, internal total reflexions may occur on the extraordinary ray paths.

II-2b: case $\alpha\neq 0$.

When the optical axis differs from the geometrical axis, the ray paths are generally non plane curves, so that we shall restrict our study to the plane ray paths located in the plane $\left(\vec{e}_x,\vec{e}_z\right)$, for which the path integral is, when using the $U$ variable $U=tg\,\xi\neq 0$ :

$$\frac{\dot{U}}{\eta U^2+1}=\frac{\partial}{\partial X}\left[Log\left(\frac{\eta\sqrt{\varepsilon'_\perp}}{\sqrt{\eta U^2+1}}\right)\right]-U\frac{\partial}{\partial Z}\left[Log\left(\frac{\eta\sqrt{\varepsilon'_\perp}}{\sqrt{\eta U^2+1}}\right)\right], \qquad (27)$$

$\xi$ being the angle between the optical axis and the tangent to the path; for zero $U$ solutions, one has from Eq. (20) $\dot{U}=\frac{\partial}{\partial X}\left[Log\left(\eta\sqrt{\varepsilon'_\perp}\right)\right]$; but $u=x'=\frac{dx}{dz}=\frac{U\cos\alpha+\sin\alpha}{\cos\alpha-U\sin\alpha}$ and $u'=\frac{\dot{U}}{\left(\cos\alpha-U\sin\alpha\right)^3}$, so that if $U=0$, then $u=tg\,\alpha$ and $\dot{U}=u'\cos^3\alpha$; reminding that $\frac{\partial}{\partial X}=-\sin\alpha\frac{\partial}{\partial z}$ for an one-dimensional system depending only on *z*, gives the local behaviour of the path for zero *U* solutions:

$$u=\frac{dx}{dz}=tg\,\alpha \qquad u'=\frac{d^2x}{dz^2}=-\frac{\sin\alpha}{\cos^3\alpha}\frac{d}{dz}\left[Log\left(\frac{\varepsilon'_{||}}{\sqrt{\varepsilon'_\perp}}\right)\right], \qquad (28)$$

Noting $\hat{\xi}$ the angle such that $u=tg\,(\alpha+\xi)=tg\,\hat{\xi}$, so that one shall say that a ray goes 1°) from left to right for growing *z* when $\hat{\xi}\in\left[0,\frac{\pi}{2}\right]$, 2°) from left to right for decreasing *z* when $\hat{\xi}\in\left[\frac{\pi}{2},\pi\right]$, 3°) from right to left for decreasing *z* when $\hat{\xi}\in\left[\pi,\frac{3\pi}{2}\right]$, and 4°) from right to left for increasing *z* when $\hat{\xi}\in\left[\frac{3\pi}{2},2\pi\right]$; in the general case where $U>0$, since $\frac{1}{U\left(\eta U^2+1\right)}=\frac{\partial}{\partial U}\left[Log\left(\frac{\eta U\sqrt{\varepsilon'_\perp}}{\sqrt{\eta U^2+1}}\right)\right]$, $\frac{\partial}{\partial X}=-sin\,\alpha\frac{\partial}{\partial z}$ and $\frac{\partial}{\partial Z}=cos\,\alpha\frac{\partial}{\partial z}$ for one-dimensional systems, it comes for the differential equation of the ray:

$$\dot{U}\frac{\partial}{\partial U}\left[Log\left(\frac{\eta U\sqrt{\varepsilon'_\perp}}{\sqrt{\eta U^2+1}}\right)\right]+\left(cos\,\alpha+\frac{sin\ \alpha}{U}\right)\frac{\partial}{\partial z}\left[Log\left(\frac{\eta U\sqrt{\varepsilon'_\perp}}{\sqrt{\eta U^2+1}}\right)\right]=0\,, \tag{29}$$

The function $F=Log\left(\frac{\eta U\sqrt{\varepsilon'_\perp}}{\sqrt{\eta U^2+1}}\right)$ depending explicitly only on *z* and *U*, $\frac{dF}{dz}=\frac{\partial F}{\partial z}+U'\frac{\partial F}{\partial U}$, and since $\dot{U}=\left(cos\,\alpha-U\,sin\,\alpha\right)U'$, one has the equivalent ray path equation for $U\neq\frac{1}{tg\,\alpha}$:

$$\frac{d}{dz}\left[Log\left(\frac{\eta U\sqrt{\varepsilon'_\perp}}{\sqrt{\eta U^2+1}}\right)\right]=\left(1-\frac{u}{U}\right)\frac{\partial}{\partial z}\left[Log\left(\frac{\eta\sqrt{\varepsilon'_\perp}}{\sqrt{\eta U^2+1}}\right)\right], \tag{30}$$

Obviously, if $\alpha$ = *0*, one has $\frac{d}{dz}\left[Log\left(\frac{\eta u\sqrt{\varepsilon'_\perp}}{\sqrt{\eta u^2+1}}\right)\right]=0$ since *u* = *U*; in the general case, Eq. (20) can be rewritten as $\frac{U'}{u\left(\eta U^2+1\right)}+\frac{\partial}{\partial z}\left[Log\left(\frac{\eta\sqrt{\varepsilon'_\perp}}{\sqrt{\eta U^2+1}}\right)\right]=0$, such that a simple decomposition leads to:

$$\frac{1}{u\left(\eta U^2+1\right)}=\frac{1}{cos^2\alpha+\eta\,sin^2\alpha}\left[\frac{cos\,\alpha}{U\,cos\,\alpha+sin\,\alpha}-\frac{\eta U}{\eta U^2+1}-\frac{\left(1-\eta\right)sin\,\alpha\,cos\,\alpha}{\eta U^2+1}\right]$$, from which:

$$\frac{U'}{u\left(\eta U^2+1\right)}=\frac{U'}{cos^2\alpha+\eta\,sin^2\alpha}\frac{\partial}{\partial U}\left[Log\left(\frac{U\,cos\,\alpha+sin\,\alpha}{\sqrt{\eta U^2+1}}\right)-\frac{\left(1-\eta\right)sin\,\alpha\,cos\,\alpha}{\sqrt{\eta}}Arctg\left(\sqrt{\eta}\,U\right)\right], \tag{31}$$

if $U>-tg\,\alpha$; when $U<-tg\,\alpha$, $\frac{cos\,\alpha}{U\,cos\,\alpha+sin\,\alpha}<0$ so that one notes $U^*=-U>tg\,\alpha$, and:

$$\frac{\cos\alpha}{U\cos\alpha+\sin\alpha}-\frac{\eta U}{\eta U^2+1}-\frac{(1-\eta)\sin\alpha\cos\alpha}{\eta U^2+1}=-\frac{\cos\alpha}{U^*\cos\alpha-\sin\alpha}+\frac{\eta U^*}{\eta U^{*2}+1}-\frac{(1-\eta)\sin\alpha\cos\alpha}{\eta U^{*2}+1}$$
$$=\frac{\partial}{\partial U}\left\{Log\left[\frac{-(U\cos\alpha+\sin\alpha)}{\sqrt{\eta U^2+1}}\right]-\frac{(1-\eta)\sin\alpha\cos\alpha}{\sqrt{\eta}}Arctg\left(\sqrt{\eta}U\right)\right\}$$

hence one easily deduces that:

$$U'\frac{\partial}{\partial U}\left\{Log\left[\left(\frac{|U\cos\alpha+\sin\alpha|}{\sqrt{\eta U^2+1}}\right)^{\frac{1}{\cos^2\alpha+\eta\sin^2\alpha}}\right]-\frac{(1-\eta)\sin(2\alpha)Arctg\left(\sqrt{\eta}U\right)}{2\sqrt{\eta}\left(\cos^2\alpha+\eta\sin^2\alpha\right)}\right\}+\frac{\partial}{\partial z}\left[Log\left(\frac{\eta\sqrt{\varepsilon'_\perp}}{\sqrt{\eta U^2+1}}\right)\right]=0$$

then, using the formal equality $Arctg\left(\sqrt{\eta}U\right)=Log\left[\left(\frac{1+i\sqrt{\eta}U}{\sqrt{\eta U^2+1}}\right)^{-i}\right]$ and since $\frac{d}{dz}=\frac{\partial}{\partial z}+U'\frac{\partial}{\partial U}$, one obtains:

$$\frac{d}{dz}\left\{Log\left[\eta\left(\frac{|U\cos\alpha+\sin\alpha|}{\sqrt{\eta U^2+1}}\right)^{\frac{1}{\cos^2\alpha+\eta\sin^2\alpha}}\left(\frac{1+i\sqrt{\eta}U}{\sqrt{\eta U^2+1}}\right)^{i\frac{(1-\eta)\sin(2\alpha)}{2\sqrt{\eta}\left(\cos^2\alpha+\eta\sin^2\alpha\right)}}\sqrt{\varepsilon'_\perp}\right]\right\},\qquad(32)$$
$$=\frac{\partial}{\partial z}\left\langle Re\left\{Log\left[\left(1+i\sqrt{\eta}U\right)^{i\frac{(1-\eta)\sin\alpha}{\sqrt{\eta}\left(\cos\alpha-i\sqrt{\eta}\sin\alpha\right)}}\right]\right\}\right\rangle$$

so that employing the angular variable $\xi(z)$, angle between the optical axis and the tangent to the ray path at point z finally leads to:

$$\frac{d}{dz}\left\langle Log\left\{\eta\sqrt{\varepsilon'_\perp}\left[\frac{|\sin(\alpha+\xi)|}{\sqrt{\cos^2\xi+\eta\sin^2\xi}}e^{-\frac{1-\eta}{2\sqrt{\eta}}\sin(2\alpha)Arctg\left(\sqrt{\eta}\,tg\,\xi\right)}\right]^{\frac{1}{\cos^2\alpha+\eta\sin^2\alpha}}\right\}\right\rangle,\qquad(33)$$
$$=\frac{\partial}{\partial z}\left\{Log\left[\frac{e^{-\cos\alpha\,Arctg\left(\sqrt{\eta}\,tg\,\xi\right)}}{\sqrt{\cos^2\xi+\eta\sin^2\xi}^{\sqrt{\eta}\sin\alpha}}\right]^{\frac{(1-\eta)\sin\alpha}{\cos^2\alpha+\eta\sin^2\alpha}}\right\}$$

This fundamental 1st order equation has an immediate solution when the anisotropy factor $\eta$ is constant (which allows $\varepsilon'_\perp$ to be spatially variable), for the rhs member is 0; then the solution is the "generalized" Descartes' law, defined as:

$$\eta\sqrt{\varepsilon'_\perp}\left[\frac{|sin(\alpha+\xi)|}{\sqrt{cos^2\xi+\eta\, sin^2\xi}}\, e^{-\frac{1-\eta}{2\sqrt{\eta}}sin(2\alpha)Arctg\left(\sqrt{\eta}\, tg\,\xi\right)}\right]^{\frac{1}{cos^2\alpha+\eta\, sin^2\alpha}} = constant\,, \tag{34}$$

Obviously, for isotropic media, Eq. (34) leads to $\sqrt{\varepsilon'_\perp}\, sin\xi = constant$ which is the classical Descartes' law, and when the optical axis coincides with the geometrical one (i. e. $\alpha = 0$), it comes $\frac{\eta\sqrt{\varepsilon'_\perp}\, sin\xi}{\sqrt{cos^2\xi+\eta\, sin^2\xi}} = constant$ which is Eq. (25); we shall from now focus our study to uniaxial crystals for which $\varepsilon'_{||}(z) = \eta\,\varepsilon'_\perp(z)$ where $\eta$ = *constant*, so that the generalized Descartes' law remains applicable and constitutes the 1st order differential equation with $\frac{dx}{dz} = tg(\alpha+\xi)$.

From Eq. (34) one deduces the transcendent equation to solve in $\xi$ to obtain the complete extraordinary ray path inside the crystal:

$$\frac{|sin(\alpha+\xi)|\, e^{-\frac{1-\eta}{2\sqrt{\eta}}sin(2\alpha)Arctg\left(\sqrt{\eta}\, tg\,\xi\right)}}{\sqrt{cos^2\xi+\eta\, sin^2\xi}} = \sqrt{\frac{\varepsilon'_\perp(z_0)}{\varepsilon'_\perp(z)}}^{\,cos^2\alpha+\eta\, sin^2\alpha}\ \frac{|sin(\alpha+\xi_0)|\, e^{-\frac{1-\eta}{2\sqrt{\eta}}sin(2\alpha)Arctg\left(\sqrt{\eta}\, tg\,\xi_0\right)}}{\sqrt{cos^2\xi_0+\eta\, sin^2\xi_0}}\,, \tag{35}$$

where $z_0$ is the initial point of the path and $\xi_0$ the angle between the optical axis and the unit initial tangent to the trajectory; let us notice that for $\xi_0 = -\alpha$, that is for an initial tangent parallel to the slab's geometrical axis, the constant of the generalized Descartes' law is zero, so that the ray path remains a straight line parallel to the geometrical axis. The previous Eq. (35) shows a special behaviour for $\xi_0 = \frac{\pi}{2}[\pi]$, since $|tg\,\xi_0| = +\infty$; for $\xi_0 = \frac{\pi}{2} + h$ with $|h| \to 0$, a simple development leads to:

$$\lim_{h\to 0^+}\frac{|sin(\alpha+\xi_0)|\, e^{-\frac{1-\eta}{2\sqrt{\eta}}sin(2\alpha)Arctg\left(\sqrt{\eta}\, tg\,\xi_0\right)}}{\sqrt{cos^2\xi_0+\eta\, sin^2\xi_0}} = \frac{cos\,\alpha\, e^{\frac{\pi(1-\eta)}{4\sqrt{\eta}}sin(2\alpha)}}{\sqrt{\eta}} = S_1$$

$$\lim_{h\to 0^-}\frac{|sin(\alpha+\xi_0)|\, e^{-\frac{1-\eta}{2\sqrt{\eta}}sin(2\alpha)Arctg\left(\sqrt{\eta}\, tg\,\xi_0\right)}}{\sqrt{cos^2\xi_0+\eta\, sin^2\xi_0}} = \frac{cos\,\alpha\, e^{-\frac{\pi(1-\eta)}{4\sqrt{\eta}}sin(2\alpha)}}{\sqrt{\eta}} = S_2$$

with $\frac{S_1}{S_2}=e^{\frac{\pi(1-\eta)}{2\sqrt{\eta}}\sin(2\alpha)}$: one concludes that $S_1>S_2$ for a negative crystal (i. e. $\eta$ < 1), while $S_1<S_2$ for a positive crystal (i. e. $\eta$ > 1); hence the Descartes' law is not continuous at $\xi=\frac{\pi}{2}[\pi]$: the straight line perpendicular to the optical axis, representing the two directions $\xi=\frac{\pi}{2}$ (a ray travelling from left to right for decreasing *z*) and $\xi=\frac{3\pi}{2}$ (a ray travelling from right to left for increasing *z*), is a cut line of discontinuity, excepted for $\alpha=\frac{\pi}{2}$: if the initial tangent at the trajectory is exactly $\xi_0=\frac{\pi}{2}[\pi]$, the trajectory remains a straight line parallel to the cut line which is an internal line of total reflection: hence if the trajectory "reaches" a cut line at *z*, its tangent being almost parallel to the cut line, if $\xi(z)\to\left(\frac{\pi}{2}\right)^-$ or $\xi(z)\to\left(\frac{3\pi}{2}\right)^+$ (the incident trajectory is above the cut line), the reflected part of the trajectory remains above the cut line, but since $\varepsilon'_\perp=\varepsilon'_\perp(z)$, the reflected trajectory is no longer symmetric to the incident one relatively to the optical axis, and if $\xi(z)\to\left(\frac{\pi}{2}\right)^+$ or $\xi(z)\to\left(\frac{3\pi}{2}\right)^-$ (the incident trajectory is below the cut line), the reflected part of the trajectory remains below the cut line: the cut lines are the locations of possible "true" internal total reflections. Hence the point *z* locations at which a total true reflection may arise are solutions of $\varepsilon'_\perp(z)=C_i^{\frac{2}{\cos^2\alpha+\eta\sin^2\alpha}}$, where the two constants $C_1$ and $C_2$ are given by $C_i=\sqrt{\varepsilon'_\perp(z_0)}^{\cos^2\alpha+\eta\sin^2\alpha}\frac{\left|\sin(\alpha+\xi_0)\right|e^{-\frac{1-\eta}{2\sqrt{\eta}}\sin(2\alpha)Arctg\left(\sqrt{\eta}\,tg\,\xi_0\right)}}{S_i\sqrt{\cos^2\xi_0+\eta\sin^2\xi_0}}$, with $S_i=S_1$ for $\frac{\pi}{2}<\xi_0<\pi$ or $\pi<\xi_0<\frac{3\pi}{2}$, the admissible solution being such that $0<z<z_0$, and $S_i=S_2$ for $0<\xi_0<\frac{\pi}{2}$ or $\frac{3\pi}{2}<\xi_0<2\pi$, the admissible solution being such that $z_0<z<d$.

Noting $f(\xi;\alpha,\eta)=\frac{\left|\sin(\alpha+\xi)\right|e^{-\frac{1-\eta}{2\sqrt{\eta}}\sin(2\alpha)Arctg\left(\sqrt{\eta}\,tg\,\xi\right)}}{\sqrt{\cos^2\xi+\eta\sin^2\xi}}\geq 0$ the function in $\xi$ of period $\pi$, discontinuous at $\xi=\frac{\pi}{2}[\pi]$, with $\left|\sin(\alpha+\xi)\right|=\sin(\alpha+\xi)$ for $\alpha+\xi\in[0,\pi]$ and $f(\xi;\alpha,\eta)\underset{\alpha+\xi\to 0,\pi}{\to}0$; the f derivative calculation easily leads to $f'(\xi;\alpha,\eta)=\frac{\left(\cos^2\alpha+\eta\sin^2\alpha\right)\cos(\alpha+\xi)}{\left(\cos^2\xi+\eta\sin^2\xi\right)^{\frac{3}{2}}}e^{-\frac{1-\eta}{2\sqrt{\eta}}\sin(2\alpha)Arctg\left(\sqrt{\eta}\,tg\,\right)}$, also discontinuous at

$\xi=\frac{\pi}{2}[\pi]$, from which one deduces that at maximal value is reached at $\xi=\frac{\pi}{2}-\alpha$, with $f\left(\frac{\pi}{2}-\alpha;\alpha,\eta\right)=\frac{e^{-\frac{1-\eta}{2\sqrt{\eta}}\sin(2\alpha)Arctg\left(\frac{\sqrt{\eta}}{tg\,\alpha}\right)}}{\sqrt{\sin^2\alpha+\eta\cos^2\alpha}}=S_3$; hence, a pseudo total internal reflexion may occur at the point $z$ defined by $\varepsilon'_\perp(z)=C_3^{\frac{2}{\cos^2\alpha+\eta\sin^2\alpha}}$, where the constant $C_3$ is given by:

$$C_3=\sqrt{\varepsilon'_\perp(z_0)}^{\cos^2\alpha+\eta\sin^2\alpha}\left|\sin(\alpha+\xi_0)\right|\sqrt{\frac{\sin^2\alpha+\eta\cos^2\alpha}{\cos^2\xi_0+\eta\sin^2\xi_0}}\,e^{\frac{1-\eta}{2\sqrt{\eta}}\sin(2\alpha)Arctg\left[\frac{2\sqrt{\eta}\cos(\alpha+\xi_0)}{(1+\eta)\sin(\alpha+\xi_0)+(1-\eta)\sin(\alpha-\xi_0)}\right]}, \qquad (36)$$

Then, if there exists a given $z_M\in[0,d]$ such that $\varepsilon'_\perp(z_M)=C_3^{\frac{2}{\cos^2\alpha+\eta\sin^2\alpha}}$, there is at least one pseudo total reflexion on the extraordinary ray path and Eq. (35) has to be solved for all $z\in[z_0,z_M]$; note that if $z_M$ exists, if $z_M\in[z_0,d]$ the internal pseudo total reflexion appears for initial directions $\xi_0\in\left[-\alpha,\frac{\pi}{2}-\alpha\right]$, while if $z_M\in[0,z_0]$, the pseudo total reflexion appears for initial directions $\xi_0\in\left[\frac{\pi}{2}-\alpha,\pi-\alpha\right]$.

Note that $\frac{S_1}{S_3}=\frac{\cos\alpha\sqrt{\sin^2\alpha+\eta\cos^2\alpha}}{\sqrt{\eta}}e^{\frac{1-\eta}{2\sqrt{\eta}}\sin(2\alpha)\left[\pi-Arctg\left(\frac{tg\,\alpha}{\sqrt{\eta}}\right)\right]}$, so that $S_1=S_2=S_3$ if and only if $\alpha=0$; for a positive crystal, $\eta>1$ and $\frac{S_1}{S_3}<\frac{\cos\alpha\sqrt{\sin^2\alpha+\eta\cos^2\alpha}}{\sqrt{\eta}}<1$ for all $\alpha$: hence for a positive crystal one has $S_1<S_2$ and $S_1<S_3$; similarly, for a negative crystal, $\eta<1$ and $\frac{S_1}{S_3}>\frac{\cos\alpha\sqrt{\sin^2\alpha+\eta\cos^2\alpha}}{\sqrt{\eta}}>1$ for all $\alpha$, hence for a negative crystal one has $S_1>S_2$ and $S_1>S_3$; note also that whatever the positivity of the crystal is, $S_2<S_3$ since $f$ is continuous on the set $\left[0,\frac{\pi}{2}\right[$ and reaches its maximal (or minimal) value at $\frac{\pi}{2}-\alpha$, and since $f$ is strictly increasing on $\left[0,\frac{\pi}{2}-\alpha\right]$ one has $\sup_{\xi\in\left[0,\frac{\pi}{2}\right[} f(\xi;\alpha,\eta)=S_3\Rightarrow S_3>S_2$; the behaviour of the function $f$ is presented for a crystal such that $\varepsilon'_\perp=2.25$, and $\alpha=22.5°$; for the first case (Fig. 1a), the parameter of anisotropy is $\eta_1=2.$ (positive crystal), while for the 2nd case (Fig. 1b) the corresponding parameter is $\eta_2=0.5$ (negative crystal)

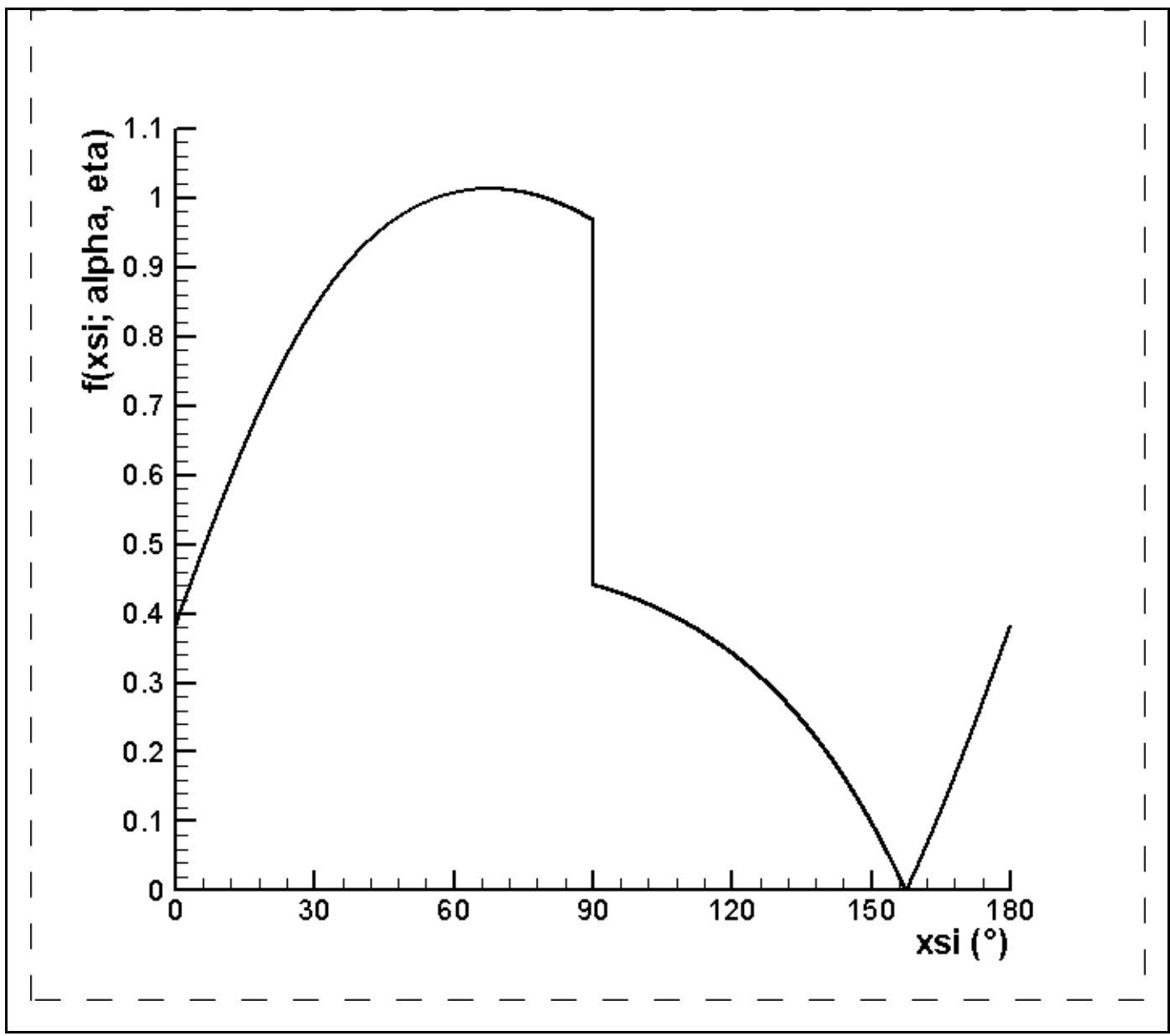

Fig. 1a: $\eta_1=2.$

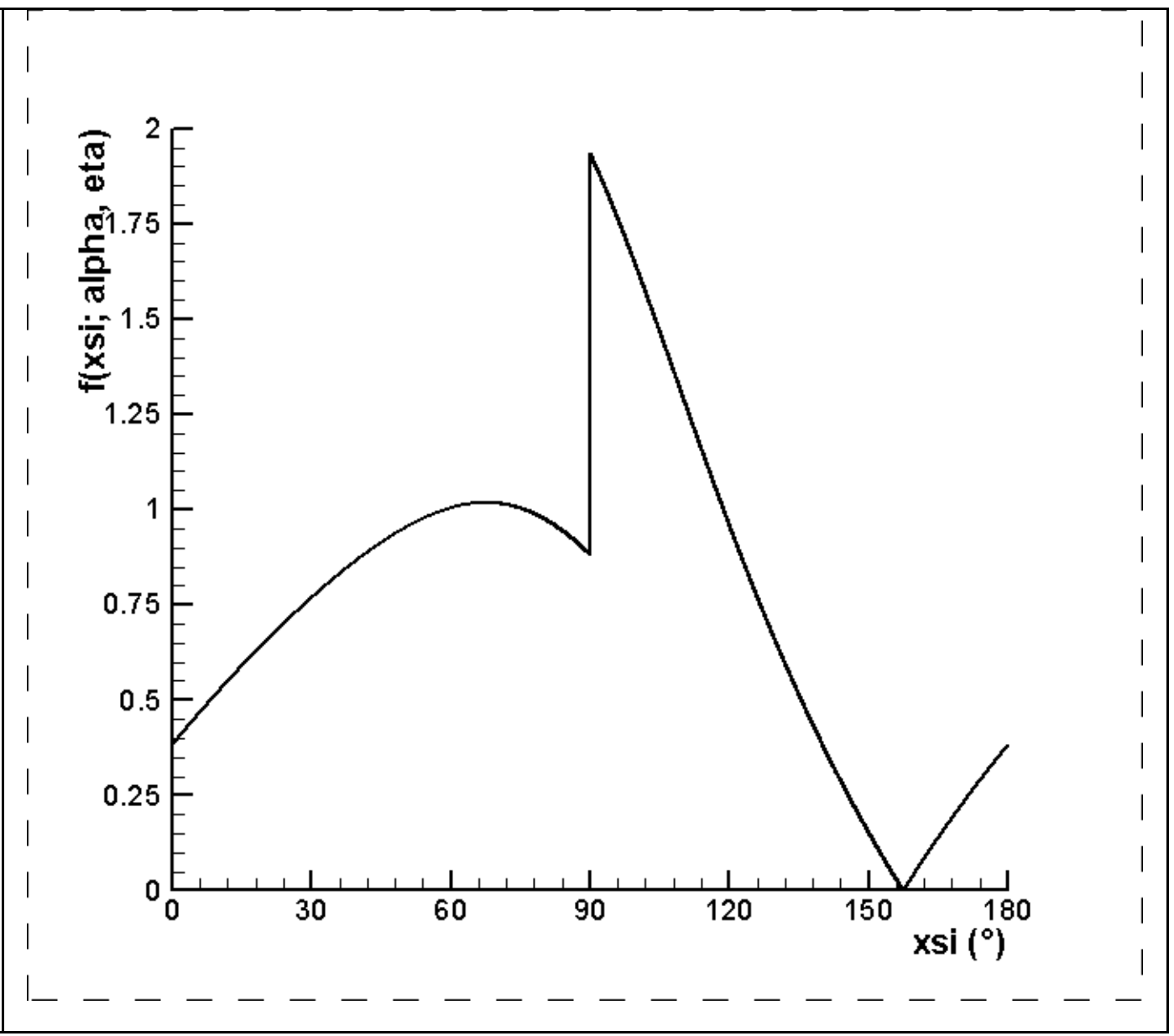

Fig. 1b: $\eta_2=0.5$

## III – REFLEXION AND TRANSMISSION PHENOMENA FOR OPTICAL RAYS AT AN INTERFACE BETWEEN TWO UNIAXIAL CRYSTALS

Let us now consider two homogeneous uniaxial crystals 1, located at $z < 0$, and 2, for $z > 0$, characterized by their optical tensors and axis such that $\alpha_i \in \left[0, \frac{\pi}{2}\right]$, both separated by a perfectly thin specular interface at $z = 0$. From what precedes, one shall examine the different possible situations that may arise when a physical optical extraordinary ray travelling in the crystal 1 falls down to the interface separating the two media, thanks to the generalised Descartes' law.

### III-1 Ordinary/ordinary reflexion/transmission phenomena:

This very well-known case is such that $\hat{\xi}_{1r} = \pi - \hat{\xi}_{1i}$ and $\sqrt{\varepsilon'_{\perp 2}}\, sin\ \hat{\xi}_{2t} = \sqrt{\varepsilon'_{\perp 1}}\, sin\ \hat{\xi}_{1i}$ where $\hat{\xi}_{1i}$, $\hat{\xi}_{1r}$ and $\hat{\xi}_{2t}$ denote the incident, reflected and transmitted angles between the geometrical axis and the associated rays; a total reflexion on the interface is possible if and only if $\sqrt{\varepsilon'_{\perp 2}} \le \sqrt{\varepsilon'_{\perp 1}}$, and in this case, the symmetric cone of total reflexion is defined by $\hat{\xi}_{1i} \in \left[ Arc\,sin\left(\sqrt{\frac{\varepsilon'_{\perp 2}}{\varepsilon'_{\perp 1}}}\right), \frac{\pi}{2}\right]$

### III-2 Extraordinary/extraordinary reflexion/transmission phenomena:

Since the physical interface at $z = 0$ is not a discontinuity line (i.e. a natural reflection line for rays in crystal 1), the incident rays shall obey at the impact point to the general reflection rules relatively to the impact discontinuity line of crystal 1, as illustrated one Fig. 2a below: the reflected angle at the impact point on the interface equals the incident one, relatively to the optical axis of crystal 1. It is to note that since $\xi_{1i}=\frac{\pi}{2}[\pi]$ is a discontinuity line, it is impossible for a ray to travel the discontinuity line; a major consequence is that the incident rays above the cut line cannot be reflected in crystal 1 since the virtual reflected ray is in crystal 2; one can then define a virtual interface inside crystal 1 which is reflected in the direction of the real interface at $\xi_{1r}=\frac{3\pi}{2}-\alpha_1$, for which obviously $\xi_{1i}=\frac{3\pi}{2}+\alpha_1$, so that all the incident rays (from right to left) between the virtual interface and the real interface, that is $\xi_{1i}\in\left]\frac{3\pi}{2}-\alpha_1,\frac{3\pi}{2}+\alpha_1\right[$, cannot be reflected in crystal 1 in a symmetric way relatively to the optical axis; similarly, for incident rays from left to right, the real interface is reflected in the virtual interface, from which one deduces that the reflection area for a given impact point on the real interface, is located between the left real interface and the virtual interface; note that the cut line of crystal 1 is always between the virtual interface and the right real interface: it may then exist an angular area of complete "virtual reflection" between the virtual interface and the cut line.

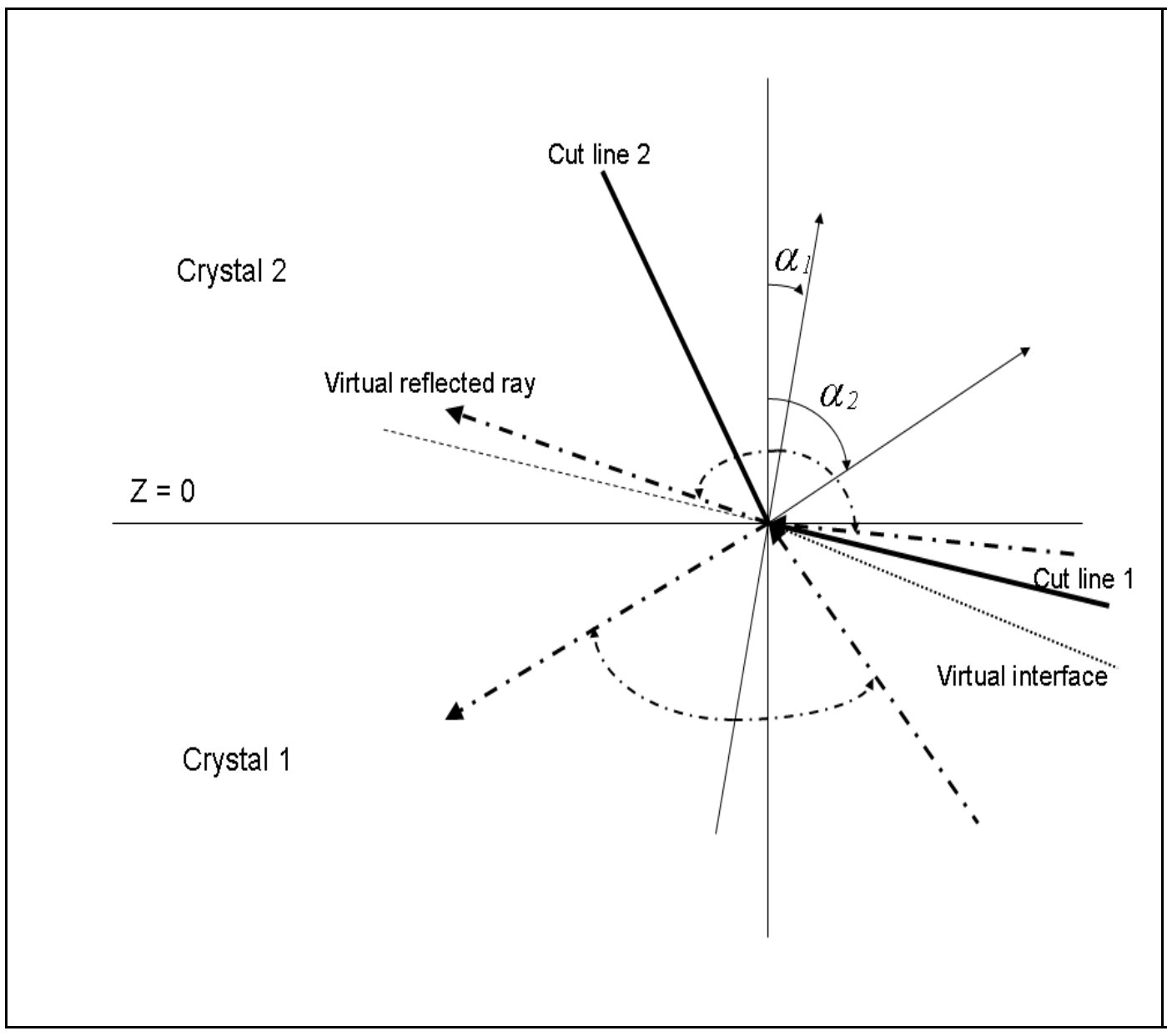

Fig. 2a: reflection rule

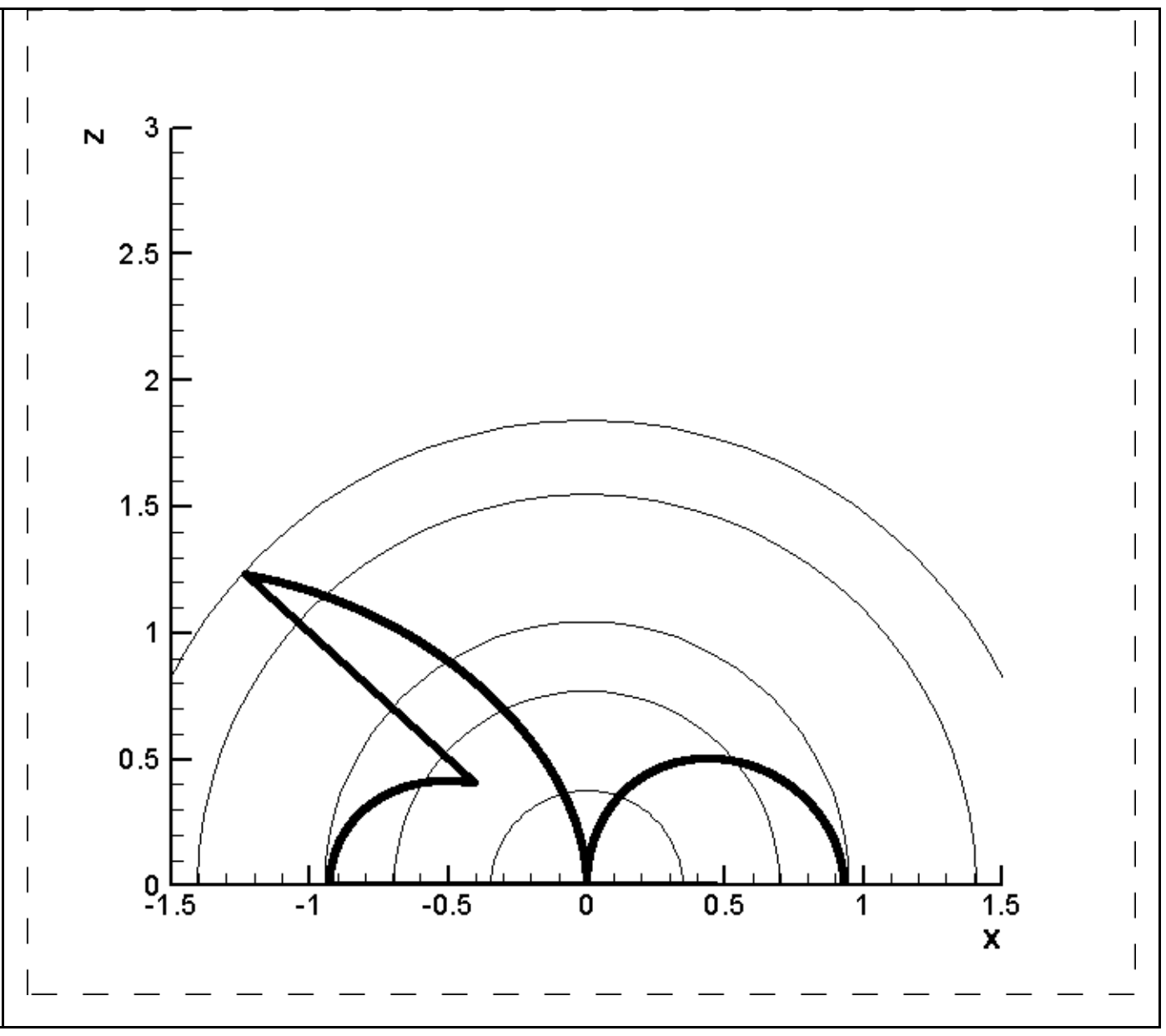

Fig. 2b: transmission rule

For the extraordinary transmission rule, using the generalized Descartes' law leads to, for the transmitted ray $\xi_{2t}$ when $\xi_{1i}\neq\frac{\pi}{2}$:

$$\eta_2\sqrt{\varepsilon'_{\perp 2}}\left[f\left(\xi_{2t};\alpha_2,\eta_2\right)\right]^{\frac{1}{\cos^2\alpha_2+\eta_2\sin^2\alpha_2}}=\eta_1\sqrt{\varepsilon'_{\perp 1}}\left[f\left(\xi_{1i};\alpha_1,\eta_1\right)\right]^{\frac{1}{\cos^2\alpha_1+\eta_1\sin^2\alpha_1}}, \tag{37}$$

or, noting $F_j\left(\hat{\xi}\right)=\eta_j\left[f\left(\xi;\alpha_j,\eta_j\right)\right]^{\frac{1}{\cos^2\alpha_j+\eta_j\sin^2\alpha_j}}$ with $\hat{\xi}=\xi+\alpha_j$ the angle between the ray and the geometrical axis, the equivalent sine function in the case of isotropic media:

$$\sqrt{\varepsilon'_{\perp 2}}\,F_2\left(\hat{\xi}_{2t}\right)=\sqrt{\varepsilon'_{\perp 1}}\,F_1\left(\hat{\xi}_{1i}\right), \tag{38}$$

<u>Particular cases:</u> in the two particular cases $\alpha_j=0$ or $\alpha_j=\frac{\pi}{2}$, $\sin\left(2\alpha_j\right)=0$ and :

$$f\left(\xi;\alpha,\eta\right)=\frac{\left|\sin\left(\alpha+\xi\right)\right|}{\sqrt{\cos^2\xi+\eta\sin^2\xi}}=\begin{cases}\dfrac{\left|\sin\hat{\xi}\right|}{\sqrt{\cos^2\hat{\xi}+\eta\sin^2\hat{\xi}}} & \text{for } \alpha=0\Rightarrow F\left(\hat{\xi}\right)=\dfrac{\eta\sin\hat{\xi}}{\sqrt{\cos^2\hat{\xi}+\eta\sin^2\hat{\xi}}}\\[2ex] \dfrac{\left|\sin\hat{\xi}\right|}{\sqrt{\sin^2\hat{\xi}+\eta\cos^2\hat{\xi}}} & \text{for } \alpha=\dfrac{\pi}{2}\Rightarrow F\left(\hat{\xi}\right)=\eta\left(\dfrac{\left|\sin\hat{\xi}\right|}{\sqrt{\sin^2\hat{\xi}+\eta\cos^2\hat{\xi}}}\right)^{\frac{1}{\eta}}\end{cases}$$

The function *f* being symmetric in those two cases, the left and right sides separated by the geometrical axis are equivalent and it is enough to consider only angles such that $\hat{\xi}\in\left[0,\frac{\pi}{2}\right]$ (i.e. incident rays from left to right) since function *F* is continuous on $\hat{\xi}\in\left]0,\frac{\pi}{2}\right]$ for both cases, with $F_{\alpha=0}\left(\hat{\xi}=0\right)=0$, $F_{\alpha=0}\left(\hat{\xi}=\frac{\pi}{2}\right)=\sqrt{\eta}$, $\lim_{\hat{\xi}\to 0}F_{\alpha=\frac{\pi}{2}}\left(\hat{\xi}\right)=0$ and $F_{\alpha=\frac{\pi}{2}}\left(\hat{\xi}=\frac{\pi}{2}\right)=\eta$, and we shall consider here the three distinct cases 1) $\alpha_1=\alpha_2=0$, 2) $\alpha_1=\alpha_2=\frac{\pi}{2}$, and 3) $\alpha_1=0$ and $\alpha_2=\frac{\pi}{2}$ (respectively $\alpha_1=\frac{\pi}{2}$ and $\alpha_2=0$)

* case $\alpha_1=\alpha_2=0$: this situation is extremely similar to a pure ordinary transmission/reflection, the transmission equation being $\frac{\eta_2\sqrt{\varepsilon'_{\perp 2}}\sin\hat{\xi}_{2t}}{\sqrt{\cos^2\hat{\xi}_{2t}+\eta_2\sin^2\hat{\xi}_{2t}}}=\frac{\eta_1\sqrt{\varepsilon'_{\perp 1}}\sin\hat{\xi}_{1i}}{\sqrt{\cos^2\hat{\xi}_{1i}+\eta_1\sin^2\hat{\xi}_{1i}}}$; hence the transmitted ray angle expressed by:

$$\sin\hat{\xi}_{2t}=\frac{\eta_1\sqrt{\varepsilon'_{\perp 1}}}{\eta_2\sqrt{\varepsilon'_{\perp 2}}}\frac{\sin\hat{\xi}_{1i}}{\sqrt{\cos^2\hat{\xi}_{1i}+\eta_1\left[1+\frac{\eta_1\left(1-\eta_2\right)}{\eta_2^2}\frac{\varepsilon'_{\perp 1}}{\varepsilon'_{\perp 2}}\right]\sin^2\hat{\xi}_{1i}}},$$

always exists for $\eta_1\varepsilon'_{\perp 1}<\eta_2\varepsilon'_{\perp 2}$, of transmitted cone upper boundary $\sin\hat{\xi}_{2m}=\sqrt{\dfrac{\eta_1\varepsilon'_{\perp 1}}{\eta_2^2\varepsilon'_{\perp 2}+(1-\eta_2)\eta_1\varepsilon'_{\perp 1}}}$, while a total reflection occurs when $\eta_1\varepsilon'_{\perp 1}\geq\eta_2\varepsilon'_{\perp 2}$ and $\hat{\xi}_{1i}>\hat{\xi}_{1m}=Arc\sin\left[\sqrt{\dfrac{\eta_2\varepsilon'_{\perp 2}}{\eta_1^2\varepsilon'_{\perp 1}+(1-\eta_1)\eta_2\varepsilon'_{\perp 2}}}\right]$; the reflected ray is symmetric to the incident one relatively to the geometrical axis of the system,

* case $\alpha_1=\alpha_2=\dfrac{\pi}{2}$: in this situation the transmission equation is:

$$\frac{\sin\hat{\xi}_{2t}}{\sqrt{\sin^2\hat{\xi}_{2t}+\eta_2\cos^2\hat{\xi}_{2t}}}=\left(\frac{\eta_1\sqrt{\varepsilon'_{\perp 1}}}{\eta_2\sqrt{\varepsilon'_{\perp 2}}}\right)^{\eta_2}\left(\frac{\sin\hat{\xi}_{1i}}{\sqrt{\sin^2\hat{\xi}_{1i}+\eta_1\cos^2\hat{\xi}_{1i}}}\right)^{\frac{\eta_2}{\eta_1}}$$

with $0<\left(\dfrac{\eta_1\sqrt{\varepsilon'_{\perp 1}}}{\eta_2\sqrt{\varepsilon'_{\perp 2}}}\right)^{\eta_2}\left(\dfrac{\sin\hat{\xi}_{1i}}{\sqrt{\sin^2\hat{\xi}_{1i}+\eta_1\cos^2\hat{\xi}_{1i}}}\right)^{\frac{\eta_2}{\eta_1}}\leq\left(\dfrac{\eta_1\sqrt{\varepsilon'_{\perp 1}}}{\eta_2\sqrt{\varepsilon'_{\perp 2}}}\right)^{\eta_2}$ and $0<\dfrac{\sin\hat{\xi}_{2t}}{\sqrt{\sin^2\hat{\xi}_{2t}+\eta_2\cos^2\hat{\xi}_{2t}}}\leq 1$; hence there is always a transmitted ray in the case where $\eta_1\sqrt{\varepsilon'_{\perp 1}}<\eta_2\sqrt{\varepsilon'_{\perp 2}}$, inside the transmitted cone of upper boundary $\sin\hat{\xi}_{2m}=\sqrt{\dfrac{\eta_2\left(\eta_1^2\varepsilon'_{\perp 1}\right)^{\eta_2}}{\left(\eta_2^2\varepsilon'_{\perp 2}\right)^{\eta_2}-(1-\eta_2)\left(\eta_1^2\varepsilon'_{\perp 1}\right)^{\eta_2}}}$, the transmitted ray angle being expressed by:

$$\sin\hat{\xi}_{2t}=\frac{\sqrt{\eta_2}\left(\eta_1\sqrt{\varepsilon'_{\perp 1}}\right)^{\eta_2}\left(\sin\hat{\xi}_{1i}\right)^{\frac{\eta_2}{\eta_1}}}{\sqrt{\left(\eta_2^2\varepsilon'_{\perp 2}\right)^{\eta_2}\left(\sin^2\hat{\xi}_{1i}+\eta_1\cos^2\hat{\xi}_{1i}\right)^{\frac{\eta_2}{\eta_1}}-(1-\eta_2)\left(\eta_1^2\varepsilon'_{\perp 1}\right)^{\eta_2}\left(\sin^2\hat{\xi}_{1i}\right)^{\frac{\eta_2}{\eta_1}}}},$$

while a pseudo total reflection (we denote a pseudo total reflection and not a real total reflection the case $\hat{\xi}_{2t}=\dfrac{\pi}{2}$ at the interface, since the real interface is not a natural line of reflection, except the case $\alpha_i=0$) occurs when $\eta_1\sqrt{\varepsilon'_{\perp 1}}\geq\eta_2\sqrt{\varepsilon'_{\perp 2}}$ and $\hat{\xi}_{1i}>\hat{\xi}_{1m}=Arc\sin\left[\sqrt{\dfrac{\eta_1\left(\eta_2^2\varepsilon'_{\perp 2}\right)^{\eta_1}}{\left(\eta_1^2\varepsilon'_{\perp 1}\right)^{\eta_1}-(1-\eta_1)\left(\eta_2^2\varepsilon'_{\perp 1}\right)^{\eta_1}}}\right]$; note that in this case no incident ray can be reflected inside the crystal 1 since the line of reflection is parallel to the geometrical axis, hence if $\eta_1\sqrt{\varepsilon'_{\perp 1}}\geq\eta_2\sqrt{\varepsilon'_{\perp 2}}$, the transmission is possible for $\hat{\xi}_{1i}\leq\hat{\xi}_{1m}$ and there no reflection in medium 1 and transmission in medium 2 for $\hat{\xi}_{1i}>\hat{\xi}_{1m}$: in the case of pure extraordinary reflection/transmission phenomena, the incident ray must be “virtually reflected” in medium 2, the direction of the “virtual reflected” ray in medium 2 being symmetric (relatively to the real interface) to the incident direction in medium 1; one may think that the incident ray must then be absorbed at the interface.

* case $\alpha_1=0$ and $\alpha_2=\frac{\pi}{2}$: in this situation the transmission equation is:

$$\frac{\sin\hat{\xi}_{2t}}{\sqrt{\sin^2\hat{\xi}_{2t}+\eta_2\cos^2\hat{\xi}_{2t}}}=\left(\frac{\eta_1\sqrt{\varepsilon'_{\perp 1}}}{\eta_2\sqrt{\varepsilon'_{\perp 2}}}\right)^{\eta_2}\left(\frac{\sin\hat{\xi}_{1i}}{\sqrt{\cos^2\hat{\xi}_{1i}+\eta_1\sin^2\hat{\xi}_{1i}}}\right)^{\eta_2}$$

hence it exists a transmitted cone of upper boundary determined by $\sin\hat{\xi}_{1i}=1$ for $\sqrt{\eta_1}\sqrt{\varepsilon'_{\perp 1}}<\eta_2\sqrt{\varepsilon'_{\perp 2}}$, the upper boundary being given by $\sin\hat{\xi}_{2m}=\sqrt{\frac{\eta_2\left(\eta_1\varepsilon'_{\perp 1}\right)^{\eta_2}}{\left(\eta_2^2\varepsilon'_{\perp 2}\right)^{\eta_2}-(1-\eta_2)\left(\eta_1\varepsilon'_{\perp 1}\right)^{\eta_2}}}$, and the transmitted ray angle being expressed by:

$$\sin\hat{\xi}_{2t}=\frac{\sqrt{\eta_2}\left(\eta_1\sqrt{\varepsilon'_{\perp 1}}\right)^{\eta_2}\left(\sin\hat{\xi}_{1i}\right)^{\eta_2}}{\sqrt{\left(\eta_2^2\varepsilon'_{\perp 2}\right)^{\eta_2}\left(\cos^2\hat{\xi}_{1i}+\eta_1\sin^2\hat{\xi}_{1i}\right)^{\eta_2}-(1-\eta_2)\left(\eta_1^2\varepsilon'_{\perp 1}\right)^{\eta_2}\left(\sin^2\hat{\xi}_{1i}\right)^{\eta_2}}}$$

There is a “true” total reflection for $\sin\hat{\xi}_{2t}=1$, leading to $\frac{\sin\hat{\xi}_{1i}}{\sqrt{\cos^2\hat{\xi}_{1i}+\eta_1\sin^2\hat{\xi}_{1i}}}=\frac{\eta_2\sqrt{\varepsilon'_{\perp 2}}}{\eta_1\sqrt{\varepsilon'_{\perp 1}}}$ or equivalently to $\hat{\xi}_{1i}>\hat{\xi}_{1m}=Arc\sin\left[\frac{\eta_2\sqrt{\varepsilon'_{\perp 2}}}{\sqrt{\eta_1^2\varepsilon'_{\perp 1}+(1-\eta_1)\eta_2^2\varepsilon'_{\perp 2}}}\right]$, possible if and only if $\sqrt{\eta_1}\sqrt{\varepsilon'_{\perp 1}}\geq\eta_2\sqrt{\varepsilon'_{\perp 2}}$; in this case, the reflected ray obviously verifies $\sin\hat{\xi}_{1r}=\sin\hat{\xi}_{1i}$

One considers now the most general case $\alpha_{i,j}\notin\left\{0,\frac{\pi}{2}\right\}$; the behaviour of function $f$ (or equivalently function $F$) in polar coordinates is presented on Fig. 2b for incident rays from left to right (i. e. $\hat{\xi}_{1i}\in\left[0,\frac{\pi}{2}\right]$) on the right side of the figure, and for incident rays from right to left (i. e. $\hat{\xi}_{1i}\in\left[\frac{3\pi}{2},\frac{3\pi}{2}+\alpha_1\right[\cup\left]\frac{3\pi}{2}+\alpha_1,2\pi\right]$) on the left side, where the function $F$ is non continuous at the given value $\hat{\xi}_{1i}=\frac{3\pi}{2}+\alpha_1$; this description is analogous to the classical Fresnel's diagram of transmission for uniaxial media, and shows here, contrarily to the classical electromagnetic description, that for right to left incident rays, there may or not exist a solution above the cut line (i.e. the line of natural reflection). From what precedes, function $F$ is continuously strictly increasing for $\hat{\xi}\in\left[0,\frac{\pi}{2}\right]$, so that $\varepsilon'_{\perp 1}<\varepsilon'_{\perp 2}\Rightarrow\hat{\xi}_{2t}<\hat{\xi}_{1i}$ and $\varepsilon'_{\perp 1}>\varepsilon'_{\perp 2}\Rightarrow\hat{\xi}_{2t}>\hat{\xi}_{1i}$: the transmission behaviour for incident rays from left to right is similar to the isotropic transmission behaviour; then it may exist a pseudo total reflection in crystal 1 (i. e. the transmitted ray is located on the real interface) for $\left|sin\left(\alpha_2+\xi_{2t}\right)\right|=1$, or

$$\sqrt{\varepsilon'_{\perp 1}}\,F_1\left(\hat{\xi}_{1i}\right)=\hat{\Omega}_2\sqrt{\varepsilon'_{\perp 2}}=\eta_2\sqrt{\varepsilon'_{\perp 2}}\left[\frac{e^{-\frac{1-\eta_2}{2\sqrt{\eta_2}}sin(2\alpha_2)Arctg\left(\frac{\sqrt{\eta_2}}{tg\,\alpha_2}\right)}}{\sqrt{sin^2\alpha_2+\eta_2 cos^2\alpha_2}}\right]^{\frac{1}{cos^2\alpha_2+\eta_2 sin^2\alpha_2}}$$

, if and only if $\hat{\Omega}_1\sqrt{\varepsilon'_{\perp 1}}>\hat{\Omega}_2\sqrt{\varepsilon'_{\perp 2}}$; let us remark that for $\alpha_i=0$ this leads to $\hat{\Omega}_i=\sqrt{\eta_i}$ and for $\alpha_i=\frac{\pi}{2}$ one has $\hat{\Omega}_i=\eta_i$, so that the general relation $\hat{\Omega}_1\sqrt{\varepsilon'_{\perp 1}}>\hat{\Omega}_2\sqrt{\varepsilon'_{\perp 2}}$ to have a pseudo total reflection is obviously verified for the three particular cases previously detailed; one notes similarly for the discontinuity line the associated values

$$\breve{\Omega}_i=\eta_i\left[\frac{cos\,\alpha_i\,e^{\frac{\pi(1-\eta_i)sin(2\alpha_i)}{4\sqrt{\eta_i}}}}{\sqrt{\eta_i}}\right]^{\frac{1}{cos^2\alpha_i+\eta_i sin^2\alpha_i}}$$ and $$\overset{\frown}{\Omega}_i=\eta_i\left[\frac{cos\,\alpha_i\,e^{-\frac{\pi(1-\eta_i)sin(2\alpha_i)}{4\sqrt{\eta_i}}}}{\sqrt{\eta_i}}\right]^{\frac{1}{cos^2\alpha_i+\eta_i sin^2\alpha_i}}$$

, with $\overset{\frown}{\Omega}_i<\hat{\Omega}_i$ for all crystals; then for positive crystals $\breve{\Omega}_i<\overset{\frown}{\Omega}_i<\hat{\Omega}_i$, while for negative crystals the order relation is $\overset{\frown}{\Omega}_i<\hat{\Omega}_i<\breve{\Omega}_i$; hence, since the maximal value of $F$ is reached at $\hat{\xi}=\frac{\pi}{2}$ for incident rays from left to right, one deduces that if $\hat{\Omega}_1\sqrt{\varepsilon'_{\perp 1}}>\hat{\Omega}_2\sqrt{\varepsilon'_{\perp 2}}$, it exists a left transmission cone whose left boundary $\hat{\xi}_{1m}$ is solution of $F_1\left(\hat{\xi}_{1m}\right)=\hat{\Omega}_2\sqrt{\frac{\varepsilon'_{\perp 2}}{\varepsilon'_{\perp 1}}}$, such that if $\hat{\xi}_{1i}\in\left[\hat{\xi}_{1m},\frac{\pi}{2}\right]$ no ray can be transmitted in crystal 2 (area of "total reflection"), and if $\hat{\xi}_{1i}\in\left[0,\hat{\xi}_{1m}\right]$, the incident ray is partially reflected in crystal 1 and partially transmitted in crystal 2; similarly to isotropic media, if $\hat{\Omega}_1\sqrt{\varepsilon'_{\perp 1}}\le\hat{\Omega}_2\sqrt{\varepsilon'_{\perp 2}}$, all the (left to right) incident rays in crystal 1 are partially reflected in crystal 1 and transmitted in crystal 2.

For incident rays from right to left, that is $\hat{\xi}\in\left[\frac{3\pi}{2},2\pi\right]$, from what precedes, if $\hat{\xi}\in\left[\frac{3\pi}{2},\frac{3\pi}{2}+2\alpha_1\right]$ then the incident ray cannot be reflected in crystal 1, while if $\hat{\xi}\in\left[\frac{3\pi}{2}+2\alpha_1,2\pi\right]$ the incident ray is partially reflected and transmitted. The function $F$ is strictly decreasing on the two sets $\hat{\xi}\in\left[\frac{3\pi}{2},\frac{3\pi}{2}+\alpha_1\right[$ and $\hat{\xi}\in\left]\frac{3\pi}{2}+\alpha_1,2\pi\right]$, so that introducing for convenience $\hat{\zeta}=2\pi-\hat{\xi}$, function $F$ is strictly increasing on the two sets $\hat{\zeta}_{1i}\in\left[0,\frac{\pi}{2}-\alpha_1\right[$ (incident rays in crystal 1 below the right cut line of crystal 1, containing the virtual interface at $\hat{\zeta}_{1i}=\frac{\pi}{2}-2\alpha_1$: for increasing $\hat{\zeta}_{1i}$, the incident ray is displaced from the geometrical axis of the slab to the right cut line of crystal 1) and $\hat{\zeta}_{1i}\in\left]\frac{\pi}{2}-\alpha_1,\frac{\pi}{2}\right]$ (incident rays in crystal 1 above the right cut line of crystal 1); hence if an incident ray 1 is above an incident ray 2 in crystal1, the transmitted ray 1 will be below the transmitted ray 2 in crystal 2: one deduces from that result that an incident ray below the right cut line of crystal 1, between the geometrical axis and the cut line, will be transmitted above the left cut line of crystal 2, between the geometrical axis and the cut line, that is, if $\hat{\xi}_{1i}\in\left]\frac{3\pi}{2}+\alpha_1,2\pi\right]$ then $\sqrt{\varepsilon'_{\perp 2}}\,F_2\left(\hat{\xi}_{2t}\right)=\sqrt{\varepsilon'_{\perp 1}}\,F_1\left(\hat{\xi}_{1i}\right)$ has an unique solution $\hat{\xi}_{2t}\in\left]\frac{3\pi}{2}+\alpha_2,2\pi\right]$, while an incident ray above the right cut line in crystal 1, between the real right interface and the cut line, should be transmitted below the left cut line of crystal 2, between the real left interface and the cut line; for $\hat{\xi}_{1i}\to\left(\frac{3\pi}{2}+\alpha_1\right)^+$, the equation to solve on $\hat{\xi}_{2t}\in\left]\frac{3\pi}{2}+\alpha_2,2\pi\right]$ is $\sqrt{\varepsilon'_{\perp 2}}\,F_2\left(\hat{\xi}_{2t}\right)=\sqrt{\varepsilon'_{\perp 1}}\,\breve{\Omega}_1$, with $F_2\left(\hat{\xi}_{2t}\right)\in\left[0,\breve{\Omega}_2\right[$: hence if $\sqrt{\varepsilon'_{\perp 2}}\,\breve{\Omega}_2\geq\sqrt{\varepsilon'_{\perp 1}}\,\breve{\Omega}_1$, the previous equation has an unique solution $\breve{\xi}_{2t}$ which defines the boundary of the left transmitted cone in crystal 2 above the left cut line in crystal 2 for incident rays (in crystal 1) between the cut line and the geometrical axis; if $\sqrt{\varepsilon'_{\perp 2}}\,\breve{\Omega}_2<\sqrt{\varepsilon'_{\perp 1}}\,\breve{\Omega}_1$, the equation $\sqrt{\varepsilon'_{\perp 2}}\,F_2\left(\hat{\xi}_{2t}\right)=\sqrt{\varepsilon'_{\perp 1}}\,\breve{\Omega}_1$ has no solution, but there exists similarly an unique $\breve{\xi}_{1i}\in\left]\frac{3\pi}{2}+\alpha_1,2\pi\right]$ such that $\sqrt{\varepsilon'_{\perp 1}}\,F_1\left(\breve{\xi}_{1i}\right)=\sqrt{\varepsilon'_{\perp 2}}\,\breve{\Omega}_2$ which defines the boundary of the incident right transmission cone below the right cut line of crystal 1, so that if and incident ray is between the cut line of crystal 1 and the boundary of the incident transmission cone below the cut line, the ray cannot be transmitted in crystal 2: hence if the virtual interface is above the previous incident transmission cone, it may exist a finite angular area in crystal 1 for which incident rays cannot be transmitted in crystal 2 and reflected in crystal1, from which the conclusion, followed by a simple illustration of each case:

* if $\hat{\Omega}_2\sqrt{\varepsilon'_{\perp 2}}<\hat{\Omega}_1\sqrt{\varepsilon'_{\perp 1}}$ and $\breve{\Omega}_2\sqrt{\varepsilon'_{\perp 2}}<\breve{\Omega}_1\sqrt{\varepsilon'_{\perp 1}}$ (illustrated on Fig. 3a), the incident allowed rays are $\hat{\xi}_{1i}\in\left[0,\hat{\xi}_{1m}\right]\cup\left[\breve{\xi}_{1i},2\pi\right]$ and the transmitted allowed rays are $\hat{\xi}_{2t}\in\left[0,\frac{\pi}{2}\right]\cup\left]\frac{3\pi}{2}+\alpha_2,2\pi\right]$,

* if $\hat{\Omega}_2\sqrt{\varepsilon'_{\perp 2}}<\hat{\Omega}_1\sqrt{\varepsilon'_{\perp 1}}$ and $\breve{\Omega}_2\sqrt{\varepsilon'_{\perp 2}}\geq\breve{\Omega}_1\sqrt{\varepsilon'_{\perp 1}}$ (illustrated on Fig. 3b), the incident allowed rays are $\hat{\xi}_{1i}\in\left[0,\hat{\xi}_{1m}\right]\cup\left]\frac{3\pi}{2}+\alpha_1,2\pi\right]$ and the transmitted allowed rays are $\hat{\xi}_{2t}\in\left[0,\frac{\pi}{2}\right]\cup\left[\breve{\xi}_{2t},2\pi\right]$,

* if $\hat{\Omega}_2\sqrt{\varepsilon'_{\perp 2}} \geq \hat{\Omega}_1\sqrt{\varepsilon'_{\perp 1}}$ and $\breve{\Omega}_2\sqrt{\varepsilon'_{\perp 2}} < \breve{\Omega}_1\sqrt{\varepsilon'_{\perp 1}}$ (illustrated on Fig. 3c), the incident allowed rays are $\hat{\xi}_{1i} \in \left[0,\frac{\pi}{2}\right] \cup \left[\breve{\xi}_{1i},2\pi\right]$ and the transmitted allowed rays are $\hat{\xi}_{2t} \in \left[0,\hat{\xi}_{2m}\right] \cup \left]\frac{3\pi}{2}+\alpha_2,2\pi\right]$,

* if $\hat{\Omega}_2\sqrt{\varepsilon'_{\perp 2}} \geq \hat{\Omega}_1\sqrt{\varepsilon'_{\perp 1}}$ and $\breve{\Omega}_2\sqrt{\varepsilon'_{\perp 2}} \geq \breve{\Omega}_1\sqrt{\varepsilon'_{\perp 1}}$ (illustrated on Fig. 3d), the incident allowed rays are $\hat{\xi}_{1i} \in \left[0,\frac{\pi}{2}\right] \cup \left]\frac{3\pi}{2}+\alpha_1,2\pi\right]$ and the transmitted allowed rays are $\hat{\xi}_{2t} \in \left[0,\hat{\xi}_{2m}\right] \cup \left[\breve{\xi}_{2t},2\pi\right]$

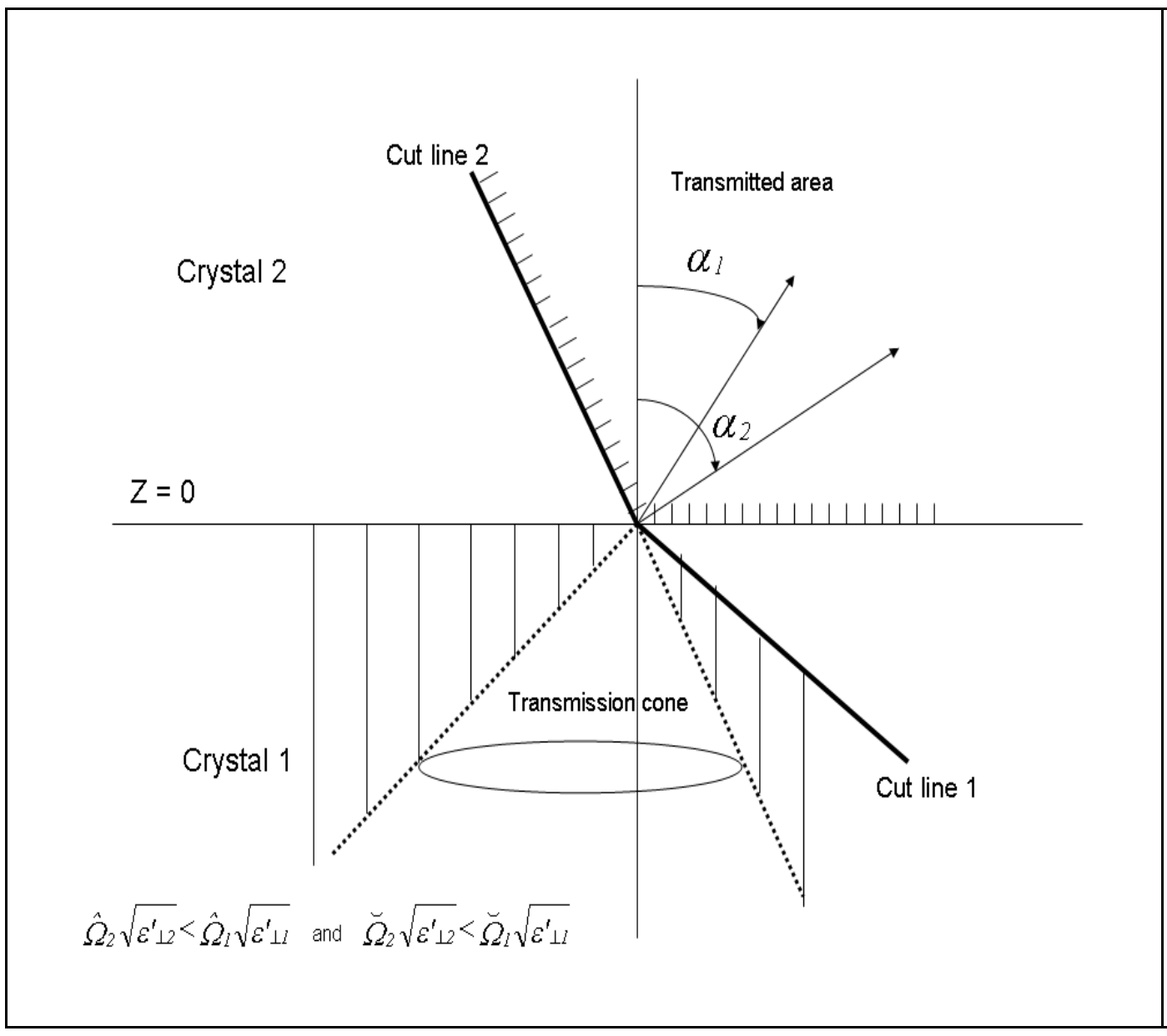

Fig. 3a

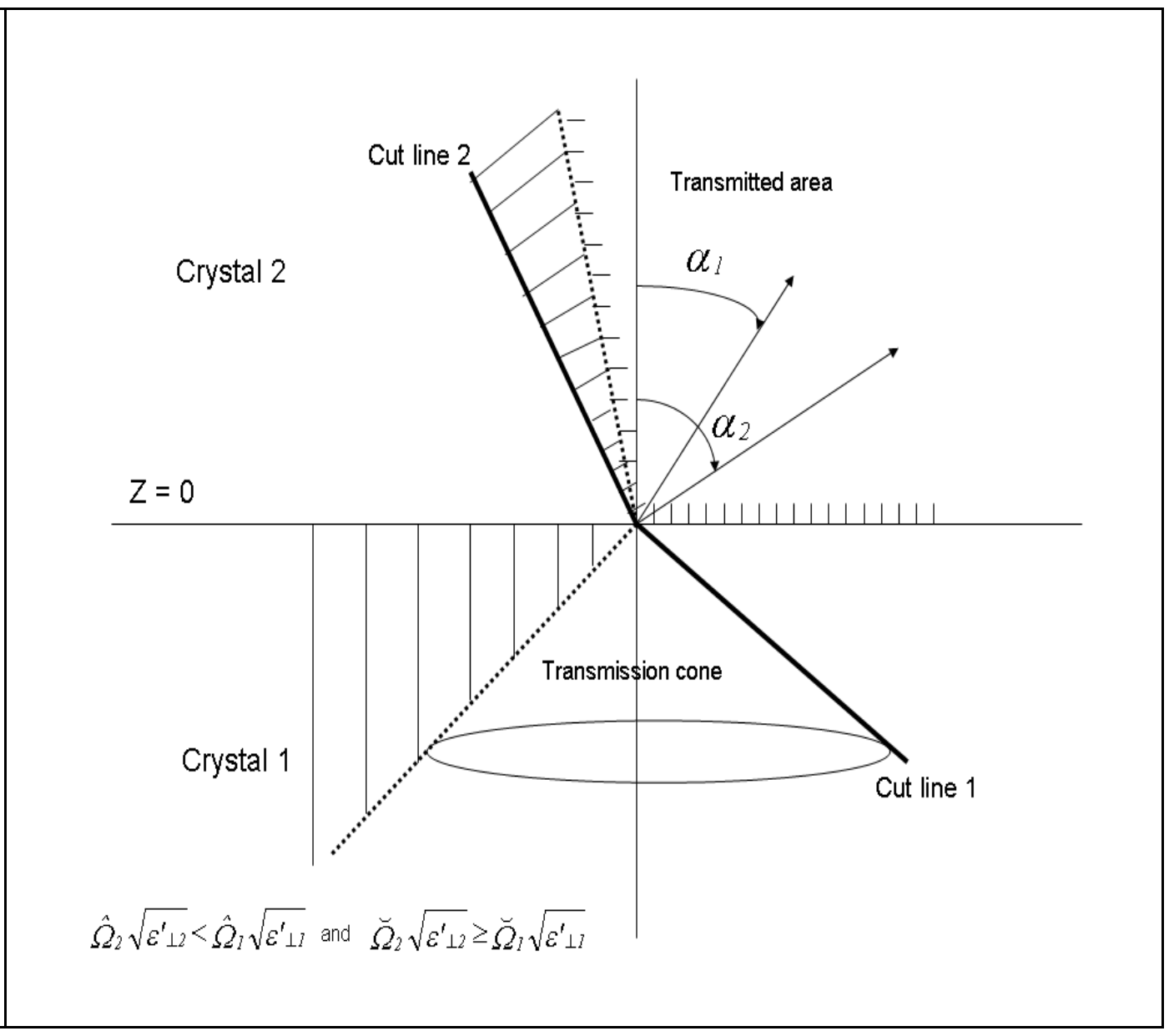

Fig. 3b

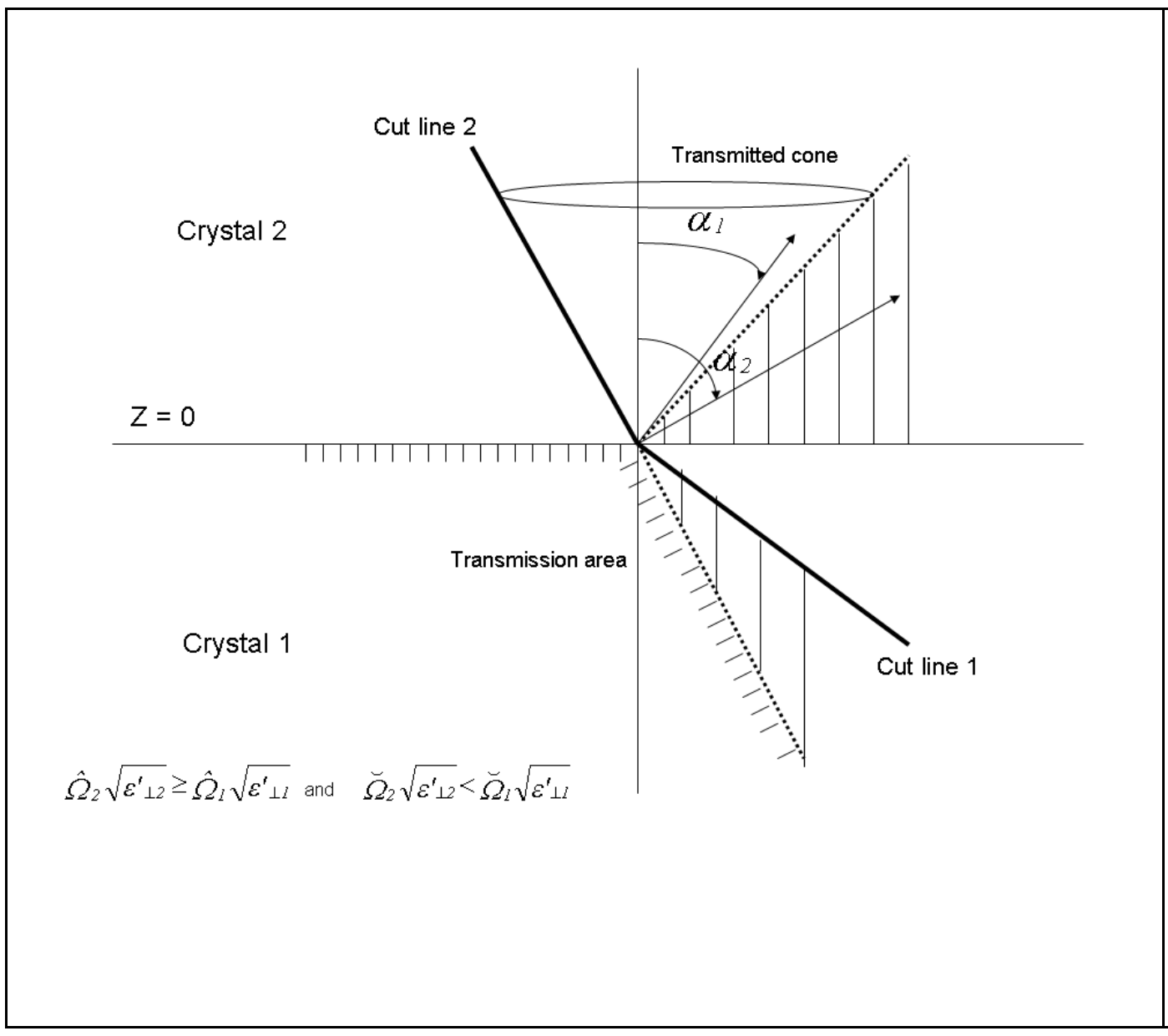

Fig. 3c

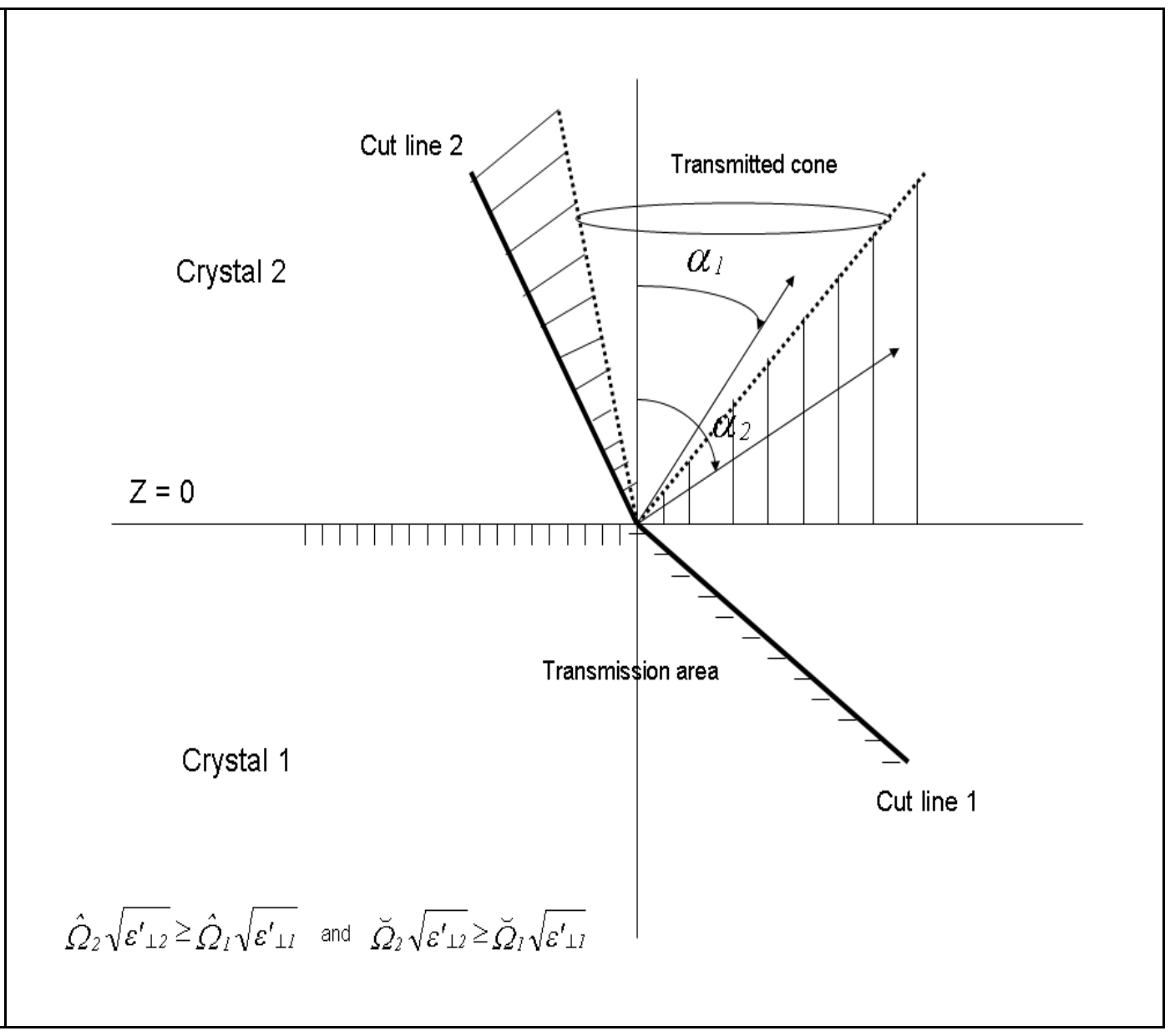

Fig. 3d

One presents hereafter on the two following Figs. 4a-b the transmitted angle (between the geometrical axis and the transmitted ray) function of the incident angle (between the geometrical axis and the incident ray), corresponding to the schematic case 3b, for crystals whose characteristics are $\sqrt{\varepsilon'_{\perp 1}}=1.5$, $\eta_1=2.0$ and $\alpha_1=22.5°$ for crystal 1, $\sqrt{\varepsilon'_{\perp 2}}=2.0$, $\eta_2=0.5$ and $\alpha_2=45.0°$ for crystal 2; in this situation, there is a pseudo total reflection on the left, the boundary of the transmission cone being at $\hat{\xi}_{1i}=15.47°$, and for incident rays from right to left, they can tend towards $\dfrac{3\pi}{2}+\alpha_1\equiv 2\pi-\hat{\xi}_{1i}=67.5°$, the corresponding transmitted value being $2\pi-\hat{\xi}_{2t}=36.23°$; compared to an isotropic transmission law, for which there is no total reflection but a symmetric transmitted cone of boundary $\hat{\xi}_{2t}=48.59°$, one can easily consider here the important differences between the ordinary and extraordinary behaviours.

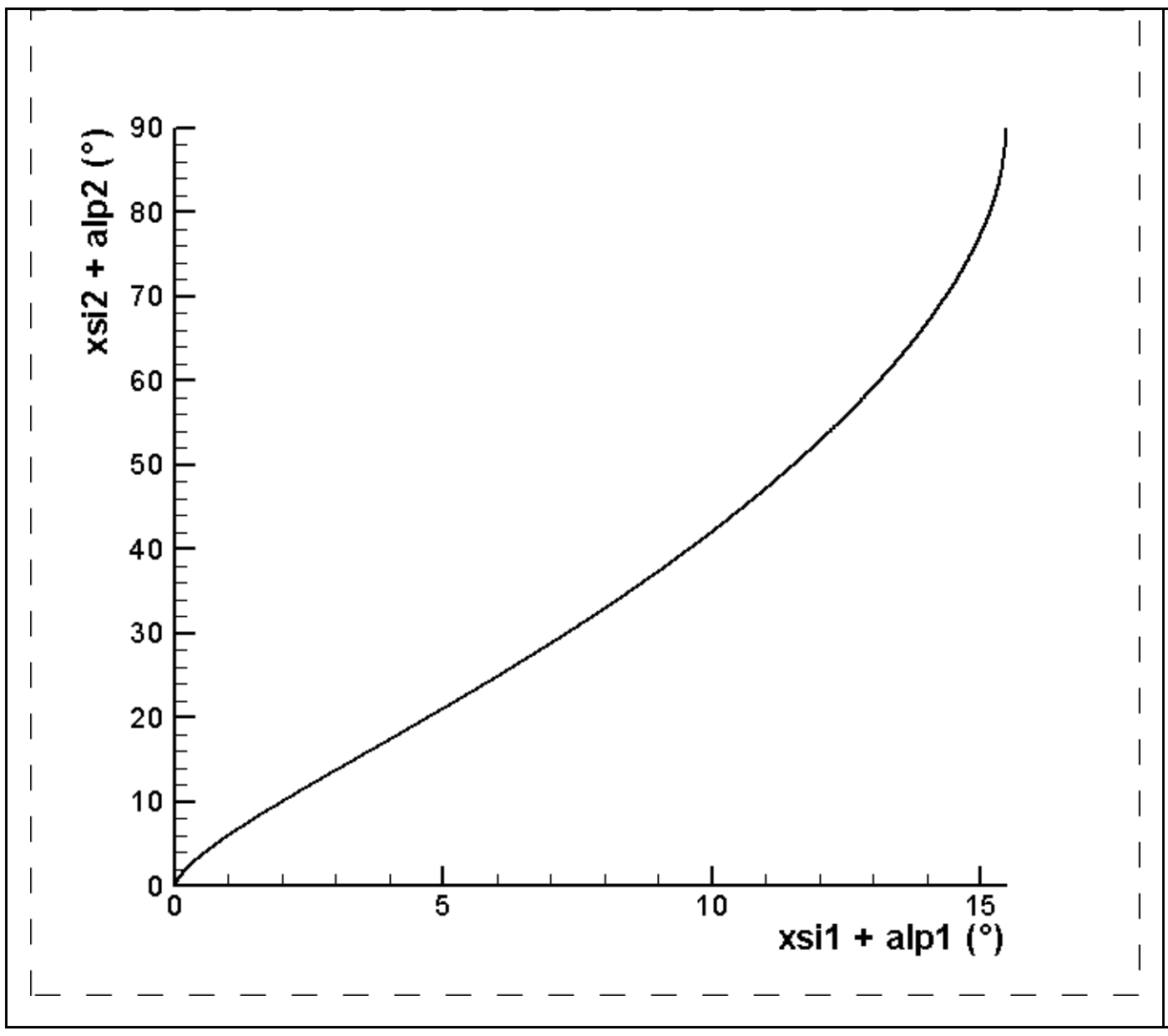

Fig. 4a: incident rays from left to right

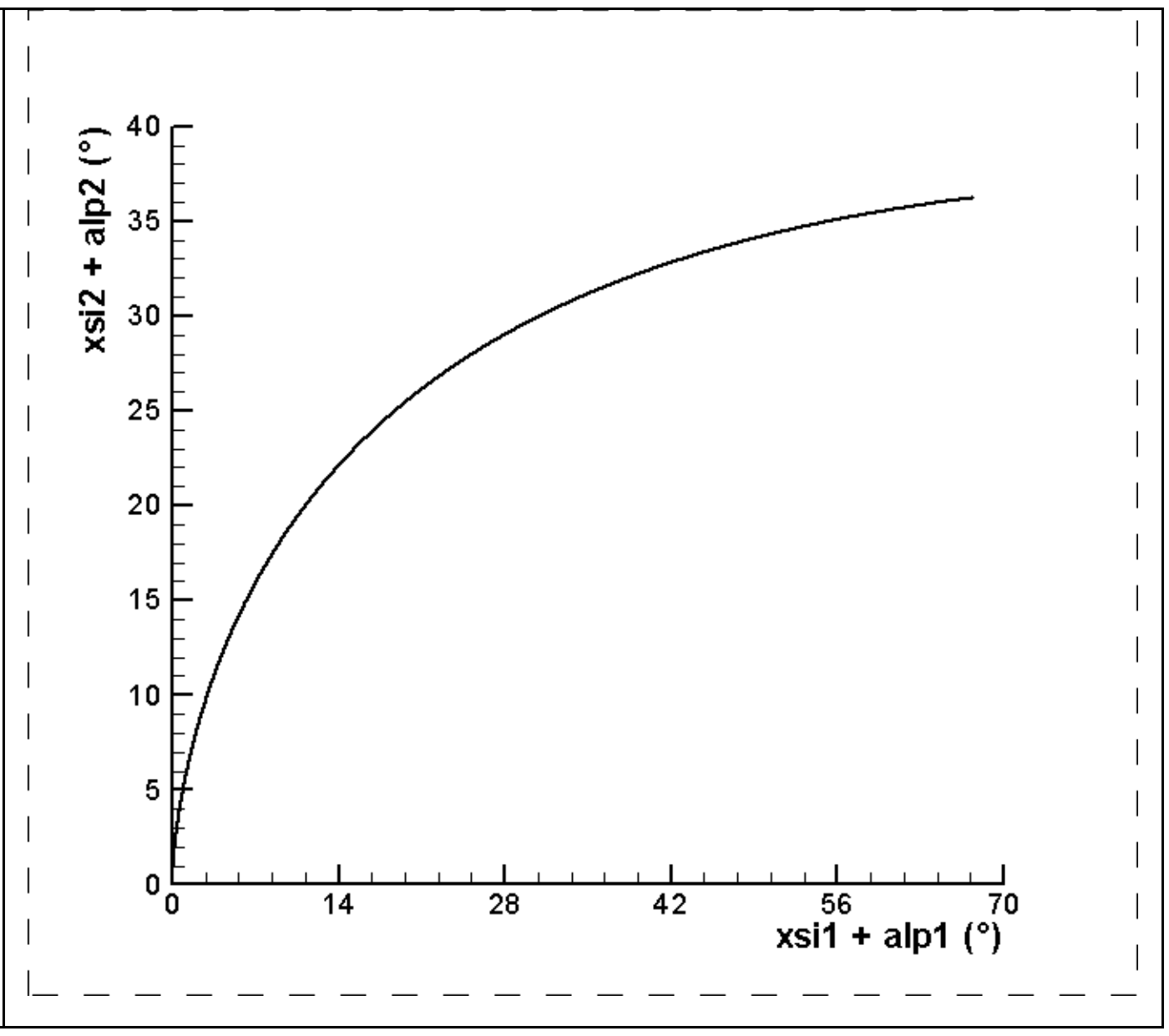

Fig. 4b: right to left incident rays below the cut line

Similarly as before, an incident ray between the right cut line in crystal 1 and the real interface can be transmitted in crystal 2 if there exists $\hat{\xi}_{2t}\in\left[\dfrac{3\pi}{2},\dfrac{3\pi}{2}+\alpha_2\right[$ such that $\sqrt{\varepsilon'_{\perp 1}}\,\check{\Omega}_1\leq\sqrt{\varepsilon'_{\perp 2}}\,F_2\left(\hat{\xi}_{2t}\right)\leq\sqrt{\varepsilon'_{\perp 1}}\,\hat{\Omega}_1$, so that the boundaries of the transmitted cone (if they exist) must be solutions of $\sqrt{\varepsilon'_{\perp 2}}\,F_2\left(\hat{\xi}_{2t}\right)=\sqrt{\varepsilon'_{\perp 1}}\,\check{\Omega}_1$ for the lower one, and $\sqrt{\varepsilon'_{\perp 2}}\,F_2\left(\hat{\xi}_{2t}\right)=\sqrt{\varepsilon'_{\perp 1}}\,\hat{\Omega}_1$ for the upper one; they correspond to the boundaries of the transmission cone (if they exist) defined by $\sqrt{\varepsilon'_{\perp 1}}\,F_1\left(\hat{\xi}_{1i}\right)=\sqrt{\varepsilon'_{\perp 2}}\,\check{\Omega}_2$ and $\sqrt{\varepsilon'_{\perp 1}}\,F_1\left(\hat{\xi}_{1i}\right)=\sqrt{\varepsilon'_{\perp 2}}\,\hat{\Omega}_2$, where $\hat{\xi}_{1i}\in\left[\dfrac{3\pi}{2},\dfrac{3\pi}{2}+\alpha_1\right[$; then, if a pseudo total reflection exists for the left side, the relation $\hat{\Omega}_1\sqrt{\varepsilon'_{\perp 1}}>\hat{\Omega}_2\sqrt{\varepsilon'_{\perp 2}}$ is verified and a pseudo total reflection may obviously exist also for the right side, but the pseudo total reflection can arise if and only if $\check{\Omega}_1\sqrt{\varepsilon'_{\perp 1}}<\check{\Omega}_2\sqrt{\varepsilon'_{\perp 2}}$ : hence if $\hat{\Omega}_1\sqrt{\varepsilon'_{\perp 1}}>\hat{\Omega}_2\sqrt{\varepsilon'_{\perp 2}}$ and $\check{\Omega}_1\sqrt{\varepsilon'_{\perp 1}}\geq\check{\Omega}_2\sqrt{\varepsilon'_{\perp 2}}$, no incident ray between the right cut line and the right real interface in crystal 1 can be transmitted in crystal 2, while if $\hat{\Omega}_1\sqrt{\varepsilon'_{\perp 1}}>\hat{\Omega}_2\sqrt{\varepsilon'_{\perp 2}}$ and if $\overset{\frown}{\Omega}_1\sqrt{\varepsilon'_{\perp 1}}<\hat{\Omega}_2\sqrt{\varepsilon'_{\perp 2}}$, it exists one unique $\tilde{\xi}_{1i}\in\left[\frac{3\pi}{2},\frac{3\pi}{2}+\alpha_1\right[$ such that $\sqrt{\varepsilon'_{\perp 1}}\,F_1\left(\tilde{\xi}_{1i}\right)=\sqrt{\varepsilon'_{\perp 2}}\,\hat{\Omega}_2$, that is incident rays in crystal 1 such that $\hat{\xi}_{1i}\in\left[\frac{3\pi}{2},\tilde{\xi}_{1i}\right[$ cannot be transmitted in crystal 2; in the case of a possible pseudo total reflection, there is a "real" total reflection when $\hat{\xi}_{2t}=\frac{3\pi}{2}$, so that $\sqrt{\varepsilon'_{\perp 1}}\,F_1\left(\hat{\xi}_{1i}\right)=\sqrt{\varepsilon'_{\perp 2}}\,\overset{\frown}{\Omega}_2$, possible if and only if $\overset{\frown}{\Omega}_1\sqrt{\varepsilon'_{\perp 1}}<\overset{\frown}{\Omega}_2\sqrt{\varepsilon'_{\perp 2}}$; then, if $\overset{\frown}{\Omega}_1\sqrt{\varepsilon'_{\perp 1}}<\overset{\frown}{\Omega}_2\sqrt{\varepsilon'_{\perp 2}}$ it exists one unique $\ddot{\xi}_{1i}\in\left[\tilde{\xi}_{1i},\frac{3\pi}{2}+\alpha_1\right[$ such that $\sqrt{\varepsilon'_{\perp 1}}\,F_1\left(\ddot{\xi}_{1i}\right)=\sqrt{\varepsilon'_{\perp 2}}\,\overset{\frown}{\Omega}_2$, that is for incident rays $\hat{\xi}_{1i}\in\left\lfloor\tilde{\xi}_{1i},\ddot{\xi}_{1i}\right\rfloor$ in crystal 1 the transmitted ray in crystal 2 is $\hat{\xi}_{2t}\in\left[\frac{3\pi}{2},\frac{3\pi}{2}+\alpha_2\right[$, and for $\hat{\xi}_{1i}\in\left[\frac{3\pi}{2},\tilde{\xi}_{1i}\right[\cup\left]\ddot{\xi}_{1i},\frac{3\pi}{2}+\alpha_1\right[$, no ray can be transmitted in crystal 2; if $\overset{\frown}{\Omega}_1\sqrt{\varepsilon'_{\perp 1}}\geq\overset{\frown}{\Omega}_2\sqrt{\varepsilon'_{\perp 2}}$, there exists from what precedes an unique $\ddot{\xi}_{2t}\in\left[\frac{3\pi}{2},\frac{3\pi}{2}+\alpha_2\right[$ such that $\sqrt{\varepsilon'_{\perp 2}}\,F_2\left(\hat{\xi}_{2t}\right)=\sqrt{\varepsilon'_{\perp 1}}\,\overset{\frown}{\Omega}_1$, that is for incident rays $\hat{\xi}_{1i}\in\left[\tilde{\xi}_{1i},\frac{3\pi}{2}+\alpha_1\right[$ the transmitted rays are in the set $\hat{\xi}_{2t}\in\left[\frac{3\pi}{2},\ddot{\xi}_{2t}\right]$. Similarly, if no pseudo total reflection can exist because $\hat{\Omega}_1\sqrt{\varepsilon'_{\perp 1}}\leq\hat{\Omega}_2\sqrt{\varepsilon'_{\perp 2}}$, which means that incident rays $\hat{\xi}_{1i}\in\left[\frac{3\pi}{2},\frac{3\pi}{2}+\alpha_1\right[$ may be transmitted in crystal 2, it may however exist a "real" total reflection analogous to what happens in the previous case, i.e. $\hat{\xi}_{2t}=\frac{3\pi}{2}$, possible if and only if $\overset{\frown}{\Omega}_1\sqrt{\varepsilon'_{\perp 1}}<\overset{\frown}{\Omega}_2\sqrt{\varepsilon'_{\perp 2}}$; then, if $\overset{\frown}{\Omega}_1\sqrt{\varepsilon'_{\perp 1}}<\overset{\frown}{\Omega}_2\sqrt{\varepsilon'_{\perp 2}}$ it exists one unique $\ddot{\xi}_{1i}\in\left[\frac{3\pi}{2},\frac{3\pi}{2}+\alpha_1\right[$ such that $\sqrt{\varepsilon'_{\perp 1}}\,F_1\left(\ddot{\xi}_{1i}\right)=\sqrt{\varepsilon'_{\perp 2}}\,\overset{\frown}{\Omega}_2$, that is for incident rays $\hat{\xi}_{1i}\in\left[\frac{3\pi}{2},\ddot{\xi}_{1i}\right]$ in crystal 1 the transmitted ray in crystal 2 is $\hat{\xi}_{2t}\in\left[\frac{3\pi}{2},\frac{3\pi}{2}+\alpha_2\right[$, and for $\hat{\xi}_{1i}\in\left]\ddot{\xi}_{1i},\frac{3\pi}{2}+\alpha_1\right[$ no ray can be transmitted in crystal 2; note that for $\hat{\xi}_{1i}=\frac{3\pi}{2}$, the transmitted boundary cone verifies $\sqrt{\varepsilon'_{\perp 2}}\,F_2\left(\hat{\xi}_{2t}\right)=\sqrt{\varepsilon'_{\perp 1}}\,\hat{\Omega}_1$: hence if $\hat{\Omega}_1\sqrt{\varepsilon'_{\perp 1}}\geq\hat{\Omega}_2\sqrt{\varepsilon'_{\perp 2}}$ there is one unique valid solution $\hat{\xi}_{2t}\in\left[\frac{3\pi}{2},\frac{3\pi}{2}+\alpha_2\right[$ verifying the previous equation, while if $\hat{\Omega}_1\sqrt{\varepsilon'_{\perp 1}}<\hat{\Omega}_2\sqrt{\varepsilon'_{\perp 2}}$ no incident ray $\hat{\xi}_{1i}\in\left[\frac{3\pi}{2},\frac{3\pi}{2}+\alpha_1\right[$ can be transmitted in crystal 2, from which the conclusion, followed by a simple illustration of each case (for possible transmission)

* if $\hat{\Omega}_2\sqrt{\varepsilon'_{\perp 2}}<\hat{\Omega}_1\sqrt{\varepsilon'_{\perp 1}}$ or $\hat{\Omega}_1\sqrt{\varepsilon'_{\perp 1}}<\hat{\Omega}_2\sqrt{\varepsilon'_{\perp 2}}$, no incident ray $\hat{\xi}_{1i}\in\left[\frac{3\pi}{2},\frac{3\pi}{2}+\alpha_1\right[$ can be transmitted in crystal 2,

* if $\hat{\Omega}_1\sqrt{\varepsilon'_{\perp 1}}<\hat{\Omega}_2\sqrt{\varepsilon'_{\perp 2}}<\hat{\Omega}_2\sqrt{\varepsilon'_{\perp 2}}<\hat{\Omega}_1\sqrt{\varepsilon'_{\perp 1}}$ (illustrated on Fig. 5a), the incident allowed rays are $\hat{\xi}_{1i}\in\left[\tilde{\xi}_{1i},\ddot{\xi}_{1i}\right]$ and the transmitted allowed rays are $\hat{\xi}_{2t}\in\left[\frac{3\pi}{2},\frac{3\pi}{2}+\alpha_2\right[$,

* if $\hat{\Omega}_2\sqrt{\varepsilon'_{\perp 2}}<\hat{\Omega}_1\sqrt{\varepsilon'_{\perp 1}}<\hat{\Omega}_2\sqrt{\varepsilon'_{\perp 2}}<\hat{\Omega}_1\sqrt{\varepsilon'_{\perp 1}}$ (illustrated on Fig. 5b), the incident allowed rays are $\hat{\xi}_{1i}\in\left[\tilde{\xi}_{1i},\frac{3\pi}{2}+\alpha_1\right[$ and the transmitted allowed rays are $\hat{\xi}_{2t}\in\left[\frac{3\pi}{2},\ddot{\xi}_{2t}\right]$,

* if $\hat{\Omega}_1\sqrt{\varepsilon'_{\perp 1}}<\hat{\Omega}_2\sqrt{\varepsilon'_{\perp 2}}<\hat{\Omega}_1\sqrt{\varepsilon'_{\perp 1}}<\hat{\Omega}_2\sqrt{\varepsilon'_{\perp 2}}$ (illustrated on Fig. 5c), the incident allowed rays are $\hat{\xi}_{1i}\in\left[\frac{3\pi}{2},\ddot{\xi}_{1i}\right]$ and the transmitted allowed rays are $\hat{\xi}_{2t}\in\left]\tilde{\xi}_{2t},\frac{3\pi}{2}+\alpha_2\right[$,

* if $\hat{\Omega}_2\sqrt{\varepsilon'_{\perp 2}}<\hat{\Omega}_1\sqrt{\varepsilon'_{\perp 1}}<\hat{\Omega}_1\sqrt{\varepsilon'_{\perp 1}}<\hat{\Omega}_2\sqrt{\varepsilon'_{\perp 2}}$ (illustrated on Fig. 5d), the incident allowed rays are $\hat{\xi}_{1i}\in\left[\frac{3\pi}{2},\frac{3\pi}{2}+\alpha_1\right[$ and the transmitted allowed rays are $\hat{\xi}_{2t}\in\left[\tilde{\xi}_{2t},\ddot{\xi}_{2t}\right]$

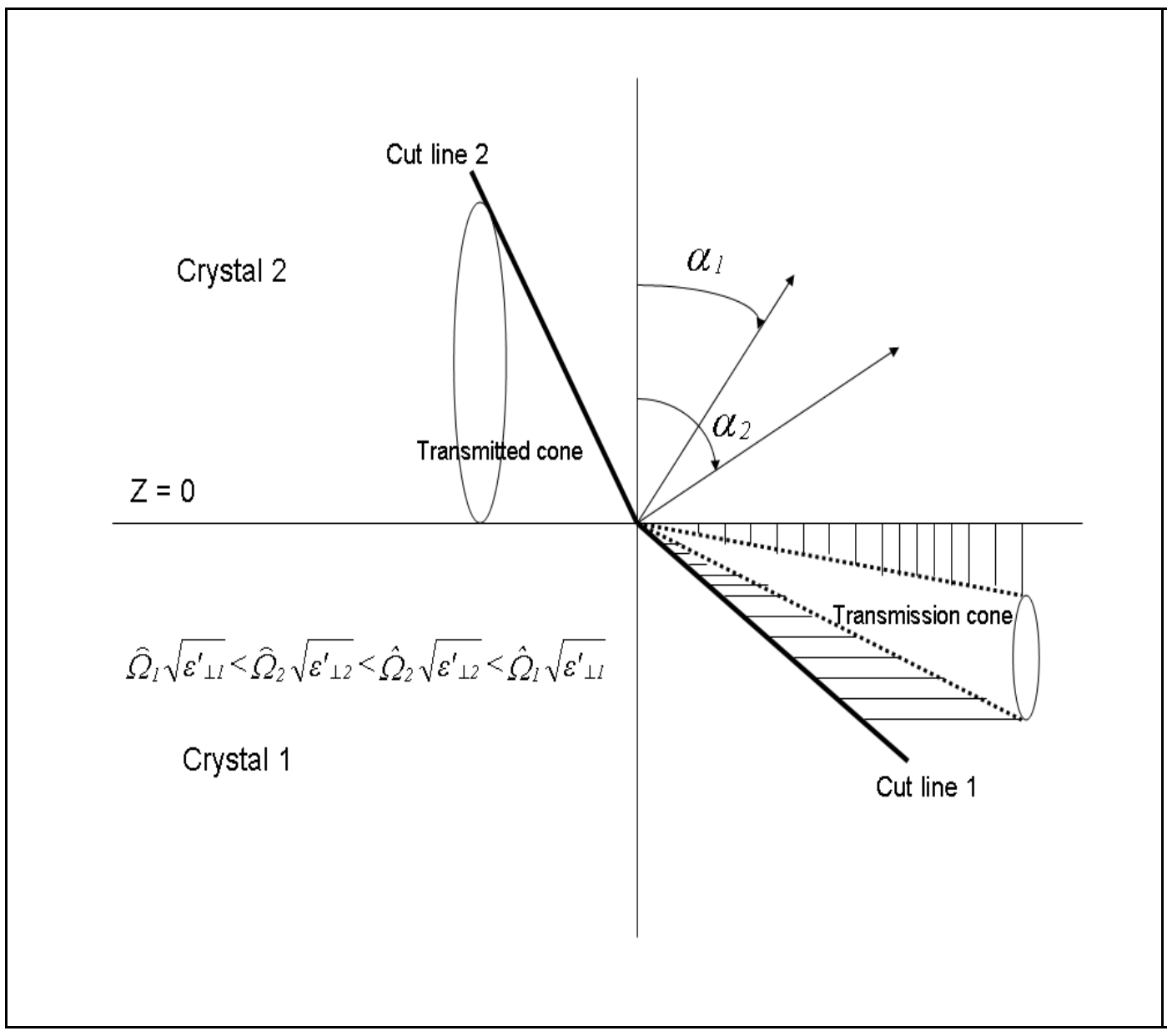

Fig. 5a

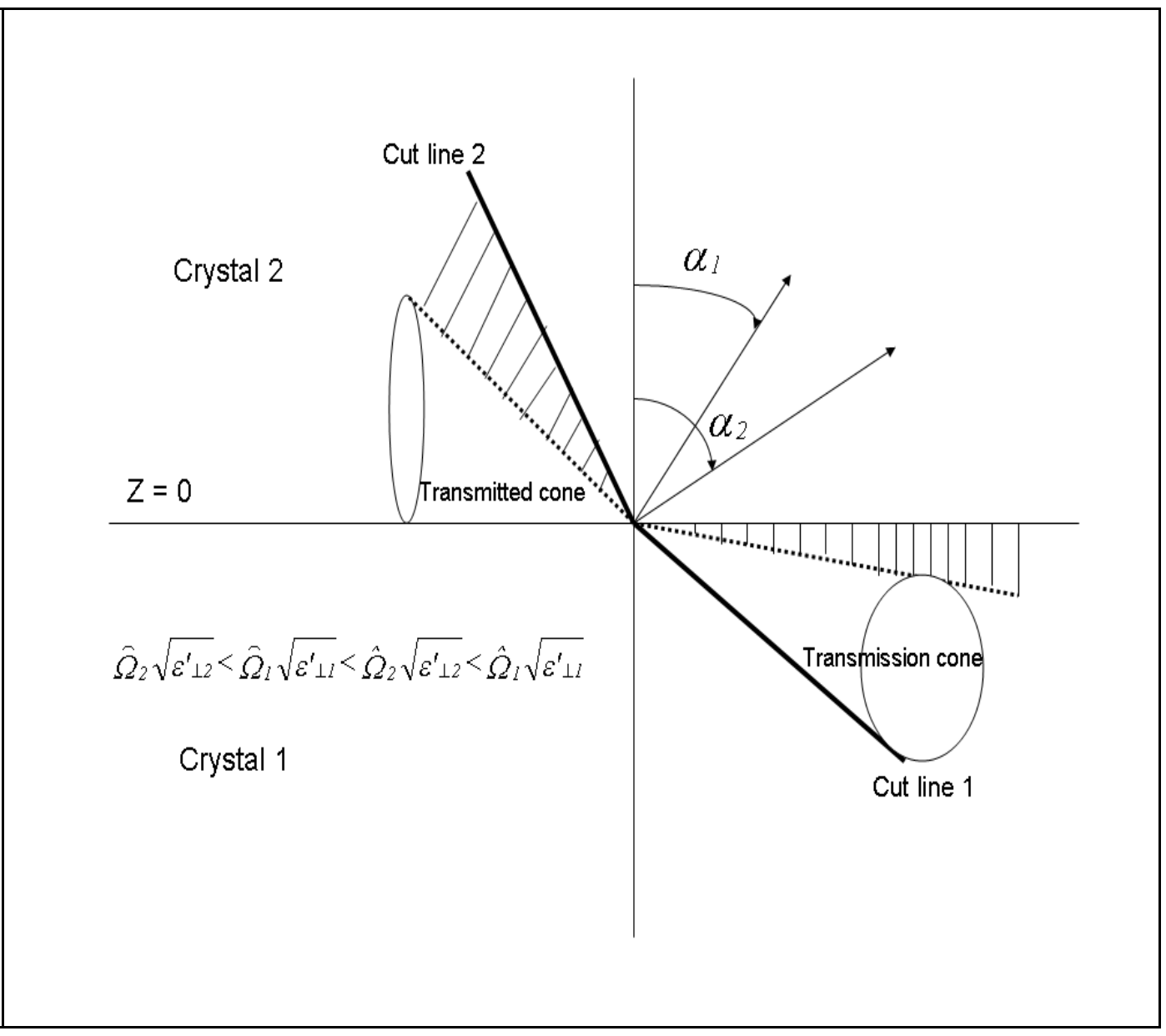

Fig. 5b

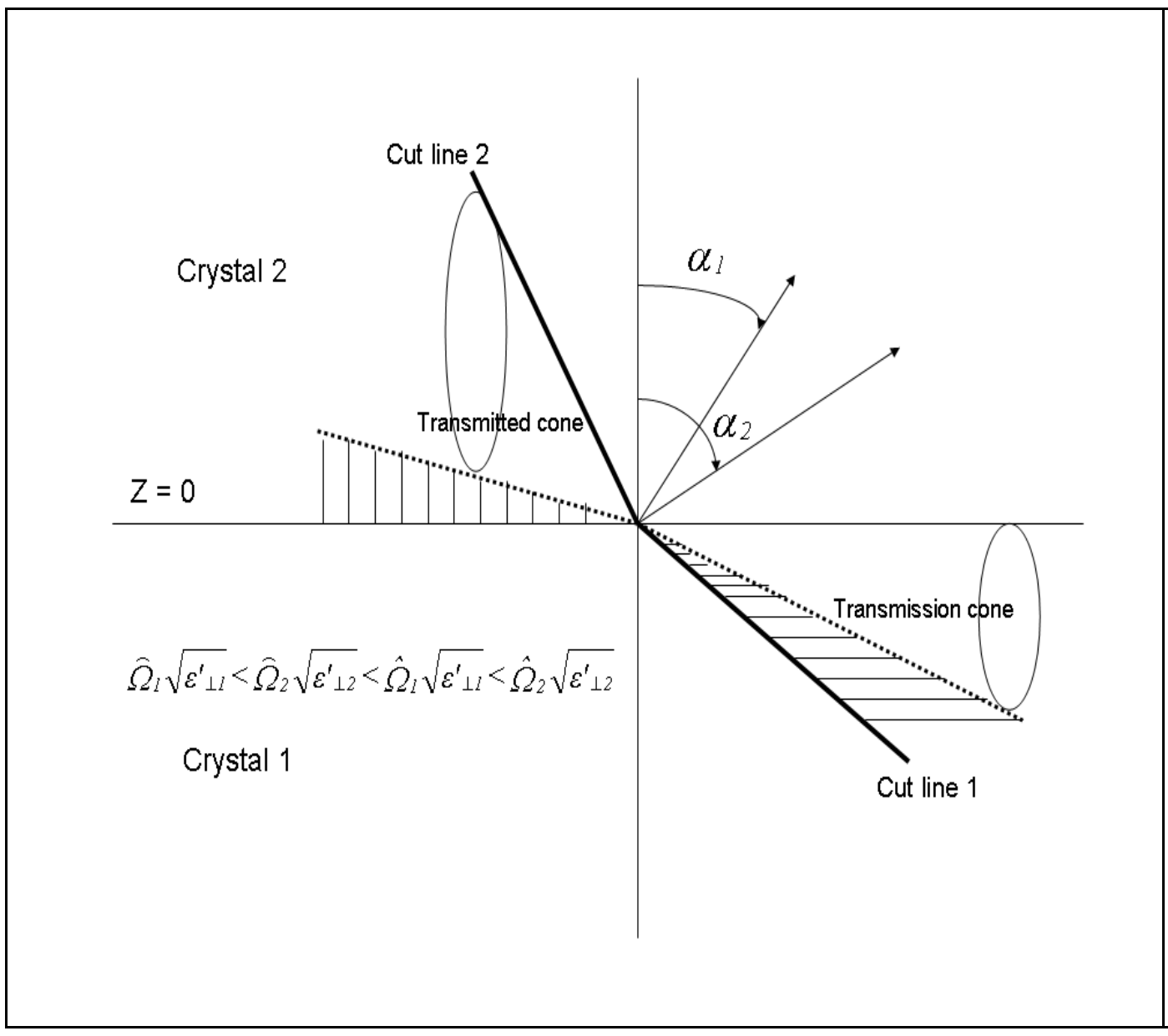

Fig. 5c

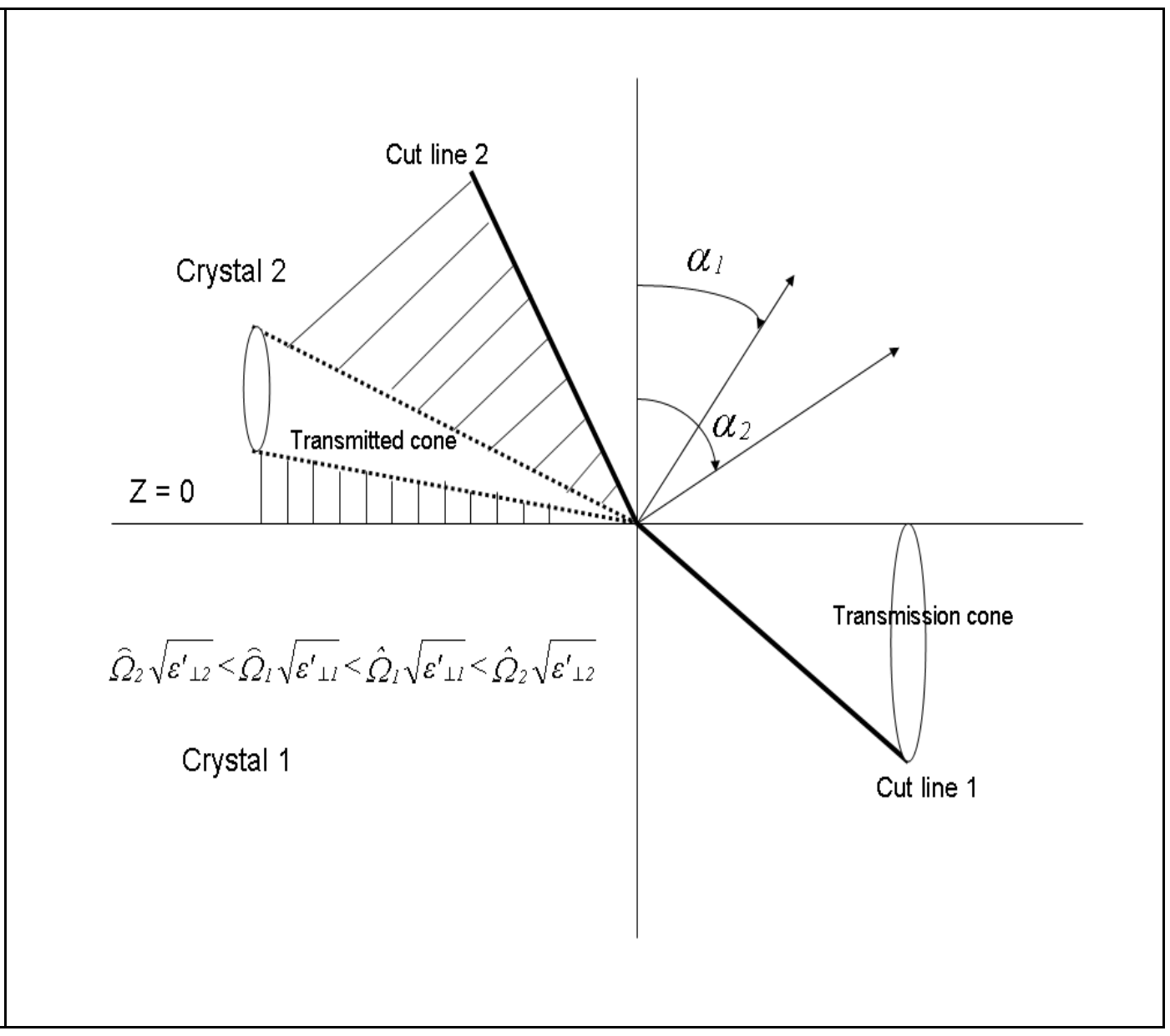

Fig. 5d

One presents now on the three following Figs. 6b-d the transmitted angle (between the geometrical axis and the transmitted ray) function of the incident angle (between the geometrical axis and the incident ray), illustrated on Fig. 6a and corresponding to the schematic cases 3d and 5d, for crystals whose characteristics are $\sqrt{\varepsilon'_{\perp 1}}=1.5$, $\eta_1=2.0$ and $\alpha_1=30.0°$ for crystal 1, $\sqrt{\varepsilon'_{\perp 2}}=2.0$, $\eta_2=2.0$ and $\alpha_1=60.0°$ for crystal 2; in this situation, there is no pseudo total reflection on the left, the boundary of the transmitted cone being at $\hat{\xi}_{2t}=48.75°$, and for incident rays from right to left below the discontinuity line, they can tend towards $\frac{3\pi}{2}+\alpha_1\equiv 2\pi-\hat{\xi}_{1i}=60.0°$, the corresponding transmitted value being $2\pi-\hat{\xi}_{2t}=19.80°$, while for incident rays above the discontinuity line, they also can tend towards $\frac{3\pi}{2}+\alpha_1\equiv 2\pi-\hat{\xi}_{1i}=60.0°$, the corresponding transmitted value being $2\pi-\hat{\xi}_{2t}=31.64°$, and since there is no pseudo total reflection, the corresponding transmitted value for $\hat{\xi}_{1i}=270.0°$ is $2\pi-\hat{\xi}_{2t}=37.41°$; in this case, the boundary of the right transmitted cone is very close to the one of the ordinary transmitted cone.

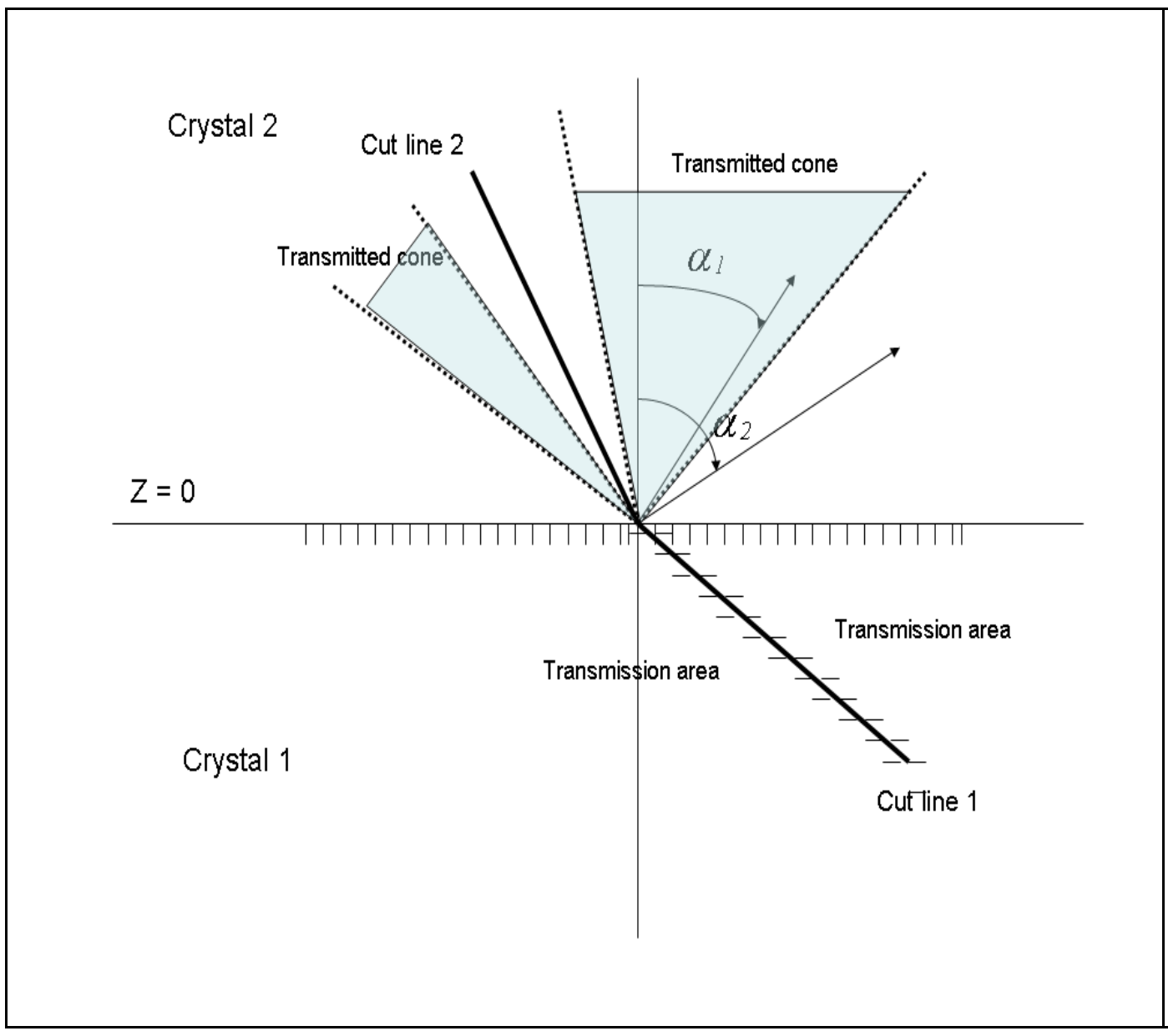

Fig. 6a: Extraordinary transmission

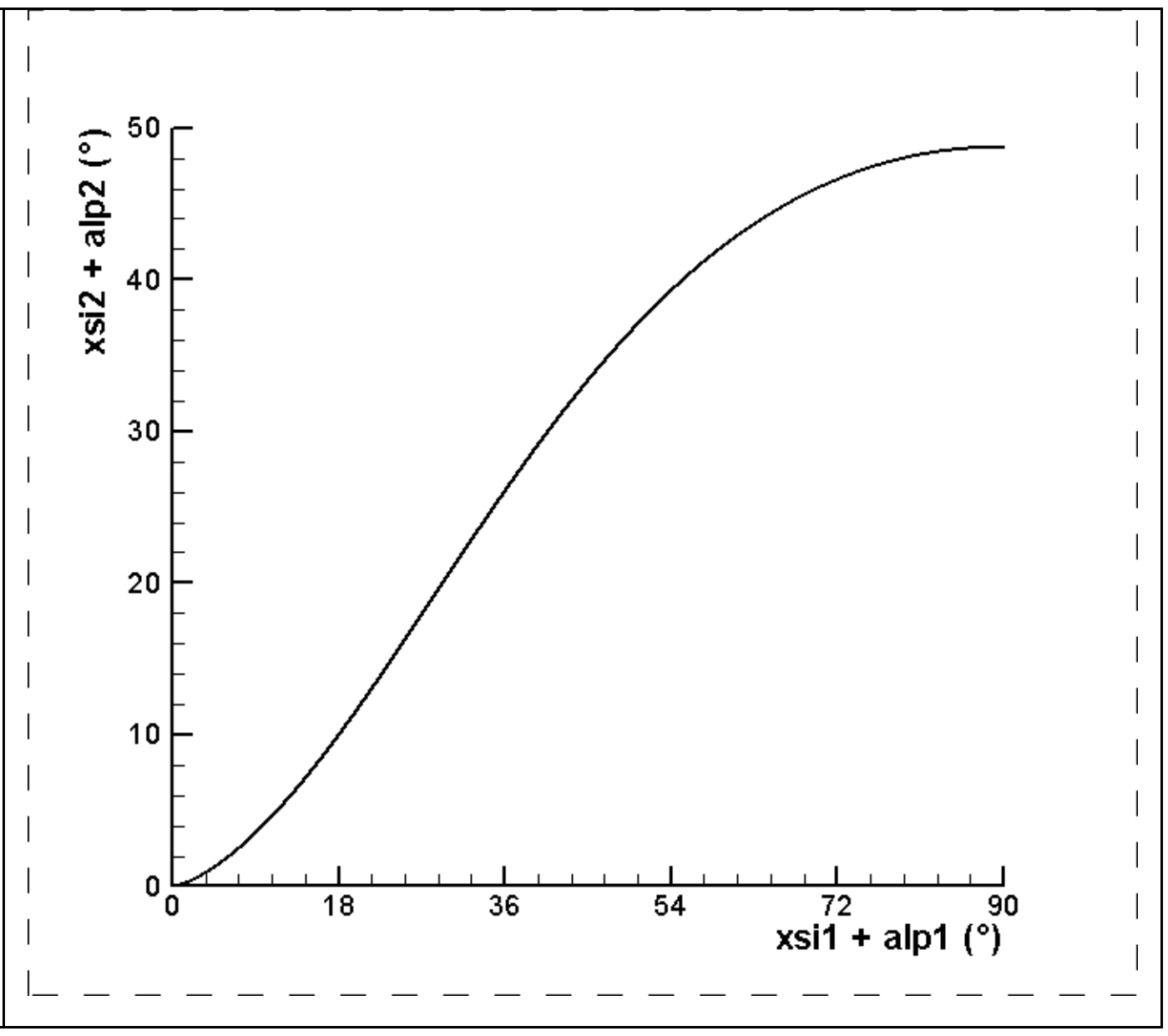

Fig. 6b: from left to right

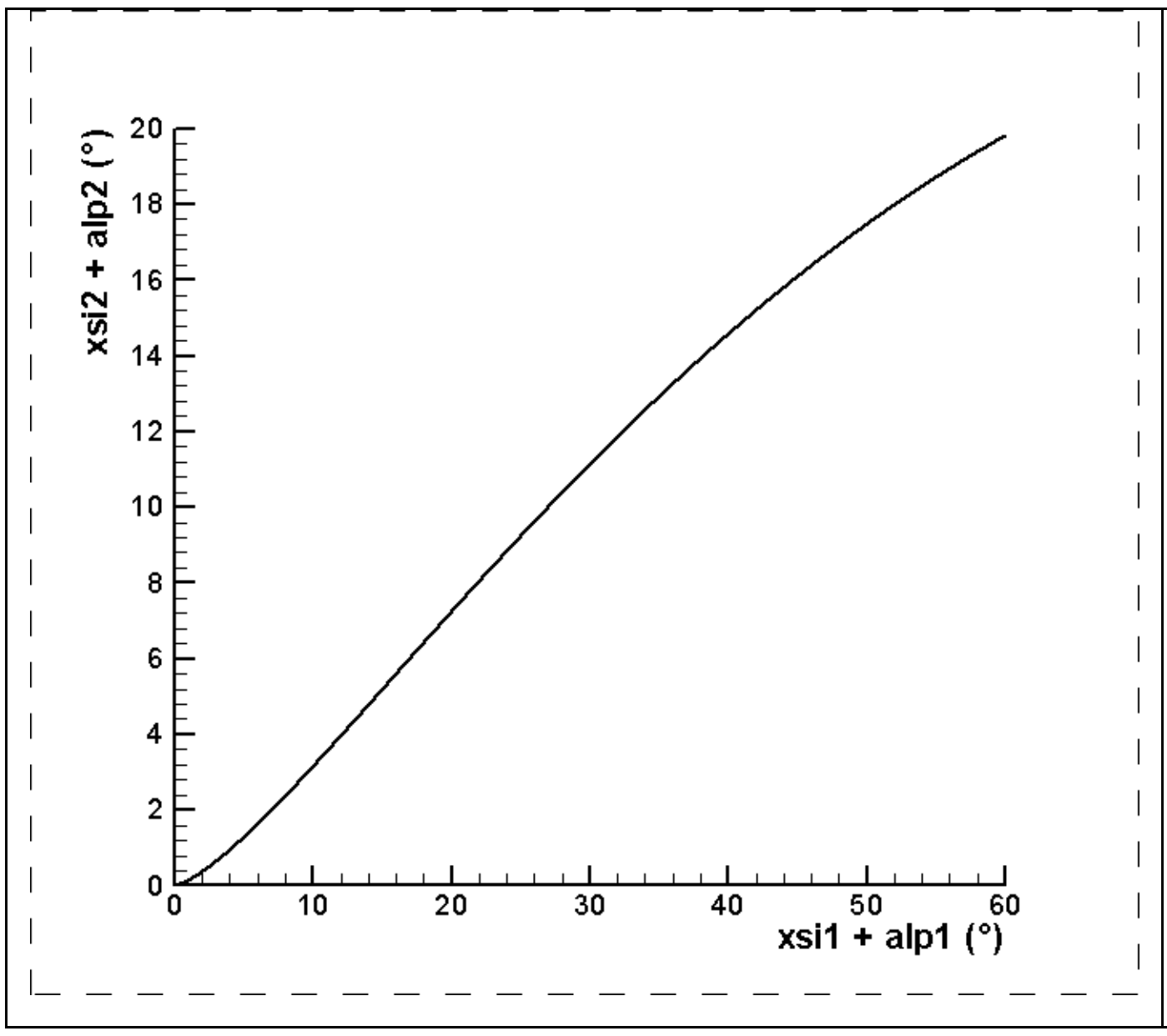

Fig. 6c: right to left, below the discontinuity line

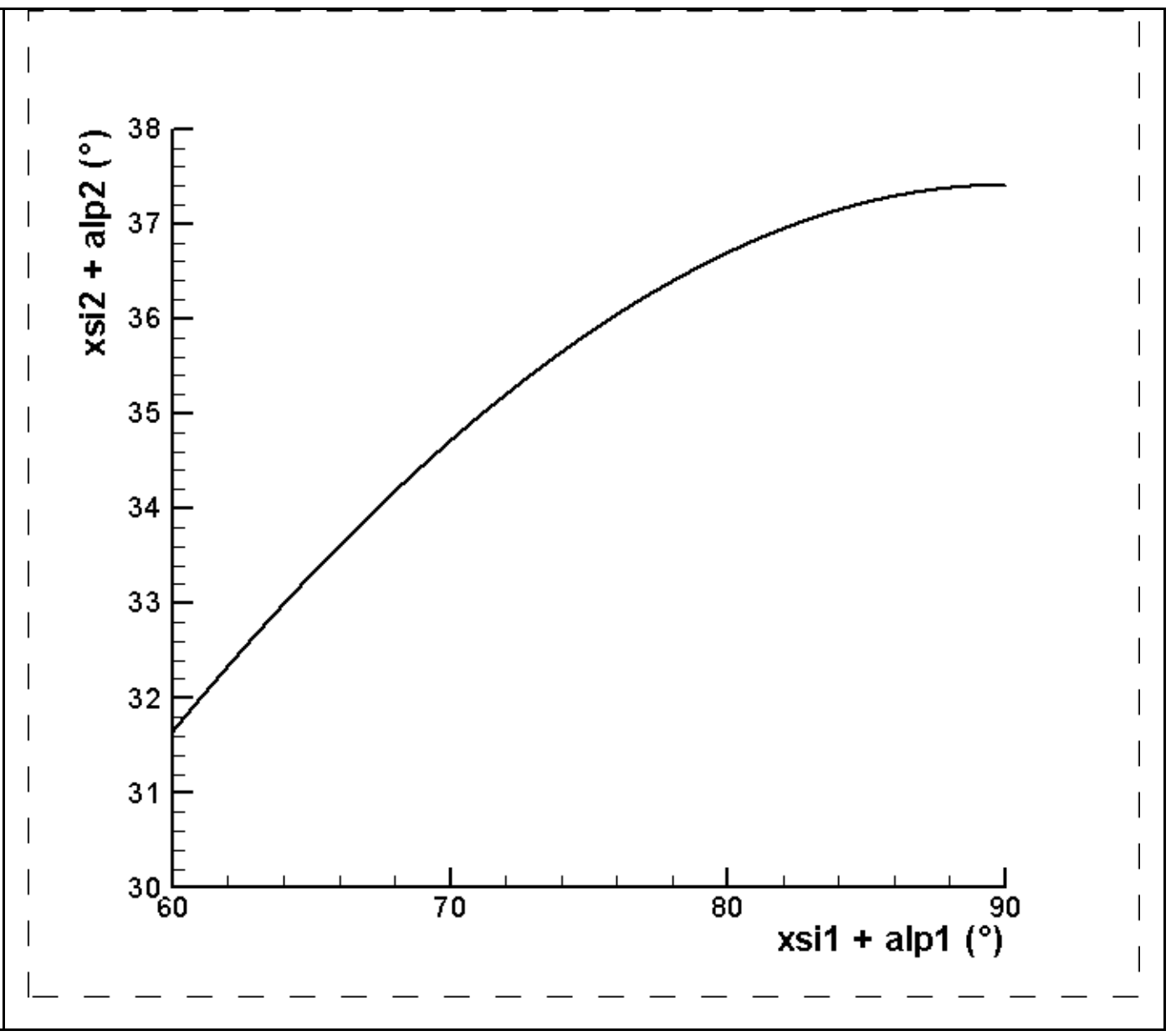

Fig. 6d: right to left, above the discontinuity line

Note that some incident rays inside the possible transmission cones are above the virtual interface of crystal 1, so that for this numerical example the transmission cone above the discontinuity line of crystal 1 is an area of transmission and "virtual reflection" or absorption, while for the (right) transmission cone below the discontinuity line of crystal 1, the part between the virtual interface and the discontinuity line is also an area of transmission and absorption, and the part between the geometrical axis and the virtual interface is an area of partial transmission and real reflection. Reciprocally, there may exist angular areas in crystal 1 for which no incident ray can be transmitted and reflected: for instance, for the numerical example where $\sqrt{\varepsilon'_{\perp 1}}=1.5$, $\eta_1=2.0$ and $\alpha_1=22.5°$ for crystal 1, $\sqrt{\varepsilon'_{\perp 2}}=2.0$, $\eta_2=0.5$ and $\alpha_1=45.0°$ for crystal 2, no incident ray between the real (right) interface and the discontinuity line in crystal 1 can be transmitted in crystal 2, so that this region is an area of no transmission and reflection since above the virtual interface: hence the “virtually reflected” ray in medium 2 corresponds to an absorbed ray at the interface, this absorbed energy being totally reemitted in the two media.
The case where the medium of transmission is isotropic shall be examined in the next subsection.

III-3 Extraordinary/ordinary reflexion/transmission phenomena:

In this case, the ordinary and extraordinary rays are related thanks to the modified Descartes’ law, still valid for ordinary rays, with $\eta_i \equiv 1$ and $\alpha_i \equiv 0$, so that the relation between the two rays is:

$$\sqrt{\varepsilon'_{\perp 2}}\, sin\,\hat{\xi}^{o}_{2t} = \sqrt{\varepsilon'_{\perp 1}}\, F_1\left(\hat{\xi}^{e}_{1i}\right), \tag{39}$$

where $\hat{\xi}^{o}_{2t}$ is the angle between the geometrical axis and the transmitted ordinary ray, $\hat{\xi}^{e}_{1i}$ being the extraordinary corresponding angle, defined as in previous subsection; note that if it exists a transmitted extraordinary ray and an ordinary one in crystal 2, the two transmitted rays are related by $sin\,\hat{\xi}^{o}_{2t} = F_2\left(\hat{\xi}^{e}_{2t}\right)$, possible if and only if $F_2\left(\hat{\xi}^{e}_{2t}\right) \le 1$; a necessary condition for an ordinary ray to be transmitted is then $\sqrt{\varepsilon'_{\perp 1}}\, F_1\left(\hat{\xi}^{e}_{1i}\right) \le \sqrt{\varepsilon'_{\perp 2}}$ for the incident extraordinary ray; the two particular cases $\alpha_j = 0$ or $\alpha_j = \frac{\pi}{2}$ are summarized as follows:

* case $\alpha_1 = \alpha_2 = 0$: the transmission equation is $\sqrt{\varepsilon'_{\perp 2}}\, sin\,\hat{\xi}^{o}_{2t} = \frac{\eta_1 \sqrt{\varepsilon'_{\perp 1}}\, sin\,\hat{\xi}^{e}_{1i}}{\sqrt{cos^2\,\hat{\xi}^{e}_{1i} + \eta_1 sin^2\,\hat{\xi}^{e}_{1i}}}$, and the transmitted angle always exists for $\eta_1 \varepsilon'_{\perp 1} < \varepsilon'_{\perp 2}$, of transmitted cone upper boundary $sin\,\hat{\xi}^{o}_{2m} = \sqrt{\frac{\eta_1 \varepsilon'_{\perp 1}}{\varepsilon'_{\perp 2}}}$, while a total extraordinary/ordinary reflection occurs when $\eta_1 \varepsilon'_{\perp 1} \ge \varepsilon'_{\perp 2}$ and $\hat{\xi}_{1i} > \hat{\xi}^{o}_{1m} = Arc\,sin\left[\sqrt{\frac{\varepsilon'_{\perp 2}}{\eta_1^2 \varepsilon'_{\perp 1} + (1-\eta_1)\varepsilon'_{\perp 2}}}\right]$,

* case $\alpha_1 = \alpha_2 = \frac{\pi}{2}$: the transmission equation is $\sqrt{\varepsilon'_{\perp 2}}\, sin\,\hat{\xi}_{2t} = \eta_1 \sqrt{\varepsilon'_{\perp 1}} \left(\frac{sin\,\hat{\xi}_{1i}}{\sqrt{sin^2\,\hat{\xi}_{1i} + \eta_1 cos^2\,\hat{\xi}_{1i}}}\right)^{\frac{1}{\eta_1}}$, and the transmitted angle always exists for $\eta_1 \sqrt{\varepsilon'_{\perp 1}} < \sqrt{\varepsilon'_{\perp 2}}$ inside the cone of upper boundary $sin\,\hat{\xi}^{o}_{2m} = \eta_1 \sqrt{\frac{\varepsilon'_{\perp 1}}{\varepsilon'_{\perp 2}}}$, while a total reflection occurs for $\eta_1 \sqrt{\varepsilon'_{\perp 1}} \ge \sqrt{\varepsilon'_{\perp 2}}$ and $\hat{\xi}_{1i} > \hat{\xi}_{1m} = Arc\,sin\left[\sqrt{\frac{\eta_1 \varepsilon'^{\eta_1}_{\perp 2}}{\left(\eta_1^{2\eta_1} + \eta_1 - 1\right)\varepsilon'^{\eta_1}_{\perp 1}}}\right]$.

Following now the previous procedure for the general case, one examines first incident rays from left to right, i.e. $\hat{\xi}^{e}_{1i}\in\left[0,\frac{\pi}{2}\right]$, with $\sup_{\hat{\xi}^{e}_{1i}\in\left[0,\frac{\pi}{2}\right]} F_1\left(\hat{\xi}^{e}_{1i}\right)=\hat{\Omega}_1$ : one immediately deduces that if $\sqrt{\varepsilon'_{\perp 1}}\,\hat{\Omega}_1\leq\sqrt{\varepsilon'_{\perp 2}}$ all the left to right incident rays can be transmitted in ordinary rays in crystal 2, while if $\sqrt{\varepsilon'_{\perp 1}}\,\hat{\Omega}_1>\sqrt{\varepsilon'_{\perp 2}}$ , there is a total reflection, the left boundary of the extraordinary/ordinary transmission cone $C^{t}_{e\to o}$ being solution of $F_1\left(\hat{\xi}^{e\to o}_{1m}\right)=\sqrt{\frac{\varepsilon'_{\perp 2}}{\varepsilon'_{\perp 1}}}$; note that $F_1\left(\hat{\xi}^{e\to e}_{1m}\right)=\hat{\Omega}_2\sqrt{\frac{\varepsilon'_{\perp 2}}{\varepsilon'_{\perp 1}}}$ is the equation of the extraordinary/extraordinary transmission cone $C^{t}_{e\to e}$, from which it comes $F_1\left(\hat{\xi}^{e\to e}_{1m}\right)=\hat{\Omega}_2 F_1\left(\hat{\xi}^{e\to o}_{1m}\right)$; hence, since $F_1$ is a strictly increasing function on $\left[0,\frac{\pi}{2}\right]$, one deduces that if $\hat{\Omega}_2<1$, $\hat{\xi}^{e\to e}_{1m}\leq\hat{\xi}^{e\to o}_{1m}$ and $C^{t}_{e\to e}\subset C^{t}_{e\to o}$, while if $\hat{\Omega}_2>1$ it comes $C^{t}_{e\to o}\subset C^{t}_{e\to e}$; this case can be summarised as:

* if $\sqrt{\varepsilon'_{\perp 2}}<\hat{\Omega}_2\sqrt{\varepsilon'_{\perp 2}}<\hat{\Omega}_1\sqrt{\varepsilon'_{\perp 1}}$ , there is two possible total reflections for left to right incident extraordinary rays, an extraordinary/extraordinary (e/e) one and an extraordinary/ordinary (e/o) one, with $C^{t}_{e\to o}\subset C^{t}_{e\to e}$,

* if $\hat{\Omega}_2\sqrt{\varepsilon'_{\perp 2}}<\sqrt{\varepsilon'_{\perp 2}}<\hat{\Omega}_1\sqrt{\varepsilon'_{\perp 1}}$ , the behaviour is similar as before with $C^{t}_{e\to e}\subset C^{t}_{e\to o}$,

* if $\sqrt{\varepsilon'_{\perp 2}}<\hat{\Omega}_1\sqrt{\varepsilon'_{\perp 1}}<\hat{\Omega}_2\sqrt{\varepsilon'_{\perp 2}}$ , there is one possible total reflection e/o and no total reflection e/e (hence $C^{t}_{e\to e}=\left[0,\frac{\pi}{2}\right]$) and $C^{t}_{e\to o}\subset C^{t}_{e\to e}$,

* if $\hat{\Omega}_2\sqrt{\varepsilon'_{\perp 2}}<\hat{\Omega}_1\sqrt{\varepsilon'_{\perp 1}}<\sqrt{\varepsilon'_{\perp 2}}$ , there is one possible total reflection e/e and no total reflection e/o (hence $C^{t}_{e\to o}=\left[0,\frac{\pi}{2}\right]$) and $C^{t}_{e\to e}\subset C^{t}_{e\to o}$,

On the other hand, when $\hat{\Omega}_1\sqrt{\varepsilon'_{\perp 1}}\leq\sqrt{\varepsilon'_{\perp 2}}$ , the boundary of the ordinary transmitted cone $\hat{C}^{t}_{e\to o}$ is simply given by $\hat{\xi}^{e\to o}_{2m}=Arc\sin\left(\frac{\hat{\Omega}_1\sqrt{\varepsilon'_{\perp 1}}}{\sqrt{\varepsilon'_{\perp 2}}}\right)$, while the boundary of the extraordinary transmitted cone $\hat{C}^{t}_{e\to e}$, which exists if and only if $\hat{\Omega}_1\sqrt{\varepsilon'_{\perp 1}}\leq\hat{\Omega}_2\sqrt{\varepsilon'_{\perp 2}}$ , is solution of $F_2\left(\hat{\xi}^{e\to e}_{2m}\right)=\frac{\hat{\Omega}_1\sqrt{\varepsilon'_{\perp 1}}}{\sqrt{\varepsilon'_{\perp 2}}}\Leftrightarrow F_2\left(\hat{\xi}^{e\to e}_{2m}\right)=\sin\hat{\xi}^{e\to o}_{2m}$; note that if $\hat{\Omega}_2<1$, the existence of $\hat{C}^{t}_{e\to e}$ obviously implies the existence of $\hat{C}^{t}_{e\to o}$, since $\hat{\Omega}_1\sqrt{\varepsilon'_{\perp 1}} \le \hat{\Omega}_2\sqrt{\varepsilon'_{\perp 2}} \le \sqrt{\varepsilon'_{\perp 2}}$, or equivalently for $\hat{\xi}^{e\to e}_{2m} \in \left[0,\frac{\pi}{2}\right]$, $sin\,\hat{\xi}^{e\to o}_{2m} = F_2\left(\hat{\xi}^{e\to e}_{2m}\right) \le \hat{\Omega}_2 \le 1$; similarly, if $\hat{\Omega}_2 > 1$, the existence of $\hat{C}^t_{e\to o}$ implies the existence of $\hat{C}^t_{e\to e}$.

For incident rays from right to left, i.e. $\hat{\xi}^e_{1i} \in \left[\frac{3\pi}{2},\frac{3\pi}{2}+\alpha_1\right[ \cup \left]\frac{3\pi}{2}+\alpha_1,2\pi\right]$, the fundamental relation still verifies $\sqrt{\varepsilon'_{\perp 2}}\,sin\,\hat{\xi}^o_{2t} = \sqrt{\varepsilon'_{\perp 1}}\,F_1\left(\hat{\xi}^e_{1i}\right)$, the transmitted ordinary ray being not affected by the discontinuity line in crystal 2; then, an incident extraordinary ray between the right cut line in crystal 1 and the real interface can be transmitted in crystal 2 under its ordinary form if there exists $\hat{\xi}_{2t} \in \left[\frac{3\pi}{2},2\pi\right]$ such that $\sqrt{\varepsilon'_{\perp 1}}\,\overset{\frown}{\Omega}_1 \le \sqrt{\varepsilon'_{\perp 2}}\,sin\,\hat{\xi}_{2t} \le \sqrt{\varepsilon'_{\perp 1}}\,\hat{\Omega}_1$, so that the boundaries of the transmitted cone (if they exist) must be solutions of $\ddot{\xi}^{e\to o}_{2t} = 2\pi - Arc\,sin\left(\frac{\overset{\frown}{\Omega}_1\sqrt{\varepsilon'_{\perp 1}}}{\sqrt{\varepsilon'_{\perp 2}}}\right)$ for the lower one, and $\tilde{\xi}^{e\to o}_{2t} = 2\pi - Arc\,sin\left(\frac{\hat{\Omega}_1\sqrt{\varepsilon'_{\perp 1}}}{\sqrt{\varepsilon'_{\perp 2}}}\right)$ for the upper one; hence one deduces for $\hat{\xi}^e_{1i} \in \left[\frac{3\pi}{2},\frac{3\pi}{2}+\alpha_1\right[$ :

* if $\sqrt{\varepsilon'_{\perp 1}}\,\overset{\frown}{\Omega}_1 < \sqrt{\varepsilon'_{\perp 1}}\,\hat{\Omega}_1 < \sqrt{\varepsilon'_{\perp 2}}$, the incident allowed rays are $\hat{\xi}_{1i} \in \left[\frac{3\pi}{2},\frac{3\pi}{2}+\alpha_1\right[$ and the transmitted allowed rays are $\hat{\xi}_{2t} \in \left[\tilde{\xi}^{e\to o}_{2t},\ddot{\xi}^{e\to o}_{2t}\right]$, the transmitted ordinary ray being simply given by $\hat{\xi}^o_{2t} = Arc\,sin\left[\frac{\sqrt{\varepsilon'_{\perp 1}}\,F_1\left(\hat{\xi}^e_{1i}\right)}{\sqrt{\varepsilon'_{\perp 2}}}\right]$,

* if $\sqrt{\varepsilon'_{\perp 1}}\,\overset{\frown}{\Omega}_1 < \sqrt{\varepsilon'_{\perp 2}} < \sqrt{\varepsilon'_{\perp 1}}\,\hat{\Omega}_1$, the incident allowed rays are $\hat{\xi}_{1i} \in \left[\tilde{\xi}^{e\to o}_{1i},\frac{3\pi}{2}+\alpha_1\right[$ and the transmitted allowed ordinary rays are $\hat{\xi}^o_{2t} \in \left[\frac{3\pi}{2},\ddot{\xi}^{e\to o}_{2t}\right]$, with $\tilde{\xi}^{e\to o}_{1i}$ solution of $F_1\left(\tilde{\xi}^{e\to o}_{1i}\right) = \sqrt{\frac{\varepsilon'_{\perp 2}}{\varepsilon'_{\perp 1}}}$,

* if $\sqrt{\varepsilon'_{\perp 2}} < \sqrt{\varepsilon'_{\perp 1}}\,\overset{\frown}{\Omega}_1 < \sqrt{\varepsilon'_{\perp 1}}\,\hat{\Omega}_1$, no incident extraordinary ray $\hat{\xi}_{1i} \in \left[\frac{3\pi}{2},\frac{3\pi}{2}+\alpha_1\right[$ can be transmitted under its ordinary form in crystal 2.

Incident rays in crystal 1 between the right cut line of crystal 1 and its geometric axis can be transmitted in crystal 2 under its ordinary form if there exists $\hat{\xi}_{2t} \in \left[\frac{3\pi}{2},2\pi\right]$ such that $\sqrt{\varepsilon'_{\perp 2}}\,sin\,\hat{\xi}_{2t} \le \sqrt{\varepsilon'_{\perp 1}}\,\breve{\Omega}_1$, so that the boundary of the transmitted cone (if it exists) must be $\tilde{\xi}^{e\to o}_{2t} = 2\pi - Arc\,sin\left(\frac{\breve{\Omega}_1\sqrt{\varepsilon'_{\perp 1}}}{\sqrt{\varepsilon'_{\perp 2}}}\right)$; since the ordinary transmitted ray is not affected by the discontinuity line, a pseudo total reflection may arise also for right to left incident rays, for which $F_1\left(\tilde{\xi}^{e\to o}_{1i}\,\hat{\xi}^e_{1i}\right) = \sqrt{\frac{\varepsilon'_{\perp 2}}{\varepsilon'_{\perp 1}}}$; hence it comes for $\hat{\xi}^e_{1i} \in \left]\frac{3\pi}{2}+\alpha_1,2\pi\right]$:

* if $\sqrt{\varepsilon'_{\perp 1}}\,\breve{\Omega}_1 < \sqrt{\varepsilon'_{\perp 2}}$ , the incident allowed rays are $\hat{\xi}_{1i} \in \left]\frac{3\pi}{2}+\alpha_1 , 2\pi\right]$ and the transmitted allowed rays are $\hat{\xi}^{o}_{2t} \in \left[\breve{\xi}^{e\to o}_{2t} , 2\pi\right]$,

* if $\sqrt{\varepsilon'_{\perp 2}} < \sqrt{\varepsilon'_{\perp 1}}\,\breve{\Omega}_1$ , the incident allowed rays are $\hat{\xi}_{1i} \in \left[\breve{\xi}^{e\to o}_{1i} , 2\pi\right]$ and the transmitted allowed rays are $\hat{\xi}^{o}_{2t} \in \left[\frac{3\pi}{2} , 2\pi\right]$.

The illustrating figures 7a-d below present the ordinary transmitted angle (between the geometrical axis and the transmitted ray) function of the extraordinary incident angle (between the geometrical axis and the incident ray), for crystals whose characteristics are $\sqrt{\varepsilon'_{\perp 1}}=1.5$ , $\eta_1=2.0$ and $\alpha_1=30.0°$ for crystal 1, $\sqrt{\varepsilon'_{\perp 2}}=2.0$ , [ $\eta_2=2.0$ and $\alpha_1=60.0°$ ] for crystal 2 taken as an isotropic medium; in this situation, there is an ordinary pseudo total reflection on the left, the boundary of the ordinary transmission cone being at $\hat{\xi}_{1i}=35.49°$, and for incident rays from right to left below the discontinuity line, they can tend towards $\frac{3\pi}{2}+\alpha_1 \equiv 2\pi - \hat{\xi}_{1i}=60.0°$ , the corresponding transmitted value being $2\pi - \hat{\xi}_{2t}=43.60°$ , while for incident rays above the discontinuity line, none of them can be transmitted in crystal 2

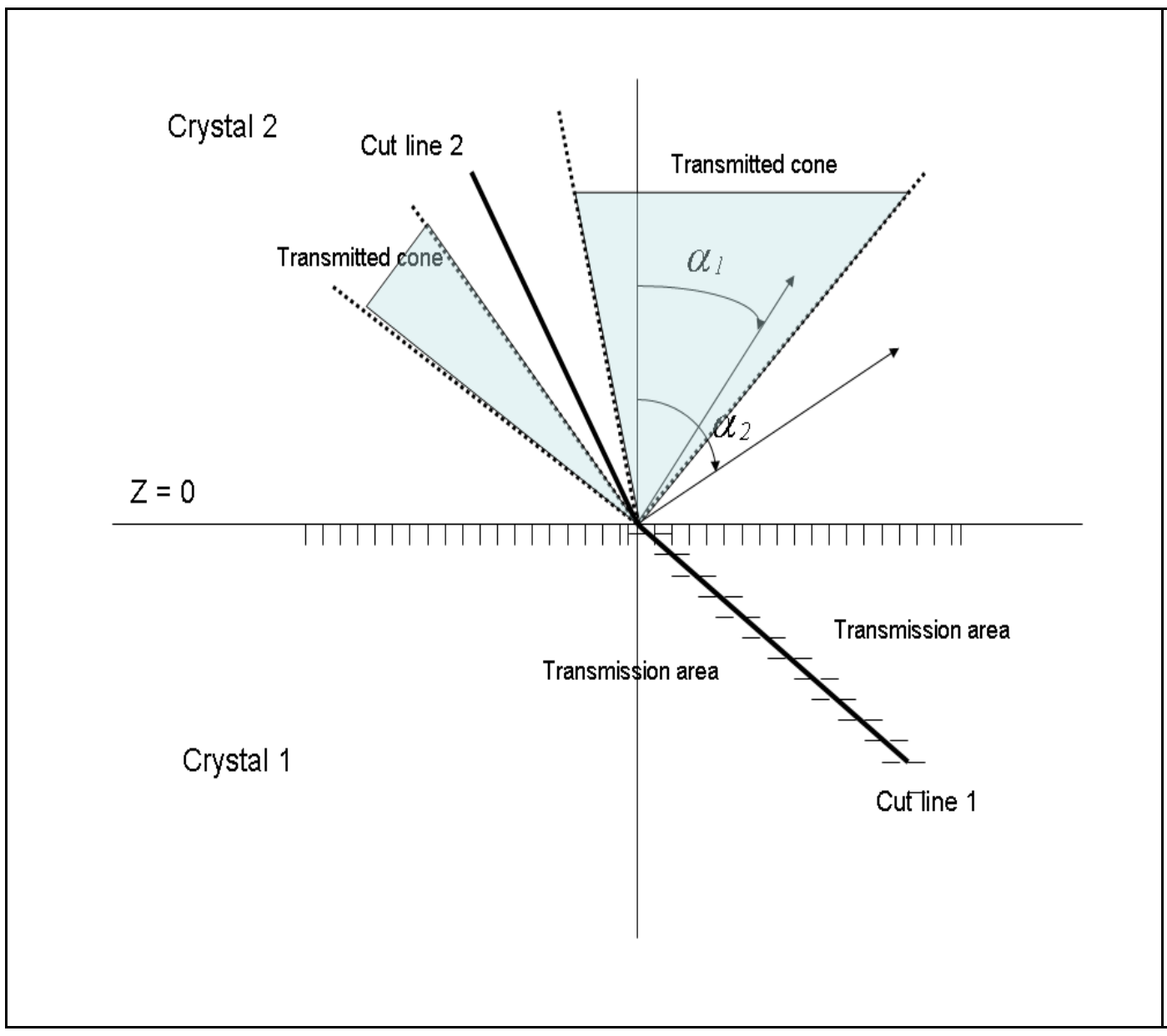

Fig. 7a: Extraordinary/extraordinary transmission

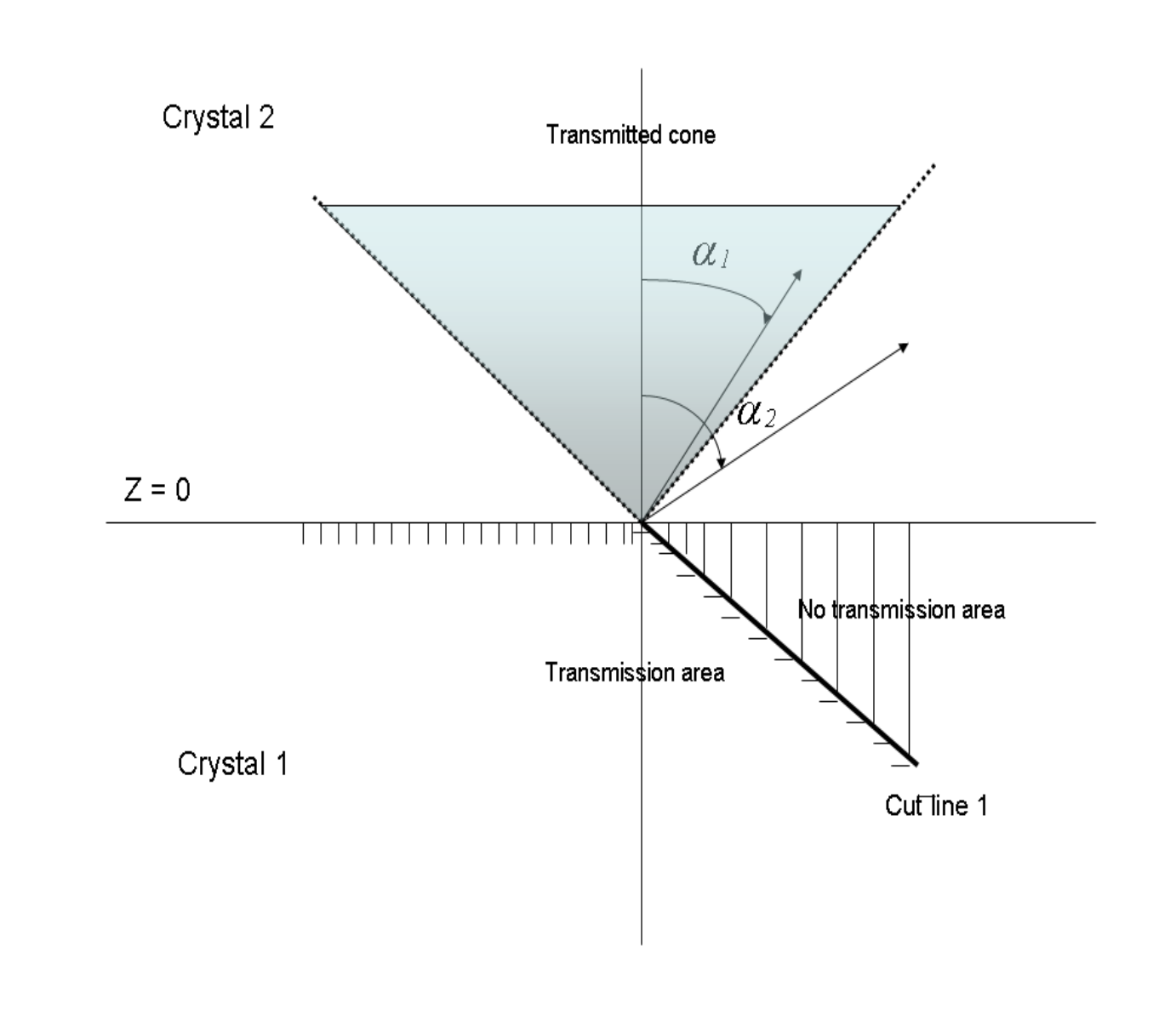

Fig. 7b: Extraordinary/ordinary transmission

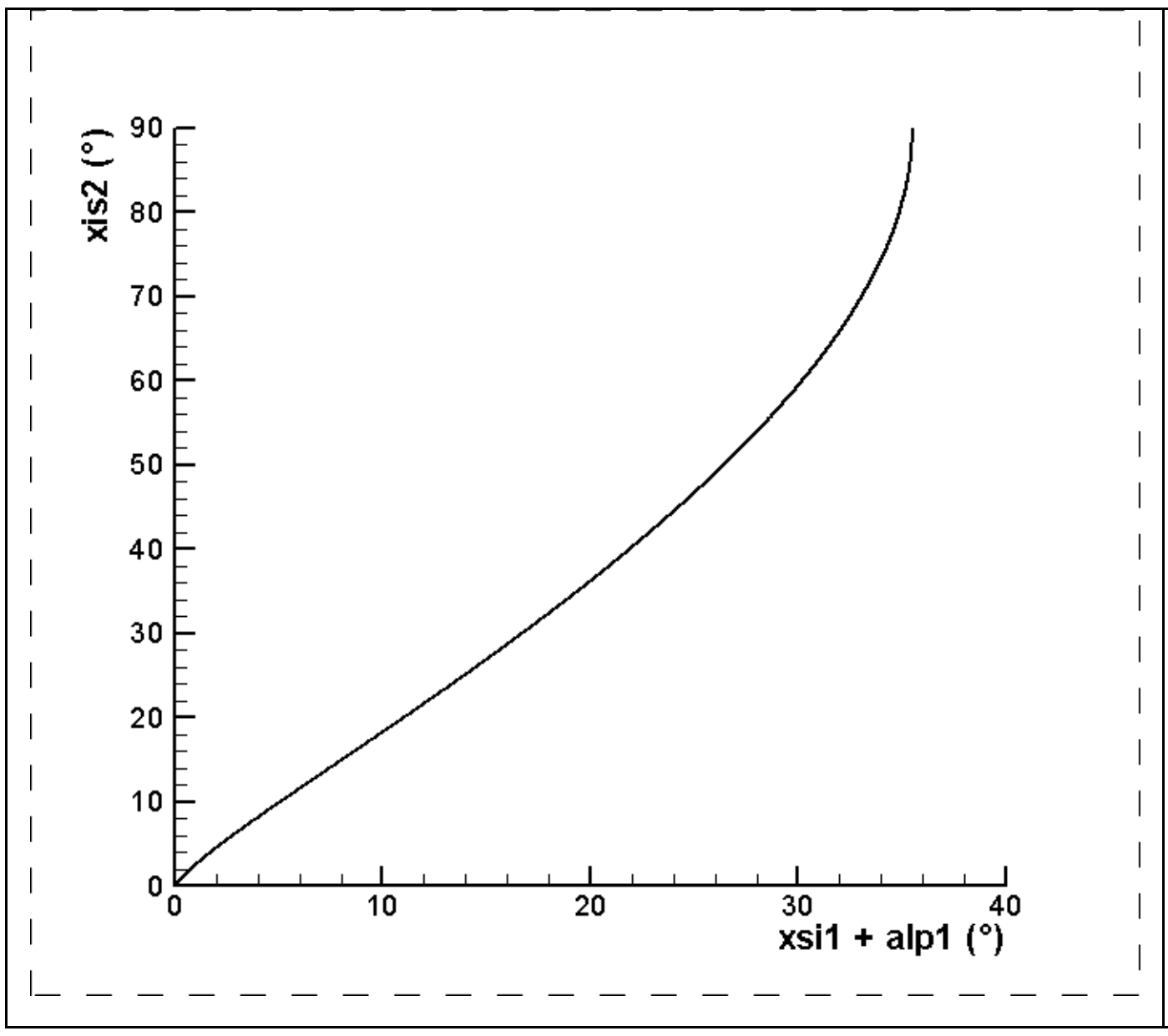

Fig. 7c: from left to right

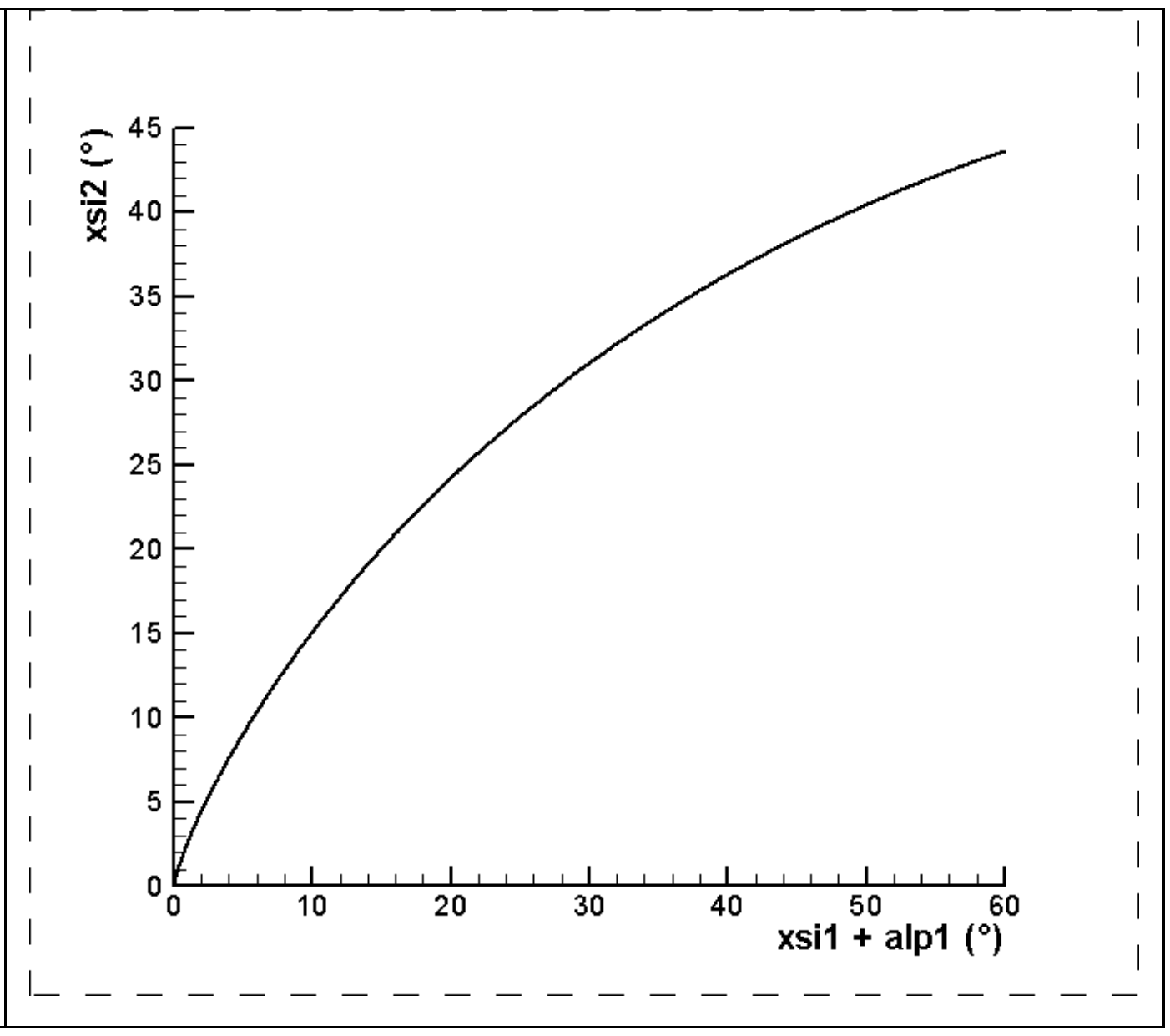

Fig. 7d: right to left, below the discontinuity line

As a second example, for crystals whose characteristics are $\sqrt{\varepsilon'_{\perp 1}}=1.5$, $\eta_1=2.0$ and $\alpha_1=22.5°$ for crystal 1, $\sqrt{\varepsilon'_{\perp 2}}=2.0$, [$\eta_2=0.5$ and $\alpha_2=45.0°$] for crystal 2 taken as an isotropic medium, there is an ordinary pseudo total reflection on the left, the boundary of the transmission cone being at $\hat{\xi}_{1i}=36.44°$, and for incident rays from right to left, they can tend towards $\frac{3\pi}{2}+\alpha_1\equiv 2\pi-\hat{\xi}_{1i}=67.5°$, the corresponding transmitted value being $2\pi-\hat{\xi}_{2t}=47.27°$; no extraordinary and ordinary transmission is possible for incident right to left rays between the discontinuity line and the real interface.

Hence it can exist right to left incident extraordinary rays above the virtual interface which cannot be reflected inside the crystal 1 under extraordinary ray, and which cannot also be transmitted in crystal2 under both extraordinary or ordinary rays: then these incident extraordinary rays must be absorbed at the interface. A similar discussion can be performed for the case ordinary/extraordinary reflexion/transmission.

After this geometrical study for uniaxial media, we shall try now to characterise from an energetic point of view, the reflection and transmission phenomena through an interface between two crystals, by calculating the energetic reflection and transmission factors.

## IV – CALCULATION OF THE REFLECTION AND TRANSMISSION EXTENDED FRESNEL'S FACTORS AT AN INTERFACE BETWEEN TWO UNIAXIAL CRYSTALS

### *IV-1 – Ordinary/ordinary reflection and transmission*

Let us first determine the reflection and transmission factors at an interface between two static homogeneous isotropic media from a corpuscular point of view, without any quantum effect: when a bundle of photons travelling in medium 1 for a given incident direction “impacts” a non moving target area made of particles of the interface between the two media, a part of the incident bundle is generally reflected back in medium 1, and it appears a transmitted bundle in medium 2, according to the symbolic equality:

$$\overline{N}_{1i}\overset{\Rightarrow}{P}_{1i}+\overline{N}_t\overset{\Rightarrow}{P}_t\rightarrow\overline{N}_{1r}\overset{\Rightarrow}{P}_{1r}+\overline{N}_{2t}\overset{\Rightarrow}{P}_{2t}+\overline{N}_m\overset{\Rightarrow}{P}_m, \quad (40)$$

where $\overline{N}_{1i}$ represents the number of incident indiscernible photons of the bundle in medium 1 taken as zero mass particles on the interface in a given direction, $\overline{N}_{1r}$ the number of reflected photons in medium 1, $\overline{N}_{2t}$ the number of transmitted photons in medium 2, $\overline{N}_t\overset{\Rightarrow}{P}_t$ the rest impulsion of the target area, and $\overline{N}_m\overset{\Rightarrow}{P}_m$ is an induced impulsion energy; $\overset{\Rightarrow}{P}_j$ is the 4-impulsion of a photon, given in vacuum by the well-know expression $\overset{\Rightarrow}{P}_j=\frac{E_j^{(0)}}{c}\begin{pmatrix}1\\ \vec{\Omega}_j\end{pmatrix}$, with $E_j^{(0)}$ the energy of the photon in vacuum, $E_j^{(0)}=h\nu_j$, and $\vec{\Omega}_j$ the unit direction of the photon of speed *c*; the rest target impulsion energy defines a mean target energy as $\overline{N}_t\overset{\Rightarrow}{P}_t=\overline{N}_{1i}\overset{\Rightarrow}{P}_{t0}=\overline{N}_{1i}\begin{pmatrix}\frac{E_{t0}}{c}\\ \vec{0}\end{pmatrix}$ since it is a non moving element, analogous to an interface-photon interaction energy, while $\overline{N}_m\overset{\Rightarrow}{P}_m$ is an induced 4-impulsion such that $\overline{N}_m\overset{\Rightarrow}{P}_m=\overline{N}_{1i}\overset{\Rightarrow}{P}_0=\overline{N}_{1i}\begin{pmatrix}\frac{E_{i0}}{c}\\ \vec{p}_0\end{pmatrix}$, $E_{i0}$ being the induced mean energy; then the conservation of the 4-impulsion energy can be symbolically rewritten as:

$$\overline{N}_{1i}\varepsilon'_{\perp 1}\frac{E_{1i}^{(0)}}{c}\begin{pmatrix}1\\ \vec{\Omega}_{1i}\end{pmatrix}+\overline{N}_{1i}\frac{E_{t0}}{c}\begin{pmatrix}1\\ \vec{0}\end{pmatrix}\rightarrow\overline{N}_{1r}\varepsilon'_{\perp 1}\frac{E_{1r}^{(0)}}{c}\begin{pmatrix}1\\ \vec{\Omega}_{1r}\end{pmatrix}+\overline{N}_{2t}\varepsilon'_{\perp 2}\frac{E_{2t}^{(0)}}{c}\begin{pmatrix}1\\ \vec{\Omega}_{2t}\end{pmatrix}+\overline{N}_{1i}\begin{pmatrix}\frac{E_{i0}}{c}\\ \vec{p}_0\end{pmatrix},$$

the latter symbolic expression being obtained for a given incident direction, i.e. a straight line; if the incident pencil of light is confined in an elementary solid angle $d\,\Omega_{1i}$ around a given direction, it comes then:

$$\varepsilon'_{\perp 1} E_{1i}^{(0)} \begin{pmatrix} 1 \\ \vec{\Omega}_{1i} \end{pmatrix} + \frac{d E_{t0}}{d \Omega_{1i}} \begin{pmatrix} 1 \\ \vec{0} \end{pmatrix} \rightarrow \frac{\overline{N}_{1r}}{\overline{N}_{1i}} \varepsilon'_{\perp 1} E_{1r}^{(0)} \begin{pmatrix} 1 \\ \vec{\Omega}_{1r} \end{pmatrix} \frac{d \Omega_{1r}}{d \Omega_{1i}} + \frac{\overline{N}_{2t}}{\overline{N}_{1i}} \varepsilon'_{\perp 2} E_{2t}^{(0)} \begin{pmatrix} 1 \\ \vec{\Omega}_{2t} \end{pmatrix} \frac{d \Omega_{2t}}{d \Omega_{1i}} + \begin{pmatrix} \frac{d E_{i0}}{d \Omega_{1i}} \\ \frac{d \vec{p}_0 c}{d \Omega_{1i}} \end{pmatrix},$$

where $d \Omega_{1r}$ and $d \Omega_{2t}$ are the reflected and transmitted solid angles, $\frac{d E_{t0}}{d \Omega_{1i}}$ is the interface-photon interaction energy density, $\frac{d E_{i0}}{d \Omega_{1i}}$ and $\frac{d \vec{p}_0}{d \Omega_{1i}}$ being the induced energy and impulsion densities; furthermore, one defines the energy reflection and transmission factors as $r = \frac{\overline{N}_{1r} E_{1r}^{(0)} d \Omega_{1r}}{\overline{N}_{1i} E_{1i}^{(0)} d \Omega_{1i}}$ and $t = \frac{\overline{N}_{2t} E_{2t}^{(0)} d \Omega_{2t}}{\overline{N}_{1i} E_{1i}^{(0)} d \Omega_{1i}}$; then, if the energy of the photons (i.e. the frequency of the photons) is not modified during the interaction with the interface, the reflection and transmission factors are simply $r = \frac{\overline{N}_{1r}}{\overline{N}_{1i}}$ for $d \Omega_{1r} = d \Omega_{1i}$, and $t = \frac{\overline{N}_{2t} d \Omega_{2t}}{\overline{N}_{1i} d \Omega_{1i}}$, which represent the fraction of the reflected photons by the incident ones, and the fraction of the transmitted photons inside the transmitted elementary cone by the incident ones inside the elementary transmission cone; noting $\hat{E}_{b0} = \frac{E_{b0}}{E_{1i}^{(0)}}$ and $\hat{p}_0 = \frac{p_0 c}{E_{1i}^{(0)}}$, one has then for the elementary reaction:

$$\varepsilon'_{\perp 1} \begin{pmatrix} 1 \\ \vec{\Omega}_{1i} \end{pmatrix} + \frac{d \hat{E}_{t0}}{d \Omega_{1i}} \begin{pmatrix} 1 \\ \vec{0} \end{pmatrix} \rightarrow r\, \varepsilon'_{\perp 1} \begin{pmatrix} 1 \\ \vec{\Omega}_{1r} \end{pmatrix} + t\, \varepsilon'_{\perp 2} \begin{pmatrix} 1 \\ \vec{\Omega}_{2t} \end{pmatrix} + \frac{d}{d \Omega_{1i}} \begin{pmatrix} \hat{E}_{i0} \\ \vec{\hat{p}}_0 \end{pmatrix}, \quad (41)$$

In an isotropic static medium of refractive index $n_j = \sqrt{\varepsilon'_{\perp j}}$, relatively to a reference observer, the 4-impulsion of a photon is, as developed in [1]:

$$\vec{\vec{P}}_j = \frac{n_j^2 h \nu_j}{c} \left[ \vec{\underline{e}}_0 + \sin \xi_j \vec{\underline{e}}_x + \cos \xi_j \vec{\underline{e}}_z \right] = \frac{n_j^2 h \nu_j}{c} \left[ \vec{e}_0 + \frac{1}{n_j} \left( \sin \xi_j \vec{e}_x + \cos \xi_j \vec{e}_z \right) \right], \quad (42)$$

where $\left(\overset{\rightarrow}{\overline{e_0}},\overset{\rightarrow}{\overline{e_x}},\overset{\rightarrow}{\overline{e_z}}\right)$ is the vacuum-like basis bound to the material medium and $\left(\overset{\rightarrow}{e_0},\overset{\rightarrow}{e_x},\overset{\rightarrow}{e_z}\right)$ is the observer basis in the flat medium associated to the metric tensor $\overline{\overline{g}}_j=\begin{pmatrix} -1 & 0 & 0 & 0 \\ 0 & n_j^2 & 0 & 0 \\ 0 & 0 & n_j^2 & 0 \\ 0 & 0 & 0 & n_j^2 \end{pmatrix}$, whence the conservation of the 4-impulsion leads to, since from the isotropic Fresnel-Descartes' law, one has $\xi_{1r}=\pi-\xi_{1i}$ and $n_1 \sin\xi_{1i}=n_2 \sin\xi_{2t}$ :

$$\begin{aligned}
&n_1^2(1-r)=n_2^2 t+\overline{\Delta E_0}\left(\xi_{1i},\theta^+,\varphi^+,\theta^-,\varphi^-\right)\\
&(1-r)n_1 \sin\xi_{1i}=t\, n_2 \sin\xi_{2t}+\frac{\overline{p_0}^+\left(\xi_{1i},\theta^+,\varphi^+\right)}{n_2}\sin\theta^+\cos\varphi^+ +\frac{\overline{p_0}^-\left(\xi_{1i},\theta^-,\varphi^-\right)}{n_1}\sin\theta^-\cos\varphi^-\\
&0=\frac{\overline{p_0}^+}{n_2}\sin\theta^+\sin\varphi^+ +\frac{\overline{p_0}^-}{n_1}\sin\theta^-\sin\varphi^-\\
&(1+r)n_1 \cos\xi_{1i}=t\, n_2 \cos\xi_{2t}+\frac{\overline{p_0}^+}{n_2}\cos\theta^+ +\frac{\overline{p_0}^-}{n_1}\cos\theta^-
\end{aligned}\quad ,\qquad (43)$$

where one has defined an impulsion $\overline{p_0}^+$ travelling in medium 2 with angle $\theta^+\in\left[0,\frac{\pi}{2}\right[$ , and an impulsion $\overline{p_0}^-$ travelling in medium 1 with angle $\theta^-\in\left]\frac{\pi}{2},\pi\right[$ , such that $\overline{\Delta E_0}=\frac{d\left(\hat{E}_{i0}^+ +\hat{E}_{i0}^- -\hat{E}_{t0}\right)}{d\,\Omega_{1i}}$ and $\overline{p_0}=\frac{d\,\hat{p}_0}{d\,\Omega_{1i}}$; assuming no creation of real particles, the two impulsions are associated to quasi-particles like phonons, so that since the pseudo-norms of the two impulsions are $\overset{\Rightarrow}{p}^2=p_{\mu\nu}\,p^{\mu\nu}=0=\begin{cases} E_{io}^{+\,2}-p_o^{+\,2}c^2 \\ E_{io}^{-\,2}-p_o^{-\,2}c^2 \end{cases}$, one deduces from that result that $E_{io}^+=p_o^+c$ and $E_{io}^-=p_o^-c$, or equivalently said, that $\overline{E_{io}^+}=\overline{p_o}^+$ and $\overline{E_{io}^-}=\overline{p_o}^-$; furthermore, due to the perfect symmetry of revolution, one chooses then isotropically emitted impulsions for any incident photon direction, i.e. $\overline{p_0}^+\left(\xi_{1i},\theta^+,\varphi^+\right)=\overline{p_0}^+\left(\xi_{1i}\right)$ and $\overline{p_0}^-\left(\xi_{1i},\theta^-,\varphi^-\right)=\overline{p_0}^-\left(\xi_{1i}\right)$, such that system (43) integrated over the whole emitted impulsions directions leads to:

$$\begin{aligned}
&n_1^2(1-r)=n_2^2 t+\overline{\Delta E_0}\left(\xi_{1i}\right)\\
&(1-r)n_1 \sin\xi_{1i}=t\, n_2 \sin\xi_{2t}\\
&(1+r)n_1 \cos\xi_{1i}=t\, n_2 \cos\xi_{2t}+\frac{1}{2}\left(\frac{\overline{p_0}^+}{n_2}-\frac{\overline{p_0}^-}{n_1}\right)
\end{aligned}\quad ,\qquad (44)$$

Hence $r+t=1$ for all incident ray, and the solution of the previous system is:

$$
\begin{aligned}
&r=\frac{n_2\cos\xi_{2t}-n_1\cos\xi_{1i}+\frac{1}{2}\left(\frac{\overline{p_0^+}}{n_2}-\frac{\overline{p_0^-}}{n_1}\right)}{n_1\cos\xi_{1i}+n_2\cos\xi_{2t}}\\
&t=\frac{2n_1\cos\xi_{1i}-\frac{1}{2}\left(\frac{\overline{p_0^+}}{n_2}-\frac{\overline{p_0^-}}{n_1}\right)}{n_1\cos\xi_{1i}+n_2\cos\xi_{2t}}\\
&\overline{\Delta E_0}=\overline{p_0^+}+\overline{p_0^-}-\overline{E_{t0}}=t\left(n_1^2-n_2^2\right)
\end{aligned}
\quad , \tag{45}
$$

with $r+t=1$ for all incident ray; note that the total reflection is $\begin{pmatrix} r=1\\ t=0\end{pmatrix}\Leftrightarrow\left(\frac{\overline{p_0^+}}{n_2}-\frac{\overline{p_0^-}}{n_1}=4n_1\cos\xi_{1i}\right)$ while similarly the complete transmission is $\begin{pmatrix} r=0\\ t=1\end{pmatrix}\Leftrightarrow\left(\frac{\overline{p_0^+}}{n_2}-\frac{\overline{p_0^-}}{n_1}=2\left(n_1\cos\xi_{1i}-n_2\cos\xi_{2t}\right)\right)$, the condition $0\le t\le 1$ implying:

$$2\left(n_1\cos\xi_{1i}-n_2\cos\xi_{2t}\right)\le\frac{\overline{p_0^+}}{n_2}-\frac{\overline{p_0^-}}{n_1}\le 4n_1\cos\xi_{1i}$$

from which it easily comes $0\le r\le 1$. Obviously, the previous expressions are valid only when a transmitted luminous ray exists in medium 2. If a total reflection occurs, the interface-photon interaction energy is from what precedes $\overline{E_{t0}}=\left(1+\frac{n_1}{n_2}\right)\overline{p_0^+}-4n_1^2\cos\xi_{1i}$; a total reflection being naturally not realized for $n_2>n_1$, the previous expression for the interface-photon interaction energy is valid only for $n_2<n_1$ and $\xi_{1i}\in\left[\xi_{1i}^M,\frac{\pi}{2}\right]$, where $\xi_{1i}^M=Arc\sin\left(\frac{n_2}{n_1}\right)$; then, since at the limit $\xi_{1i}=\frac{\pi}{2}$ there is no direct interaction between the photon and the interface, one must have $\overline{E_{t0}}\left(\frac{\pi}{2}\right)=0$, from which it comes $\overline{p_0^+}\left(\xi_{1i}=\frac{\pi}{2}\right)=0$ and $\overline{p_0^-}\left(\xi_{1i}=\frac{\pi}{2}\right)=0$; at the other limit $\xi_{1i}=\xi_{1i}^M$, $\overline{p_0^-}\left(\xi_{1i}=\xi_{1i}^M\right)=0$,and more generally for all $\xi_{1i}\in\left[\xi_{1i}^M,\frac{\pi}{2}\right]$, one has $\overline{p_0^-}\left(\xi_{1i}\right)=0$, since it is not necessary in the impulsion conservation equation: hence it comes $\overline{p_0^+}=4n_1n_2\cos\xi_{1i}$ and $\overline{p_0^-}=0$ for $\xi_{1i}\in\left[\xi_{1i}^M,\frac{\pi}{2}\right]$, with $\overline{E_{t0}}=\overline{p_0^+}$, decaying function on $\left[\xi_{1i}^M,\frac{\pi}{2}\right]$; one concludes that there is a total reflection for an interface-photon interaction lower than the potential energy $\overline{E_{to}^m}$ , i.e. $\overline{E_{t0}}\leq\overline{E_{to}^m}=4\,n_2\sqrt{n_1^2-n_2^2}$ , the transmission being possible for $\overline{E_{t0}}>\overline{E_{to}^m}$ when $\xi_{1i}<\xi_{1i}^M$ . In this case, since the transmission factor is such that $0\leq t\leq 1$, one has $\overline{\Delta E_0}>0$, from which $\overline{E_{t0}}<\overline{p_0^+}+\overline{p_0^-}$ and $\left(\overline{p_0^+}+\overline{p_0^-}\right)(\xi_{1i})>4\,n_2\sqrt{n_1^2-n_2^2}$ .

In the case where $n_2>n_1$, there is no total reflection, except for the particular situation $\xi_{1i}=\frac{\pi}{2}$ , leading to $\frac{\overline{p_0^+}}{n_2}=\frac{\overline{p_0^-}}{n_1}$: the line $\xi_{2t}^M=Arc\sin\left(\frac{n_1}{n_2}\right)$ cannot be reached and is the external boundary of the transmission cone $\xi_{2t}\in\left[0,\xi_{2t}^M\right[$ for $\xi_{1i}\in\left[0,\frac{\pi}{2}\right]$; it plays a rule analogous to the one of the cut lines described for anisotropic media; note that if $\frac{\overline{p_0^+}}{n_2}=\frac{\overline{p_0^-}}{n_1}$ for all incident direction, the condition on the transmission factor $0\leq t\leq 1$ implies $n_1\cos\xi_{1i}-n_2\cos\xi_{2t}\leq 0$, that is $n_2>n_1$: hence there is a complete reflection only for $\xi_{1i}=\frac{\pi}{2}$ when $n_2>n_1$, and the solution $\frac{\overline{p_0^+}}{n_2}=\frac{\overline{p_0^-}}{n_1}=constant$ for all incident rays is an admissible one; then one can choose $\overline{p_0^+}=\overline{p_0^-}=0$, i.e. no induced impulsion, and in this case the reflection and transmission factors are simply:

$$r=\frac{n_2\cos\xi_{2t}-n_1\cos\xi_{1i}}{n_1\cos\xi_{1i}+n_2\cos\xi_{2t}}=\frac{\sin\left(\xi_{1i}-\xi_{2t}\right)}{\sin\left(\xi_{1i}+\xi_{2t}\right)}\qquad t=\frac{2\,n_1\cos\xi_{1i}}{n_1\cos\xi_{1i}+n_2\cos\xi_{2t}}\qquad for\;\xi_{1i}\in\left[0,\frac{\pi}{2}\right],\qquad(46)$$

possible only for $n_2>n_1$: here the two impulsions need not be created, and $\overline{\Delta E_0}=-\overline{E_{t0}}<0$: the interface-photon interaction energy is simply related to the transmission factor; when $n_2<n_1$ however, $\overline{\Delta E_0}>0$ and the reflection factor defined by Eq. (46) is negative, without any clear physical significance from its definition: hence in this situation one must take into account a non zero induced 4-impulsion to insure a positive reflection factor: it is to note, that, since for $n_2>n_1$ an additional impulsion need not exist, it is efficient to choose only one induced positive impulsion in medium 2 when $n_2<n_1$: indeed $n_1\cos\xi_{1i}>n_2\cos\xi_{2t}$, which leads to $0\leq\frac{\overline{p_0^+}}{n_2}-\frac{\overline{p_0^-}}{n_1}\leq 4\,n_1\cos\xi_{1i}$; hence an induced backward impulsion is not necessary and $\overline{p_0^-}\equiv 0$, from which it comes $0\leq\overline{p_0^+}\leq 4\,n_1 n_2\cos\xi_{1i}$ and:

* if $n_2>n_1$, $\quad\overline{E_{t0}}=t\left(n_2^2-n_1^2\right)$

$$* \text{ if } n_2<n_1, \quad \overline{E_{t0}}=\begin{cases} t^*\left(n_2^2-n_1^2\right)+\overline{p_0^+} & for\ \xi_{1i}\in\left[0,\xi_{1i}^M\right] \\ 4\,n_1 n_2 \cos\xi_{1i} & for\ \xi_{1i}\in\left[\xi_{1i}^M,\frac{\pi}{2}\right] \end{cases}$$

where $t^*$ is the transmission factor of case $n_2<n_1$ (different from the transmission factor of case $n_2>n_1$) Under this form, it is obvious that $\overline{E_{t0}}(1\to 2)\neq\overline{E_{t0}}(2\to 1)$. One names $\overline{\Delta E_0}$ the activation energy, such that if $\overline{\Delta E_0}<0$ (i.e. $n_2>n_1$) the incident bundle of photons in medium 1 is reflected at the separating interface for one part in medium 1 and transmitted in medium 2 for the other part, while if $\overline{\Delta E_0}>0$ (i.e. $n_2<n_1$), a bundle of induced non zero impulsion appears, forward in medium 2, in an isotropic diffuse way; when $\overline{\Delta E_0}=0$, with $n_2\neq n_1$, a total reflection is effective, which can be realized only for $t=0$. At the total reflection angle $\xi_{1i}^M=Arc\sin\left(\frac{n_2}{n_1}\right)$, which is analogous to a cut line, one must have like previously $r=1$ and $t=0$, from which it comes for this particular direction $\overline{p_0^+}=4\,n_2\sqrt{n_1^2-n_2^2}$, and for angles between the isotropic 2[nd] cut line and the interface, one has $\overline{p_0^+}=4\,n_1 n_2\cos\xi_{1i}$; for continuously varying photon directions, one must also have for angles below the isotropic cut line, i.e. $\xi_{1i}\in\left[0,\xi_{1i}^M\right]$, $2\left(n_1\cos\xi_{1i}-n_2\cos\xi_{2t}\right)\leq\frac{\overline{p_0^+}}{n_2}\leq 4\,n_1\cos\xi_{1i}$ with $\lim_{\xi_{1i}\to\xi_{1i}^M}\overline{p_0^+}=4\,n_2\sqrt{n_1^2-n_2^2}$, from which impulsions such that $\frac{\overline{p_0^+}}{n_2}=4\left(n_1\cos\xi_{1i}-n_2\cos\xi_{2t}\right)$ verify the needed conditions: then, when $n_2<n_1$, the reflection and transmission factors from medium 1 to medium 2 are given by:

$$\begin{array}{lll} r=\dfrac{n_1\cos\xi_{1i}-n_2\cos\xi_{2t}}{n_1\cos\xi_{1i}+n_2\cos\xi_{2t}}=\dfrac{\sin\left(\xi_{2t}-\xi_{1i}\right)}{\sin\left(\xi_{1i}+\xi_{2t}\right)} & t=\dfrac{2\,n_2\cos\xi_{2t}}{n_1\cos\xi_{1i}+n_2\cos\xi_{2t}} & for\ \xi_{1i}\in\left[0,\xi_{1i}^M\right] \\ r=1 \qquad t=0 & & for\ \xi_{1i}\in\left[\xi_{1i}^M,\frac{\pi}{2}\right] \end{array}, \qquad (47)$$

Under this form, the reflection factors, so as the transmission ones, since $r+t=1$, verify the relation $r_{1\to 2}\left(\xi_{1i}\right)=r_{2\to 1}\left(\xi_{2t}\right)$, with $r_{1\to 2}\left(\xi_{1i}\right)\neq r_{2\to 1}\left(\xi_{1i}\right)$; indeed, noting $n'_1\equiv n_2$, $n'_2\equiv n_1$, $\xi'_{1i}\equiv\pi+\xi_{2t}$ and $\xi'_{2t}\equiv\pi+\xi_{1i}$ for the backward trajectory leads to, for $n_2<n_1$:

$$r_{1\to 2}=\frac{n_1\cos\xi_{1i}-n_2\cos\xi_{2t}}{n_1\cos\xi_{1i}+n_2\cos\xi_{2t}}=\frac{-n'_2\cos\xi'_{2t}+n'_1\cos\xi'_{1i}}{-\left(n'_2\cos\xi'_{2t}+n'_1\cos\xi'_{1i}\right)}=\frac{n'_2\cos\xi'_{2t}-n'_1\cos\xi'_{1i}}{n'_1\cos\xi'_{1i}+n'_2\cos\xi'_{2t}}=r_{1'\to 2'}=r_{2\to 1}$$

where $r_{1'\rightarrow 2'}$ with $n'_2 > n'_1$ is obtained by Eq. (46). Reciprocally, if for the conjugate angles $\xi_{1i}$ and $\xi_{2t}$, the reflection factors verify $r_{1\rightarrow 2}(\xi_{1i}) = r_{2\rightarrow 1}(\xi_{2t})$, it comes from what precedes, for $n_2 < n_1$:

$$r_{1\rightarrow 2} = \frac{\frac{\overline{p_0}^{+}}{2n_2} + n_2\cos\xi_{2t} - n_1\cos\xi_{1i}}{n_1\cos\xi_{1i} + n_2\cos\xi_{2t}} = r_{2\rightarrow 1} = r_{1'\rightarrow 2'} = \frac{n'_2\cos\xi'_{2t} - n'_1\cos\xi'_{1i}}{n'_1\cos\xi'_{1i} + n'_2\cos\xi'_{2t}} = \frac{n_1\cos\xi_{1i} - n_2\cos\xi_{2t}}{n_1\cos\xi_{1i} + n_2\cos\xi_{2t}}$$

from which one immediately obtains $\frac{\overline{p_0}^{+}}{n_2} = 4\left(n_1\cos\xi_{1i} - n_2\cos\xi_{2t}\right)$: this value for the induced impulsion is the only one which insures the equality of the reciprocal reflection factors for light; then one has the photon-interface interaction energy:

* if $n_2 > n_1$, $\overline{E_{t0}} = \frac{2n_1\left(n_2^2 - n_1^2\right)\cos\xi_{1i}}{n_1\cos\xi_{1i} + n_2\cos\xi_{2t}}$

* if $n_2 < n_1$, $\overline{E_{t0}} = \begin{cases} 2n_2\left[2\left(n_1\cos\xi_{1i} - n_2\cos\xi_{2t}\right) - \frac{\left(n_1^2 - n_2^2\right)\cos\xi_{2t}}{n_1\cos\xi_{1i} + n_2\cos\xi_{2t}}\right] & for\ \xi_{1i} \in \left[0, \xi_{1i}^M\right] \\ 4n_1n_2\cos\xi_{1i} & for\ \xi_{1i} \in \left[\xi_{1i}^M, \frac{\pi}{2}\right] \end{cases}$

Obviously the photon-interface interaction energy has no symmetry properties as the reflection and transmission factors, for $\overline{E_{t01\rightarrow 2}}(\xi_{1i}) \neq \overline{E_{t02\rightarrow 1}}(\xi_{2t})$: there is not a perfect symmetry relatively to the interface for the problem of transmission/reflection, which is easy to understand, since in one direction there may exist a total reflection, while in the opposite direction, no total reflection is possible.

The evolution of the directional reflection factors and photon-interface interaction energy at an interface separating two isotropic media of different refractive indices is depicted on Figs. 8a-b: in the presented example, the refractive indices are $n_{air} = 1.0$ and $n_{glass} = 1.5$; the solid line shows the directional reflection factor curve from air to glass, for which there is no total reflection, while the dotted line is the reflection factor curve from glass to air where a total reflection is located from $\xi_{1i} = 41.8°$ to 90°; note that this curve is continuous for all incident angle, and that obviously $r_{air\rightarrow glass}(\xi_{1i}) \neq r_{glass\rightarrow air}(\xi_{1i})$ but $r_{air\rightarrow glass}(\xi_{1i}) = r_{glass\rightarrow air}(\xi_{2t})$, and also the derivative (relatively to the angle $\xi_{1i}$) of the reflection factor is not continuous, since it is easy to verify that $\lim_{\xi_{1i}\to\xi_{1i}^M}\frac{dr}{d\,\xi_{1i}}=\frac{1}{2\,n_2\sqrt{n_1^2-n_2^2}}\lim_{\xi_{1i}\to\xi_{1i}^M}\frac{d\,\overline{p}_0^{-+}}{d\,\xi_{1i}}$, and $\lim_{\xi_{1i}\to\xi_{1i}^M}\frac{d\left(\cos\xi_{2t}\right)}{d\,\xi_{1i}}=-\infty$

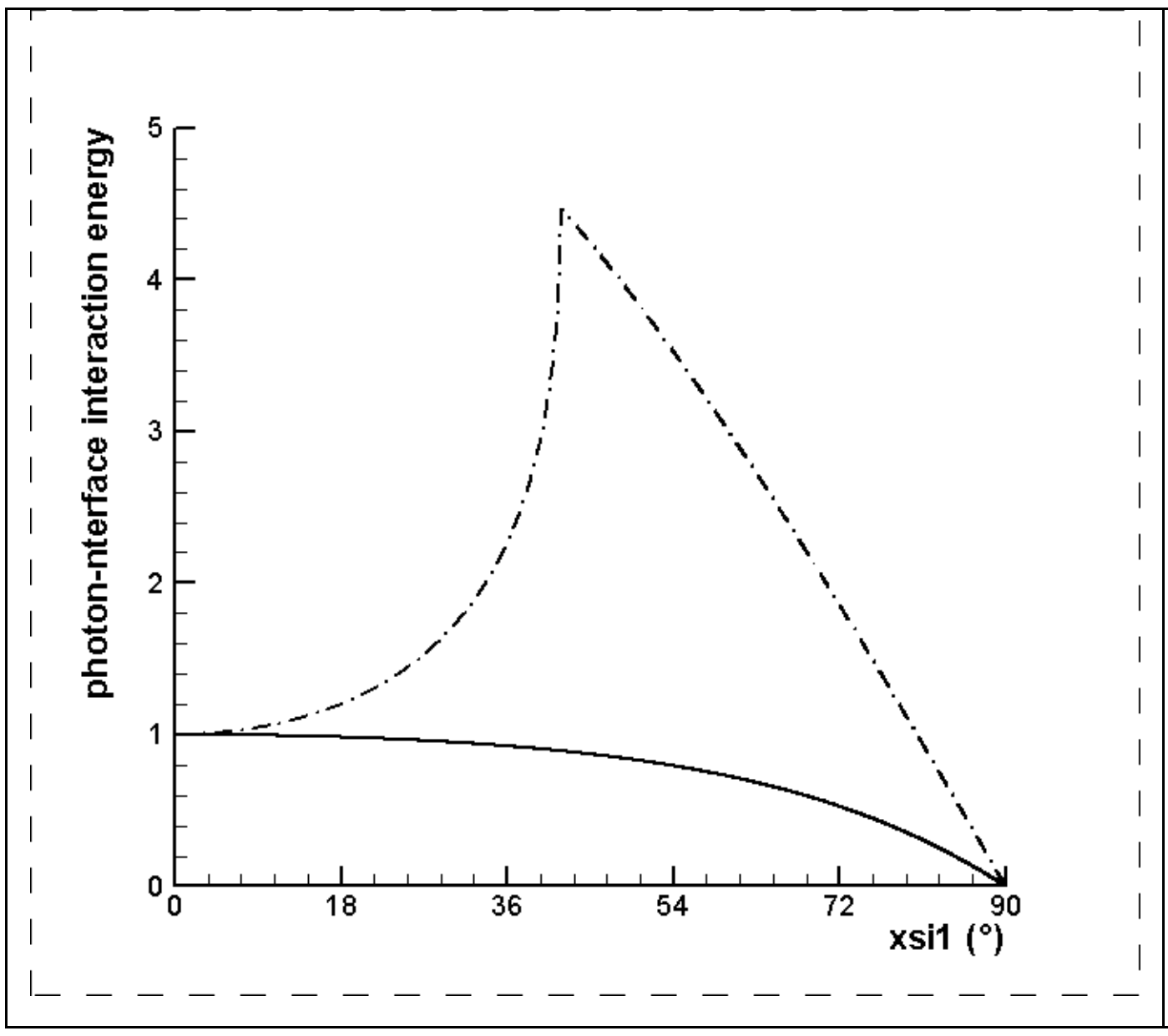

Fig. 8a: directional photon-interface interaction energy

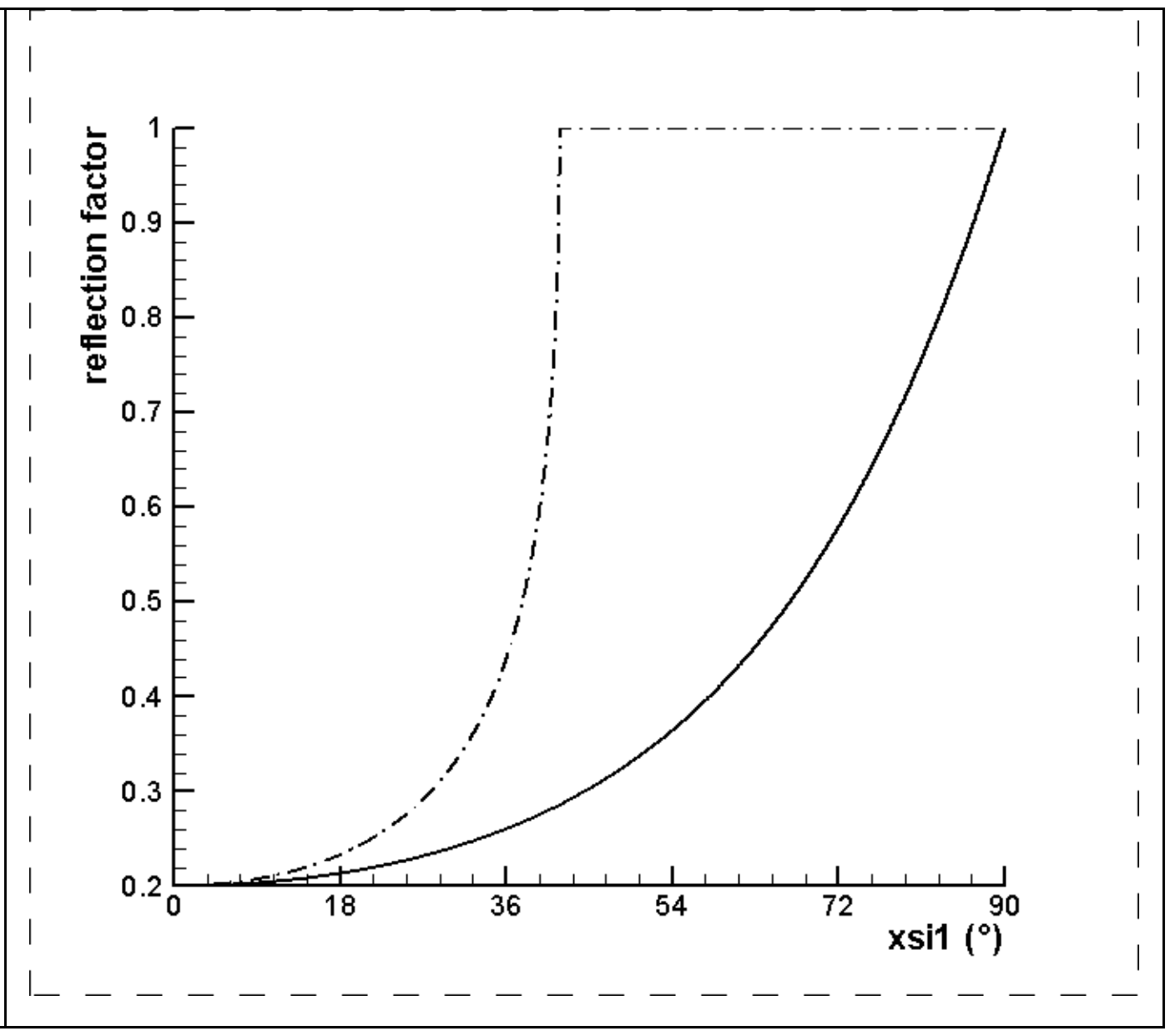

Fig. 8b: directional reflection factors

Then it comes for the hemispheric reflection factor from medium 1 to medium 2 when $n_2>n_1$:

$$R^{\cap}=\frac{1}{2\pi}\int_{\Omega=2\pi}r\,d\Omega=\int_{\xi_{1i}=0}^{\frac{\pi}{2}}r\left(\xi_{1i}\right)\sin\xi_{1i}\,d\,\xi_{1i}=\frac{2\sqrt{n_2^2-n_1^2}}{3\,n_1}-\frac{\left(n_1+2\,n_2\right)\left(n_2-n_1\right)}{3\,n_1\left(n_1+n_2\right)}$$

while the hemispheric transmission factor is such that $T^{\cap}=1-R^{\cap}$, the integrated hemispheric interaction interface-photon energy being $\overline{E}_{t0}^{\cap}=T^{\cap}\left(n_2^2-n_1^2\right)>0$: the hemispheric interaction energy is then zero if and only if $n_2=n_1$, which is obvious since there is no interface in this case, and no interaction. Similarly, it comes for the hemispheric reflection factor from medium 1 to medium 2 when $n_2<n_1$:

$$R^{\cap}=\frac{1}{2\pi}\int_{\Omega=2\pi}r\,d\Omega=\int_{\xi_{1i}=0}^{\frac{\pi}{2}}r\left(\xi_{1i}\right)\sin\xi_{1i}\,d\,\xi_{1i}=\frac{\sqrt{n_1^2-n_2^2}}{3\,n_1}-\frac{\left(n_1+2\,n_2\right)\left(n_1-n_2\right)}{3\,n_1\left(n_1+n_2\right)}+Arc\cos\left(\frac{n_2}{n_1}\right)$$

and for the hemispheric impulsion: $\overline{p}_0^{-+\cap}=2\,n_1\,n_2\left[1-\frac{n_1}{n_2}+\frac{n_1^2-n_2^2}{2\,n_2^2}Log\left(\frac{n_1+n_2}{n_1-n_2}\right)\right]$

Hence when a real interface separates two isotropic media of refractive indices $n_i$ and $n_j$, the two hemispheric reflection factors $R^{\cap}_{i\rightarrow j}$ and $R^{\cap}_{j\rightarrow i}$, with $R^{\cap}_{i\rightarrow j} \neq R^{\cap}_{j\rightarrow i}$, are:

$$\begin{aligned} R^{\cap}_{i\rightarrow j} &= \frac{2\sqrt{n_j^2 - n_i^2}}{3n_i} - \frac{(n_i + 2n_j)(n_j - n_i)}{3n_i(n_i + n_j)} \\ R^{\cap}_{j\rightarrow i} &= \frac{\sqrt{n_j^2 - n_i^2}}{3n_j} - \frac{(n_j + 2n_i)(n_j - n_i)}{3n_j(n_i + n_j)} + Arc\cos\left(\frac{n_i}{n_j}\right) \end{aligned}$$

for $n_i < n_j$, and symmetric expressions for $n_i > n_j$.

For instance as a numerical application, the hemispheric reflection factors between a slab of glass of refractive index *n* = 1.5 and its surrounding air environment of refractive index *n* = 1 are $R^{\cap}_{air\rightarrow glass} = 0.479$ and $R^{\cap}_{glass\rightarrow air} = 0.934$; similarly, the hemispheric photon-interface (non dimensional) interaction energies are $E^{\cap}_{air\rightarrow glass} = 0.652$ and $E^{\cap}_{glass\rightarrow air} = 1.435$.

Let us now remind a well-known result from electromagnetic theory for the ordinary plane waves in homogeneous media, and let us consider a linearly polarized incident ordinary wave travelling inside an isotropic crystal 1, so that the ordinary electromagnetic field inside the medium can be written under the form:

$$\begin{aligned} &\vec{E}_{1i} = E_{1i} e^{-i\varphi} \begin{pmatrix} \cos\psi_{1i}\cos\xi_{1i} \\ \sin\psi_{1i} \\ -\cos\psi_{1i}\sin\xi_{1i} \end{pmatrix} \quad \vec{D}_{1i} = \varepsilon_0 \varepsilon'_{\perp 1} \vec{E}_{1i} \quad \vec{B}_{1i} = \frac{\sqrt{\varepsilon'_{\perp 1}}}{c} E_{1i} e^{-i\varphi} \begin{pmatrix} -\sin\psi_{1i}\cos\xi_{1i} \\ \cos\psi_{1i} \\ \sin\psi_{1i}\sin\xi_{1i} \end{pmatrix} \\ &\vec{k}_{1i} = \frac{\omega_{1i}\sqrt{\varepsilon'_{\perp 1}}}{c} \begin{pmatrix} \sin\xi_{1i} \\ 0 \\ \cos\xi_{1i} \end{pmatrix} \quad \vec{\Pi}_{1i} = \frac{\sqrt{\varepsilon'_{\perp 1}}}{c} |E_{1i}|^2 \begin{pmatrix} \sin\xi_{1i} \\ 0 \\ \cos\xi_{1i} \end{pmatrix} \end{aligned}, \qquad (48)$$

where $\varphi$ is the wave phase depending on time and space, $\omega_{1i}$ is the wave angular frequency, $\vec{k}_{1i}$ is the incident wave vector and $\vec{\Pi}_{1i}$ is the incident Poynting vector, $\xi_{1i}$ being the incident propagation angle and $\psi_{1i}$ the incident polarization angle.

When impinging on the interface at $z = 0$, the incident wave can be reflected and transmitted in the isotropic crystal 2 in such a manner that the classical Descartes' law $\sin\xi_{1r} = \sin\xi_{1i}$ for the reflection and $\sqrt{\varepsilon'_{\perp 2}}\sin\xi_{2t} = \sqrt{\varepsilon'_{\perp 1}}\sin\xi_{1i}$ for the transmission is verified, with $\cos\xi_{1r} = -\cos\xi_{1i}$; furthermore, the wave phase conservation at the interface implies that the angular frequency remains constant: then the angular frequency of the reflected and transmitted waves is the angular frequency of the incident wave. To obtain the classical amplitude reflexion $r$ and transmission $t$ factors, one writes the continuity of the field components at the interface, namely:

* continuity of the tangential components of the electrical field:

$$\begin{aligned} &(49-a) \quad (\cos\psi_{1i} - r\cos\psi_{1r})\cos\xi_{1i} = t\cos\psi_{2t}\cos\xi_{2t} \\ &(49-b) \quad \sin\psi_{1i} + r\sin\psi_{1r} = t\sin\psi_{2t} \end{aligned} ,$$

* continuity of the normal component of the electrical induction:

$$(49-c) \quad \sqrt{\varepsilon'_{\perp 1}}(\cos\psi_{1i} + r\cos\psi_{1r}) = t\sqrt{\varepsilon'_{\perp 2}}\cos\psi_{2t} ,$$

* continuity of magnetic field:

$$\begin{aligned} &(49-d) \quad \sqrt{\varepsilon'_{\perp 1}}(\sin\psi_{1i} - r\sin\psi_{1r})\cos\xi_{1i} = t\sqrt{\varepsilon'_{\perp 2}}\sin\psi_{2t}\cos\xi_{2t} \\ &(49-e) \quad \sqrt{\varepsilon'_{\perp 1}}(\cos\psi_{1i} + r\cos\psi_{1r}) = t\sqrt{\varepsilon'_{\perp 2}}\cos\psi_{2t} \\ &(49-f) \quad \sin\psi_{1i} + r\sin\psi_{1r} = t\sin\psi_{2t} \end{aligned} ,$$

Hence there are 4 independent equations; the solution of this classical and well-known problem is easily obtained and can be given under the following form:

$$\begin{aligned} r\sin\psi_{1r} &= \frac{\sqrt{\varepsilon'_{\perp 1}}\cos\xi_{1i} - \sqrt{\varepsilon'_{\perp 2}}\cos\xi_{2t}}{\sqrt{\varepsilon'_{\perp 1}}\cos\xi_{1i} + \sqrt{\varepsilon'_{\perp 2}}\cos\xi_{2t}}\sin\psi_{1i} = r_{\perp}\sin\psi_{1i} \\ r\cos\psi_{1r} &= \frac{\sqrt{\varepsilon'_{\perp 2}}\cos\xi_{1i} - \sqrt{\varepsilon'_{\perp 1}}\cos\xi_{2t}}{\sqrt{\varepsilon'_{\perp 2}}\cos\xi_{1i} + \sqrt{\varepsilon'_{\perp 1}}\cos\xi_{2t}}\cos\psi_{1i} = r_{||}\sin\psi_{1i} \\ t\sin\psi_{2t} &= \frac{2\sqrt{\varepsilon'_{\perp 1}}\cos\xi_{1i}}{\sqrt{\varepsilon'_{\perp 1}}\cos\xi_{1i} + \sqrt{\varepsilon'_{\perp 2}}\cos\xi_{2t}}\sin\psi_{1i} = t_{\perp}\sin\psi_{1i} \\ t\cos\psi_{2t} &= \frac{2\sqrt{\varepsilon'_{\perp 1}}\cos\xi_{1i}}{\sqrt{\varepsilon'_{\perp 2}}\cos\xi_{1i} + \sqrt{\varepsilon'_{\perp 1}}\cos\xi_{2t}}\cos\psi_{1i} = t_{||}\sin\psi_{1i} \end{aligned} , \qquad (50)$$

where $r_{\perp}$ and $t_{\perp}$ are the amplitude reflection and transmission factors for the perpendicular polarization (i.e. $\psi_{1i} = \frac{\pi}{2}$), $r_{||}$ and $t_{||}$ are the amplitude reflection and transmission factors for the parallel polarization (i.e. $\psi_{1i} = 0$ ), so that the general reflection and transmission factors are:

$$r^2 = \rho_\perp \sin^2\psi_{1i} + \rho_{||}\cos^2\psi_{1i}$$
$$t^2 = t_\perp^2 \sin^2\psi_{1i} + t_{||}^2\cos^2\psi_{1i} \quad , \qquad (51)$$

where $\rho_\perp = r_\perp^2$ and $\rho_{||} = r_{||}^2$ are the so-called energetic perpendicular and parallel reflection factors; note that an equivalent and useful expression for the amplitude factors is $r_\perp = \frac{sin(\xi_{2t} - \xi_{1i})}{sin(\xi_{2t} + \xi_{1i})}$ and $r_{||} = \frac{tg(\xi_{2t} - \xi_{1i})}{tg(\xi_{2t} + \xi_{1i})}$, from which one deduces $tg\,\psi_{1r} = -\frac{r_\perp}{r_{||}} tg\,\psi_{1i}$ and similarly $tg\,\psi_{2t} = \frac{t_\perp}{t_{||}} tg\,\psi_{1i}$. From those results, one easily notices that if the polarization of the incident wave is perpendicular (respectively parallel), it remains perpendicular (respectively parallel) for the reflected and transmitted waves, and that for the Brewster incidence $\sin\xi_{1i} = \sqrt{\frac{\varepsilon'_{\perp 2}}{\varepsilon'_{\perp 1} + \varepsilon'_{\perp 2}}}$, the polarization of the reflected wave is always perpendicular, whatever the incident polarization is. Note also that the mean energetic reflection factor is $\bar{\rho} = \frac{1}{\pi}\int_{\psi_{1i}=0}^{\pi} \rho(\psi_{1i})\, d\,\psi_{1i} = \frac{1}{2}(\rho_\perp + \rho_{||})$, and that the global amplitude reflection and transmission factors verify the conservation of the electromagnetic flux through a closed surface, namely $\sqrt{\varepsilon'_{\perp 1}}\,(1 - r^2) cos\,\xi_{1i} = \sqrt{\varepsilon'_{\perp 2}}\, t^2 cos\,\xi_{2t}$ : note that for a total reflection, i.e. $r = 1$, the previous equation implies $t = 0$, i.e. no transmission. Obviously under this form, $\rho_{\perp(||)}(\xi_{1i}) = \rho_{\perp(||)}(\xi_{2t})$, and $|r_\perp(\xi_{1i})| = r(\xi_{1i})$, where $r$ is the energetic corpuscular reflection factor defined by Eqs. (46) or (47).

## *IV-2 – Extraordinary/extraordinary reflection and transmission*

Let us now briefly remind the major results for extraordinary modes, by examining an extraordinary electromagnetic plane wave propagating inside the homogeneous non absorbing crystal 1, along the wave vector $\vec{k_{1i}} = \frac{\omega_{1i}}{c} n_{e1}(\theta_{1i}) \vec{\Omega_{1i}}$ with $\vec{\Omega_{1i}} = \sin(\alpha_1 + \theta_{1i})\vec{e_x} + \cos(\alpha_1 + \theta_{1i})\vec{e_z}$, $\theta_{1i}$ being the incident propagation wave angle associated to the incident ray angle $\xi_{1i}$; assuming the incident electric field in a parallel polarization state (i.e. $\vec{E_{1i}}$ in the plane $(\vec{e_x}, \vec{e_z})$) with $\vec{E_{1i}} = E_{1i} e^{-i\varphi} \vec{e_{1i}}$, implies that the non zero electric induction components are in the plane $(\vec{e_x}, \vec{e_z})$, from which one deduces that $\vec{B_{1i}} = B_{1i} e^{-i\varphi} \vec{e_y}$ and $\bar{\bar{\varepsilon}}\, E_{1i} \vec{e_{1i}} = c\, B_{1i}\, n_{e1}(\theta_{1i}) \begin{bmatrix} \cos(\alpha_1 + \theta_{1i}) \\ 0 \\ -\sin(\alpha_1 + \theta_{1i}) \end{bmatrix}$; $\vec{e_{1i}}$ being a unit real vector, a simple development of the previous relations easily leads to $B_{1i}=\frac{N_{e1}(\xi_{1i})}{c}E_{1i}$, from which it comes for the representation of the extraordinary electromagnetic field:

$$\vec{E}_{1i}=\frac{n_{e1}(\theta_{1i})N_{e1}(\xi_{1i})}{\eta_1\varepsilon'_{\perp 1}}E_{1i}e^{-i\varphi}\begin{pmatrix}\eta_1\cos\alpha_1\cos\theta_{1i}-\sin\alpha_1\sin\theta_{1i}\\0\\-\eta_1\sin\alpha_1\cos\theta_{1i}-\cos\alpha_1\sin\theta_{1i}\end{pmatrix}=E_{1i}e^{-i\varphi}\begin{bmatrix}\cos(\alpha_1+\xi_{1i})\\0\\-\sin(\alpha_1+\xi_{1i})\end{bmatrix}$$

$$\vec{D}_{1i}=\varepsilon_0\varepsilon'_{\perp 1}E_{1i}e^{-i\varphi}\begin{pmatrix}\cos\alpha_1\cos\xi_{1i}-\eta_1\sin\alpha_1\sin\xi_{1i}\\0\\-\sin\alpha_1\cos\xi_{1i}-\eta_1\cos\alpha_1\sin\xi_{1i}\end{pmatrix}=\varepsilon_0 N_{e1}(\xi_{1i})n_{e1}(\theta_{1i})E_{1i}e^{-i\varphi}\begin{bmatrix}\cos(\alpha_1+\theta_{1i})\\0\\-\sin(\alpha_1+\theta_{1i})\end{bmatrix}, \quad (52)$$

$$\vec{B}_{1i}=\frac{N_{e1}(\xi_{1i})}{c}E_{1i}e^{-i\varphi}\vec{e}_y$$

the associated Poynting vector being expressed as $\vec{\Pi}_{1i}=\frac{N_{e1}(\xi_{1i})}{c}|E_{1i}|^2\begin{bmatrix}\sin(\alpha_1+\xi_{1i})\\0\\\cos(\alpha_1+\xi_{1i})\end{bmatrix}$. The next step is to show that for extraordinary electromagnetic waves propagating inside an uniaxial crystal, the only possible polarization angle is 0, or equivalently, the only possible polarization for the extraordinary waves is the parallel one; to do so, one writes for the unit vectors, since the orthogonal trihedron $\left(\vec{\mathrm{k}},\vec{\mathrm{D}},\vec{\mathrm{B}}\right)$ is direct, $\vec{e}_D=\begin{bmatrix}\cos\psi\cos(\alpha_1+\theta_{1i})\\\sin\psi\\-\cos\psi\sin(\alpha_1+\theta_{1i})\end{bmatrix}$ and $\vec{e}_B=\begin{bmatrix}-\sin\psi\cos(\alpha_1+\theta_{1i})\\\cos\psi\\\sin\psi\sin(\alpha_1+\theta_{1i})\end{bmatrix}$, from which $\vec{e}_E=\frac{D_{1i}}{\varepsilon_0\eta_1\varepsilon'_{\perp 1}E_{1i}}\begin{bmatrix}\cos\psi(\eta_1\cos\alpha_1\cos\theta_{1i}-\sin\alpha_1\sin\theta_{1i})\\\eta_1\sin\psi\\\cos\psi(-\eta_1\sin\alpha_1\cos\theta_{1i}-\cos\alpha_1\sin\theta_{1i})\end{bmatrix}$ or equivalently with the ray angle, $\vec{e}_E=\frac{D_{1i}}{\varepsilon_0\varepsilon'_{\perp 1}E_{1i}}\begin{bmatrix}\frac{\varepsilon'_{\perp 1}}{N_{e1}(\xi_{1i})n_{e1}(\theta_{1i})}\cos\psi\cos(\alpha_1+\xi_{1i})\\\sin\ \psi\\-\frac{\varepsilon'_{\perp 1}}{N_{e1}(\xi_{1i})n_{e1}(\theta_{1i})}\cos\psi\sin(\alpha_1+\xi_{1i})\end{bmatrix}$; then the general electric field and induction can be expressed as:

$$\vec{E}_{1i}=\frac{E_{1i}e^{-i\varphi}}{\sqrt{\cos^2\psi+\frac{n_{e1}^2(\theta_{1i})N_{e1}^2(\xi_{1i})}{\varepsilon'^2_{\perp 1}}\sin^2\psi}}\begin{bmatrix}\cos\psi\cos(\alpha_1+\xi_{1i})\\ \frac{n_{e1}(\theta_{1i})N_{e1}(\xi_{1i})}{\varepsilon'_{\perp 1}}\sin\psi\\ -\cos\psi\sin(\alpha_1+\xi_{1i})\end{bmatrix},$$

$$\vec{D}_{1i}=\frac{\varepsilon_0 n_{e1}(\theta_{1i})N_{e1}(\xi_{1i})E_{1i}e^{-i\varphi}}{\sqrt{\cos^2\psi+\frac{n_{e1}^2(\theta_{1i})N_{e1}^2(\xi_{1i})}{\varepsilon'^2_{\perp 1}}\sin^2\psi}}\begin{bmatrix}\cos\psi\cos(\alpha_1+\theta_{1i})\\ \sin\psi\\ -\cos\psi\sin(\alpha_1+\theta_{1i})\end{bmatrix}$$

Hence for the unit magnetic field vector $\vec{b}_{1i}=\frac{n_{e1}^2(\theta_{1i})}{\sqrt{\varepsilon'^2_{\perp 1}\cos^2\psi+n_{e1}^4(\theta_{1i})\sin^2\psi}}\begin{bmatrix}-\sin\psi\cos(\alpha_1+\theta_{1i})\\ \frac{\varepsilon'_{\perp 1}}{n_{e1}^2(\theta_{1i})}\cos\psi\\ \sin\psi\sin(\alpha_1+\theta_{1i})\end{bmatrix}$, and the 4 Maxwell equations are simultaneously verified if and only if $\psi = 0$ for all incident direction $\theta_{1i}$, which means that the extraordinary waves propagating inside an uniaxial crystal are in the parallel polarization state.

**Remark:** in this study we shall not consider possible evanescent waves, this implying that all angles and reflection/transmission factors are all pure real quantities.

The procedure to calculate the reflection and transmission factors is similar to the one previously developed for the pure ordinary case: the continuity conditions for the fields at the impinging time $t = 0$ are similarly written as:

* continuity of the tangential components of the electrical field:

$$\cos(\alpha_1+\xi_{1i})+r_{||}\cos(\alpha_1+\xi_{1r})=t_{||}\cos(\alpha_2+\xi_{2t}), \qquad \text{(53-a)}$$

* continuity of the normal component of the electrical induction:

$$N_{e1}(\xi_{1i})n_{e1}(\theta_{1i})\sin(\alpha_1+\theta_{1i})+r_{||}N_{e1}(\xi_{1r})n_{e1}(\theta_{1r})\sin(\alpha_1+\theta_{1r}) \\ =t_{||}N_{e2}(\xi_{2t})n_{e2}(\theta_{2t})\sin(\alpha_2+\theta_{2t}), \qquad \text{(53-b)}$$

* continuity of magnetic field:

$$N_{e1}(\xi_{1i})+r_{||}N_{e1}(\xi_{1r})=t_{||}N_{e2}(\xi_{2t}), \qquad \text{(53-c)}$$

Furthermore, the phase invariance along the *x* direction implies $\omega n_e(\theta)\sin(\alpha+\theta)=constant$, or expressed with the ray angle variable $\xi$, $\omega\frac{\sqrt{\varepsilon'_\perp}\,(\sin\alpha\cos\xi+\eta\cos\alpha\sin\xi)}{\sqrt{\cos^2\xi+\eta\sin^2\xi}}=constant$; hence, using the generalized Descartes' law given by Eq. (34), it comes that $\omega$ = *constant* if and only if $\eta = 1$ (i. e. isotropic medium) or $\alpha = 0$ (i. e. the optical axis coincides with the geometrical one) for all directions: hence a major consequence due to the anisotropic uniaxial Fermat's principle, is that the circular frequency of an extraordinary wave does generally not remain constant on its propagation path, except in the case of homogeneous media or crystal for which the optical axis coincides with its geometrical one; obviously, one deduces from the previous important result that the angular frequency of a wave is modified at an interface separating two uniaxial crystals when their axes do not coincide, so that it is necessary to introduce a time scale such that the time origin is determined when the incident wave impacts the interface. In other words, the electromagnetic description combined to the geometric Fermat's principle induces a wave frequency change except when the optical and geometrical axes coincide.

***Remark:*** if $\xi_{1i}=k\pi-Arctg\left(\frac{tg\,\alpha_1}{\eta_1}\right)$, one has $\frac{\omega_{1i}(\sin\alpha_1\cos\xi_{1i}+\eta_1\cos\alpha_1\sin\xi_{1i})}{\sqrt{\cos^2\xi_{1i}+\eta_1\sin^2\xi_{1i}}}=0$ from which one must have $\frac{\omega_{1r}(\sin\alpha_1\cos\xi_{1r}+\eta_1\cos\alpha_1\sin\xi_{1r})}{\sqrt{\cos^2\xi_{1r}+\eta_1\sin^2\xi_{1r}}}=0$, leading to either $\omega_{1r}=0$ or $\xi_{1r}=k'\pi-Arctg\left(\frac{tg\,\alpha_1}{\eta_1}\right)$; but as examined in subsection III-2, the reflected angle for an incident direction below the virtual interface verifies $\hat{\xi}_{1r}=\pi-\xi_{1i}+\alpha_1$, from which $\xi_{1r}=\pi-\xi_{1i}$, like for isotropic media, and one obtains $\omega_{1r}=0$ of no physical signification; for the transmitted associated ray, a similar calculation also leads to $\omega_{2t}=0$: hence one concludes that the incident direction $\xi_{1i}=[2\pi]-Arctg\left(\frac{tg\,\alpha_1}{\eta_1}\right)$ is forbidden, from an electromagnetic point of view, while it is allowed for the Fermat's principle; similarly, for the "right to left" incident direction $\xi_{1i}=\frac{3\pi}{2}^+$, the reflected direction is $\xi_{1r}=-\frac{\pi}{2}^-+\alpha_1$, and if $\sqrt{\varepsilon'_{\perp 2}}\,\breve{\Omega}_2\geq\sqrt{\varepsilon'_{\perp 1}}\,\breve{\Omega}_1$, there exists one unique solution for the transmitted angle verifying $\sqrt{\varepsilon'_{\perp 2}}\,F_2(\hat{\xi}_{2t})=\sqrt{\varepsilon'_{\perp 1}}\,\breve{\Omega}_1$, whence the phase conservation shows that the reflected and transmitted frequencies may be negative, depending on the positivity of the crystals: then the particular Fermat's principle for uniaxial crystals is not compatible with an electromagnetic formulation, and the combination of the two techniques is unable to give a coherent result for the energetic reflection/transmission factors, which must then be calculated from a corpuscular point of view.

Like for the isotropic media, the symbolic equation for purely extraordinary reflection/transmission can be rewritten as, for a given pencil around a straight licit incident direction which does not include the forbidden cut lines, and below the virtual interface:

$$\overset{\Rightarrow}{P_{1i}}+\frac{1}{c}\begin{pmatrix}\frac{d\,E_{t0}}{d\,\Omega_{1i}}\\ \vec{0}\end{pmatrix}\rightarrow\frac{\overline{N_{1r}}}{\overline{N_{1i}}}\overset{\Rightarrow}{P_{1r}}\frac{d\,\Omega_{1r}}{d\,\Omega_{1i}}+\frac{\overline{N_{2t}}}{\overline{N_{1i}}}\overset{\Rightarrow}{P_{2t}}\frac{d\,\Omega_{2t}}{d\,\Omega_{1i}}+\frac{1}{c}\frac{d}{d\,\Omega_{1i}}\begin{pmatrix}E_{i0}\\ \vec{p_0}\,c\end{pmatrix}$$

For an incident direction below the virtual interface, the reflected direction verifies $\xi_{1r}=\pi-\xi_{1i}$ in a reference location system bound to the optical axis, or equivalently $\hat{\xi}_{1r}=\pi+2\alpha_1-\hat{\xi}_{1i}$, so that the reflected elementary solid angle is such that $d\,\Omega_{1r}=d\,\Omega_{1i}$; note that if $\alpha_1\leq\frac{\pi}{4}$, there is no virtual reflection in medium 2 (i.e. a real extraordinary reflection in medium 1) for incident rays “from left to right” i.e. $\hat{\xi}_{1i}\in\left[0,\frac{\pi}{2}\right]$, and for incident rays “from right to left” such that $\hat{\xi}_{1i}\in\left\lfloor\hat{\xi}_{1v},2\pi\right\rfloor$ where $\hat{\xi}_{1v}=\frac{3\pi}{2}+2\alpha_1$, this kind of reflection being possible only for incident rays from right to left such that $\hat{\xi}_{1i}\in\left[\frac{3\pi}{2},\hat{\xi}_{1v}\right]$; however, when $\frac{\pi}{4}<\alpha_1\leq\frac{\pi}{2}$, there is no virtual reflection in medium 2 only for incident rays from left to right such that $\hat{\xi}_{1i}\in\left[\hat{\xi}_{1v},\frac{\pi}{2}\right]$ with $\hat{\xi}_{1v}=2\alpha_1-\frac{\pi}{2}$, all other incident directions being affected by a virtual reflection in medium 2; for incident rays with a real reflection in medium 1, the ray refractive indices for incident and reflected photons verify $N_{1i}\left(\vec{t_{1i}}\right)=N_{1r}\left(\vec{t_{1r}}\right)$: hence one can define like for isotropic media the reflection and transmission factors by $r=\frac{\overline{N_{1r}}}{\overline{N_{1i}}}$ and $t=\frac{\overline{N_{2t}}\,d\,\Omega_{2t}}{\overline{N_{1i}}\,d\,\Omega_{1i}}$; furthermore, the photon impulsion is $\overset{\Rightarrow}{P_j}=\frac{N_j^2\left(\vec{t_j}\right)h\nu_j}{c}\left[\vec{e_0}+\frac{1}{N_j}\left(\sin\hat{\xi}_j\vec{e_x}+\cos\hat{\xi}_j\vec{e_z}\right)\right]$ when using a Finsler metrics, from which, considering a frequency conservation for the photons, the previous conservation equality leads to:

$$\begin{aligned}&N_{1i}^2(1-r)=N_{2t}^2\,t+\overline{\Delta E_0}\left(\xi_{1i}\right)\\&\sin\hat{\xi}_{1i}=r\sin\left(\hat{\xi}_{1i}-2\alpha_1\right)+t\frac{N_{2t}}{N_{1i}}\sin\hat{\xi}_{2t}+\frac{\overline{P_x}}{n_2N_{1i}}\quad,\\&\cos\hat{\xi}_{1i}=-r\cos\left(\hat{\xi}_{1i}-2\alpha_1\right)+t\frac{N_{2t}}{N_{1i}}\cos\hat{\xi}_{2t}+\frac{\overline{P_z}}{n_2N_{1i}}\end{aligned}\qquad(54)$$

where $\overline{P}_x$ and $\overline{P}_z$ are the components of the induced impulsion, not necessarily normal to the interface; here the induced quasi-particle is not luminous and is only affected by the ordinary refractive index, and the solution of the previous system is easily given in the following form:

$$r=\frac{sin\left(\hat{\xi}_{1i}-\hat{\xi}_{2t}\right)+\dfrac{\overline{P}_z sin\,\hat{\xi}_{2t}-\overline{P}_x cos\,\hat{\xi}_{2t}}{n_2 N_{1i}}}{sin\left(\xi_{1i}+\hat{\xi}_{2t}-\alpha_1\right)}, \qquad t=\frac{N_{1i}}{N_{2t}}\frac{sin\left(2\,\xi_{1i}\right)-\dfrac{1}{n_2 N_{1i}}\left[\overline{P}_z sin\left(\xi_{1i}-\alpha_1\right)+\overline{P}_x cos\left(\xi_{1i}-\alpha_1\right)\right]}{sin\left(\xi_{1i}+\hat{\xi}_{2t}-\alpha_1\right)} \tag{55}$$

For a purely extraordinary reflection/transmission, the photons number conservation $r+t=1$ leads to:

$$\frac{\left[N_{2t}\,sin\,\hat{\xi}_{2t}-N_{1i}\,sin\left(\xi_{1i}-\alpha_1\right)\right]\overline{P}_z-\left[N_{2t}\,cos\,\hat{\xi}_{2t}+N_{1i}\,cos\left(\xi_{1i}-\alpha_1\right)\right]\overline{P}_x}{N_{1i}\,n_2}=2\left[N_{2t}\,sin\left(\hat{\xi}_{2t}-\alpha_1\right)-N_{1i}\,sin\,\xi_{1i}\right]cos\,\xi_{1i}$$

so that after a straightforward calculation, Eq. (55) expressed with the normal component of the impulsion takes the following remarkable simple form:

$$r=\frac{N_{2t}\,cos\,\hat{\xi}_{2t}-N_{1i}\,cos\,\hat{\xi}_{1i}+\dfrac{\overline{P}_z}{n_2}}{N_{1i}\,cos\left(\xi_{1i}-\alpha_1\right)+N_{2t}\,cos\,\hat{\xi}_{2t}}, \qquad t=\frac{2\,N_{1i}\,cos\,\alpha_1\,cos\,\xi_{1i}-\dfrac{\overline{P}_z}{n_2}}{N_{1i}\,cos\left(\xi_{1i}-\alpha_1\right)+N_{2t}\,cos\,\hat{\xi}_{2t}} \tag{56}$$

in the case where $N_{1i}\,cos\left(\xi_{1i}-\alpha_1\right)+N_{2t}\,cos\,\hat{\xi}_{2t}\neq 0$, the activation energy being expressed as $\overline{\Delta E_0}=\left(N_{1i}^2-N_{2t}^2\right)t$ and Eq. (56) being valid for real extraordinary reflection in medium 1, i.e. $\hat{\xi}_{1i}\in\left[0,\frac{\pi}{2}\right]\cup\left]\frac{3\pi}{2}+2\alpha_1,2\pi\right]$ when $\alpha_1\leq\frac{\pi}{4}$, and $\hat{\xi}_{1i}\in\left[2\alpha_1-\frac{\pi}{2},\frac{\pi}{2}\right]$ when $\frac{\pi}{4}<\alpha_1\leq\frac{\pi}{2}$.

One may notice that for incident rays "from left to right", i.e. $\hat{\xi}_{1i}\in\left[0,\frac{\pi}{2}\right]$ with $\xi_{1i}\in\left[-\alpha_1,\frac{\pi}{2}-\alpha_1\right]\equiv\left[0,\frac{\pi}{2}-\alpha_1\right]\cup\left[2\pi-\alpha_1,2\pi\right]$, $\hat{\xi}_{2t}\in\left[0,\frac{\pi}{2}\right]$ and $cos\,\hat{\xi}_{1i}\geq 0$, $cos\,\xi_{1i}\geq 0$, $cos\,\hat{\xi}_{2t}\geq 0$; when $\alpha_1 \leq \frac{\pi}{4}$, $\hat{\xi}_{1i}-2\alpha_1 \in \left[-2\alpha_1, \frac{\pi}{2}-2\alpha_1\right] \equiv \left[0, \frac{\pi}{2}-2\alpha_1\right] \cup [2\pi-2\alpha_1, 2\pi]$ with $cos(\xi_{1i}-\alpha_1) \geq 0$; however, for $\frac{\pi}{4} < \alpha_1 \leq \frac{\pi}{2}$, $\xi_{1i}-\alpha_1 \in \left[-2\alpha_1, \frac{\pi}{2}-2\alpha_1\right] \equiv \left[2\pi-2\alpha_1, \frac{3\pi}{2}\right] \cup \left[\frac{3\pi}{2}, \frac{5\pi}{2}-2\alpha_1\right]$, so that $cos(\xi_{1i}-\alpha_1) \geq 0$ for $\xi_{1i} \in \left[\alpha_1-\frac{\pi}{2}, \frac{\pi}{2}-\alpha_1\right]$ and $cos(\xi_{1i}-\alpha_1) < 0$ for $\xi_{1i} \in \left[-\alpha_1, \alpha_1-\frac{\pi}{2}\right]$; then for $\alpha_1 \leq \frac{\pi}{4}$, the reflection and transmission factors "without induced normal quasi-particle" are such that $0 \leq r \leq 1$ and $0 \leq t \leq 1$ if and only if $N_{1i} cos \hat{\xi}_{1i} \leq N_{2t} cos \hat{\xi}_{2t}$ for incident rays in the extraordinary transmission cone, simple generalization of what happens in the isotropic case; when $\frac{\pi}{4} < \alpha_1 \leq \frac{\pi}{2}$ however, a virtual reflection in medium 2 arises except for incident rays such that $\xi_{1i} \in \left[\alpha_1-\frac{\pi}{2}, \frac{\pi}{2}-\alpha_1\right]$, for which $cos(\xi_{1i}-\alpha_1) \geq 0$: hence like what precedes, the reflection and transmission factors without induced normal particle verify $0 \leq r \leq 1$ and $0 \leq t \leq 1$ if $N_{1i} cos \hat{\xi}_{1i} \leq N_{2t} cos \hat{\xi}_{2t}$; in other words, one can say that when there is no virtual reflection in medium 2 for incident rays from left to right inside the transmission cone, a normal component of the induced particles is not necessary for rays verifying $N_{1i} cos \hat{\xi}_{1i} \leq N_{2t} cos \hat{\xi}_{2t}$.

For incident rays from right to left, i.e. $\hat{\xi}_{1i} \in \left[\frac{3\pi}{2}, 2\pi\right]$, equivalently $\xi_{1i} \in \left[\frac{3\pi}{2}-\alpha_1, \frac{3\pi}{2}\right[ \cup \left]\frac{3\pi}{2}, 2\pi-\alpha_1\right]$, $cos \hat{\xi}_{1i} \geq 0$, $cos \hat{\xi}_{2t} \geq 0$, $cos \xi_{1i} \geq 0$ for $\xi_{1i} \in \left]\frac{3\pi}{2}, 2\pi-\alpha_1\right]$ and $cos \xi_{1i} \leq 0$ for $\xi_{1i} \in \left[\frac{3\pi}{2}-\alpha_1, \frac{3\pi}{2}\right[$; when $\alpha_1 \leq \frac{\pi}{4}$, $cos(\xi_{1i}-\alpha_1) \geq 0$ for $\xi_{1i} \in \left[\frac{3\pi}{2}+\alpha_1, 2\pi-\alpha_1\right]$, i.e. for incident rays not affected by a virtual reflection in medium 2, and $cos(\xi_{1i}-\alpha_1) < 0$ elsewhere, while for $\frac{\pi}{4} < \alpha_1 \leq \frac{\pi}{2}$, $cos(\xi_{1i}-\alpha_1) < 0$ for all incident ray: one concludes that $cos(\xi_{1i}-\alpha_1) \geq 0$ if and only if the incident ray, from left to right or from right to left, is below the virtual interface of medium 1, which is the only angular area where the quantity $N_{1i} cos(\xi_{1i}-\alpha_1) + N_{2t} cos \hat{\xi}_{2t}$ can reach a zero value and be negative. Hence Eqs. (56) can always be interpreted as a generalization of what happens in the isotropic case for incident rays below the virtual interface: these rays are really reflected in medium 1 and for incident rays inside the transmission cone, a normal component of the induced particles is not necessary for rays verifying $N_{1i} cos \hat{\xi}_{1i} \leq N_{2t} cos \hat{\xi}_{2t}$.

To illustrate this particular behaviour, the extraordinary reflection factors without induced normal component (i.e. $\overline{p}_z = 0$) are presented on the three following Figs. 9, for uniaxial crystals whose physical characteristics are $\sqrt{\varepsilon'_{\perp 1}} = 1.5$, $\eta_1 = 2.0$ and $\alpha_1 = 30.0°$ for crystal 1, $\sqrt{\varepsilon'_{\perp 2}} = 2.0$, $\eta_2 = 2.0$ and $\alpha_1 = 60.0°$ for crystal 2: for this particular case, all incident rays in medium 1 can be transmitted into medium 2, the virtual interface is located at $\hat{\xi}_{1v}=330°$ and there is no virtual reflection in medium 2 for incident rays from left to right; the reflection factor is depicted on Fig. 9a for incident rays from left to right where no virtual reflection occurs, of positive values lower than 1: contrarily to the isotropic case, the reflection factor is not here a strictly growing function; for the case corresponding to Fig. 9b, the incident rays are from right to left, below the cut line of medium 1: for rays such that $\hat{\xi}_{1i} \in [330°,360°]$, below the virtual interface, Eqs. (56) without induced normal component are valid and the *r* function is continuous at $\hat{\xi}_{1i}=0°=360°$; for incident rays between the virtual interface and the cut line ($\hat{\xi}_{1i} \in [300°,330°]$), the reflection factor remains positive and lower than 1, but without any clear signification, for the reflected rays are virtual reflected rays in medium 2; for incident rays above the cut line ($\hat{\xi}_{1i} \in [270°,300°]$), presented on Fig. 9c, the reflection factor has no longer signification, for greater than 1.

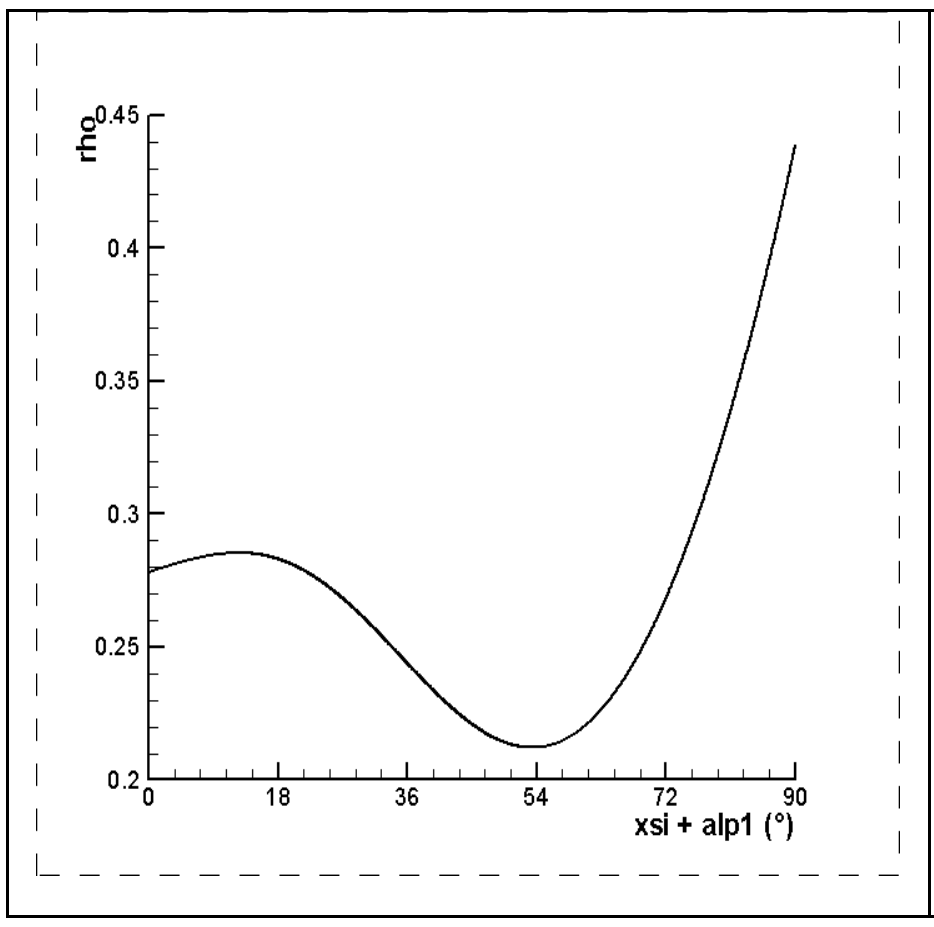

Fig. 9a

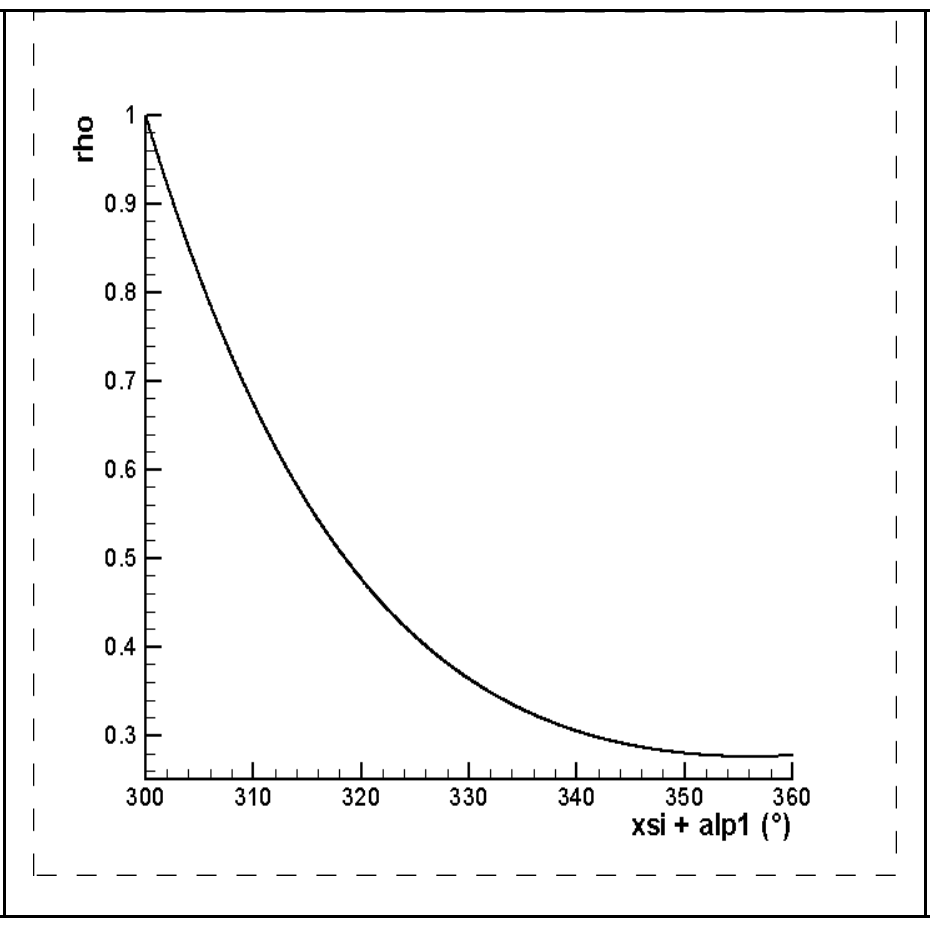

Fig. 9b

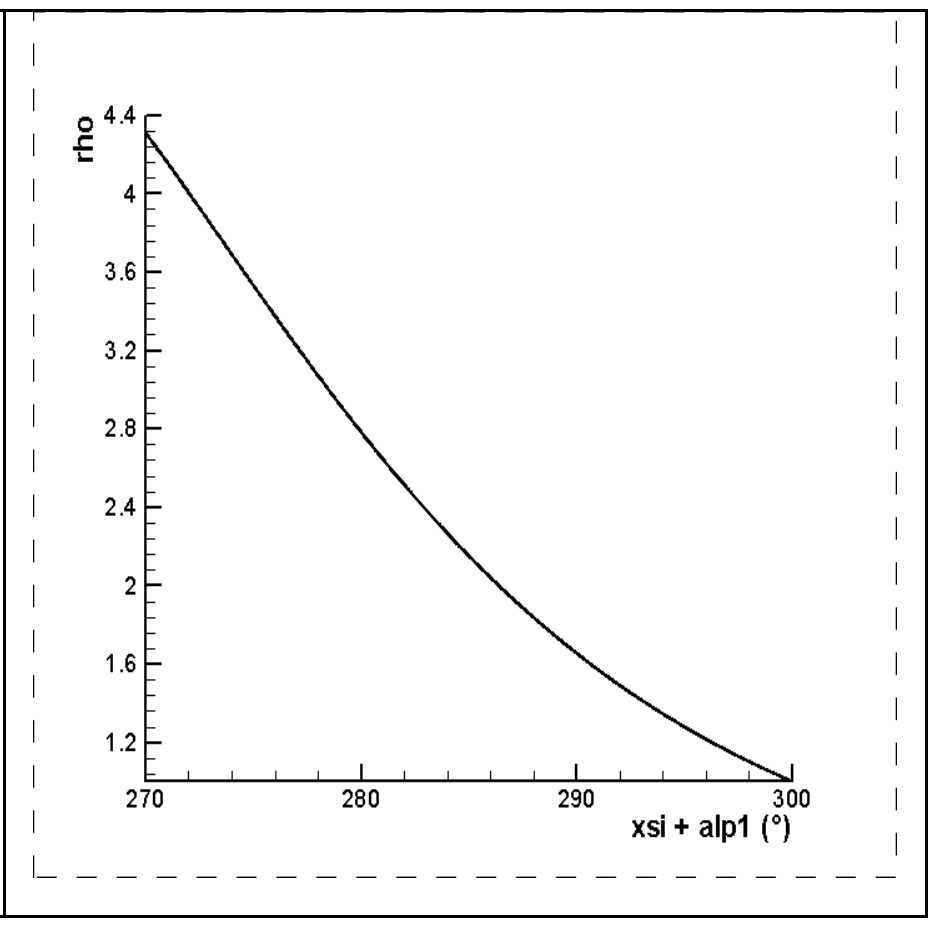

Fig. 9c

When the incident ray is above the virtual interface and below the cutline of medium 1, taking into account pure extraordinary phenomena, one may consider that the non transmitted part of incident photons which cannot be really reflected but virtually reflected, is “absorbed” at the interface, this directional absorbed energy being completely reemitted under heat: hence Fig. 9b can be interpreted as a reflection factor for the incident rays between the geometrical axis and the virtual interface, and as a directional “absorption” factor for incident rays between the virtual interface and the cut line, the transmission factor being in all case *1 - r*, corresponding to the part of incident photons really transmitted in medium 2. When the incident photons have a direction nearly parallel to, but below the cut line, which is the line of real total reflection, i.e. no transmission, they are completely absorbed by the interface at the impact point, the transmission being extremely weak. For incident rays above the cut line and below the right real interface however, a non luminous additional phenomenon must occur since the photons are above the real line of reflection, which plays here the rule of a potential barrier: then since a transmission is possible due to the particular form of the generalized Descartes’ law, and since no real reflection is possible, the non transmitted photons must be “absorbed” at the interface, and a normal induced quasi-particle must be created: indeed, for incident rays from right to left, $cos\,\xi_{1i}\leq 0$ for $\xi_{1i}\in\left[\frac{3\pi}{2}-\alpha_1,\frac{3\pi}{2}\right[$, i.e. for rays above the cut line and $2N_{1i}cos\,\alpha_1 cos\,\xi_{1i}\leq 0$, so that the transmission factor is negative without additional quasi-particle, for $-N_{1i}\left|cos\left(\xi_{1i}-\alpha_1\right)\right|+N_{2t}cos\,\hat{\xi}_{2t}>0$; similarly the reflection factor is lower than one without any induced quasi-particle if and only if $cos\,\xi_{1i}\geq 0$, which is not the case for rays above the cut-line: hence a normal component of the induced quasi-particles must exist to insure the transmission of these incident rays through medium 2: it is to note that in the considered example, the normal component must be negative, so that the induced normal quasi-particle is backwards and evolves in medium 1; hence one shall write $\frac{\overline{P_z}}{n_2}=\frac{\overline{p_z}^{+}}{n_2}-\frac{\overline{p_z}^{-}}{n_1}$ to take into account the backwards and forwards normal induced quasi-particles. This possibility of directional absorption at the interface for incident rays above the virtual interface allows the extension of Eqs. (56) in this angular area, except when the phonon multiplier $N_{1i}cos\left(\xi_{1i}-\alpha_1\right)+N_{2t}cos\,\hat{\xi}_{2t}=0$: in the following numerical presented example, $\sqrt{\varepsilon'_{\perp 1}}=2.0$, $\eta_1=2.0$ and $\alpha_1=60.0°$ for crystal 1, $\sqrt{\varepsilon'_{\perp 2}}=1.5$, $\eta_2=2.0$ and $\alpha_2=30.0°$ for crystal 2, symmetric case of the previous example, the reverse behaviour of the transmission factor without induced normal quasi-particle is exemplified on the three following Figs. 10; the transmission factor for incident rays from left to right is presented on Fig. 10a, which shows a transmission factor greater than 1: in this case, a normal quasi-particle must exist to allow a transmission lower than 1; note that the virtual interface is located at $\hat{\xi}=30°$ and only the incident rays such that $\hat{\xi}_{1i}\in\left[30°,48.75°\right]$ are really reflected and transmitted, the rays $\hat{\xi}_{1i}\in\left[0°,30°\right]$ are transmitted and absorbed, the rays $\hat{\xi}_{1i}\in\left[48.75°,90°\right]$ being totally reflected: obviously here, since $t>1$, the reflection (or absorption) factor without normal component is lower than 0, from which $N_{1i}cos\,\hat{\xi}_{1i}>N_{2t}cos\,\hat{\xi}_{2t}$: like for isotropic media, when this latter condition is realized, normal quasi-particles associated to the photons must appear. For incident rays represented on Fig. 9c, no additional normal quasi-particle need be created, since the positive transmission factor without normal component is lower than 1; for this angular sector, the incident rays are above the virtual interface and the cut line, but inside the transmission cone so that the transmission factor has a physical meaning. Fig. 10b is the illustration of what happens when $N_{1i}cos\left(\xi_{1i}-\alpha_1\right)+N_{2t}cos\,\hat{\xi}_{2t}=0$ for a given incident direction: in this particular case, the phonon multiplier is 0 for $\hat{\xi}_{1i}=350.32°$, negative for $\hat{\xi}_{1i}\in\left[340.2°,350.32°\right]$ and positive elsewhere, so that Eqs. (56) are inadequate on the angular sector $\hat{\xi}_{1i}\in\left[340.2°,360°\right]$.

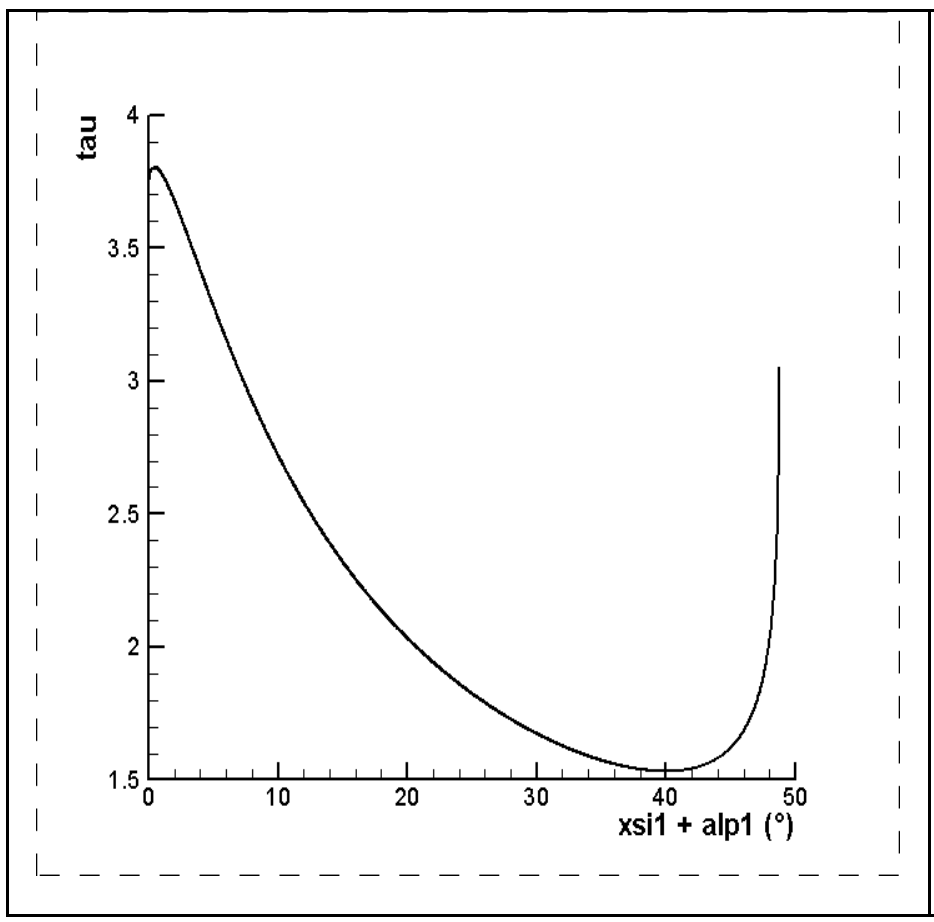

Fig. 10a

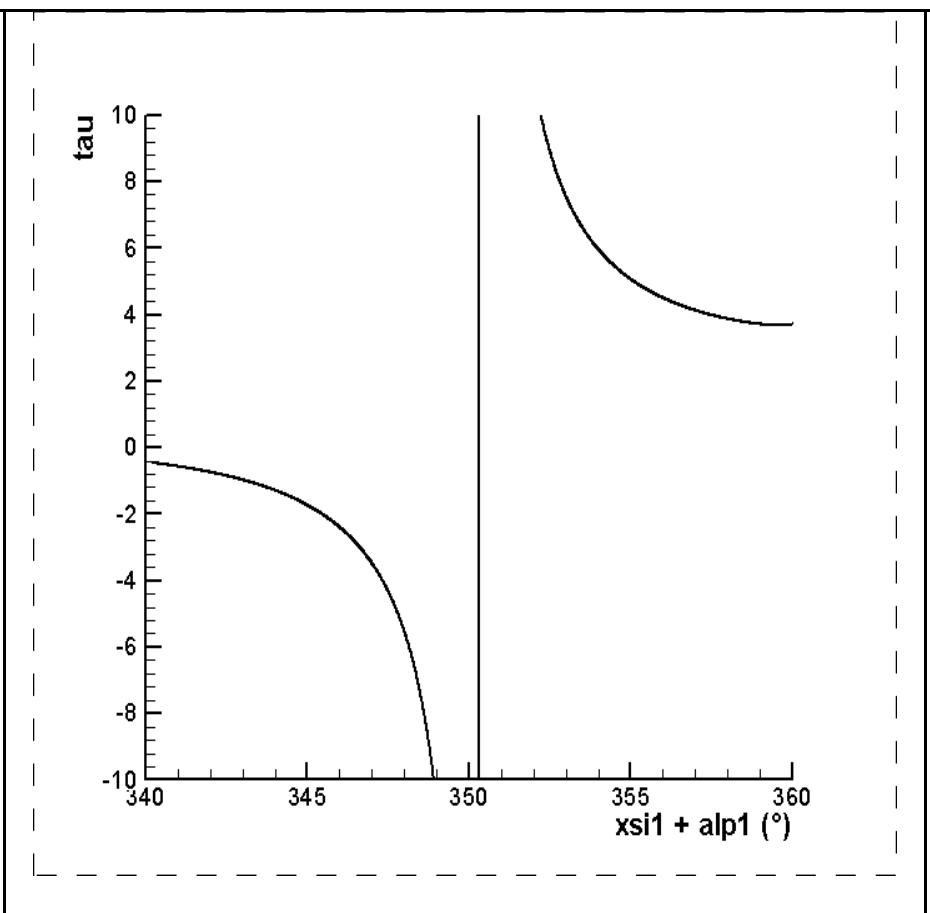

Fig. 10b

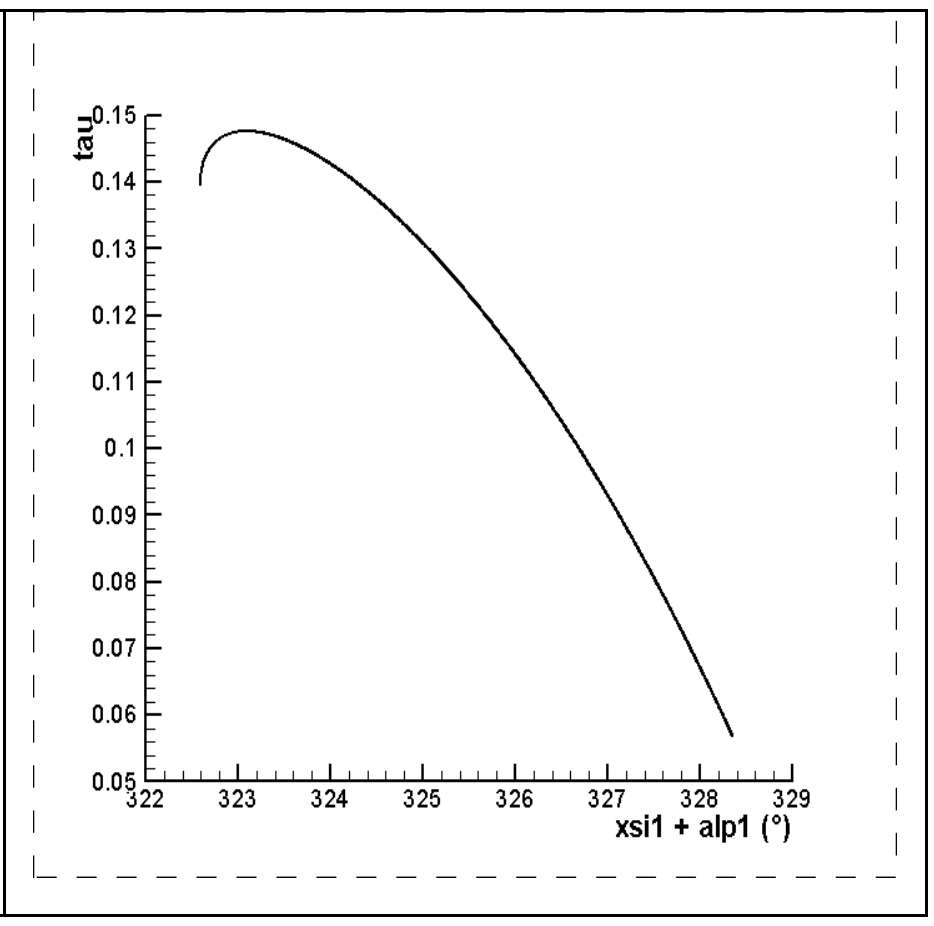

Fig. 10c

Obviously here Fig. 10a is the corresponding case of Fig. 9a, similarly Fig. 9c corresponds to Fig. 10c, which allows the calculation of the normal component of the quasi-particle: setting like for the isotropic case $n'_1 \equiv n_2$, $n'_2 \equiv n_1$, $\hat{\xi}'_{1i} \equiv \pi + \hat{\xi}_{2t}$ and $\hat{\xi}'_{2t} \equiv \pi + \hat{\xi}_{1i}$ for the backward trajectory and using the reflectivity rule $t_{1\to 2}(\xi_{1i}) = t_{2\to 1}(\xi_{2t})$ for the transmission factor leads to:

$$t_{1\to 2} = \frac{2 N_{1i} \cos\alpha_1 \cos\xi_{1i} - P^*}{N_{1i}\cos(\xi_{1i} - \alpha_1) + N_{2t}\cos\hat{\xi}_{2t}} = t_{2\to 1} = t_{1'\to 2'} = \frac{2 N_{2t}\cos\alpha_2 \cos\xi_{2t}}{N_{1i}\cos\hat{\xi}_{1i} + N_{2t}\cos(\xi_{2t} - \alpha_2)}$$

from which it comes the value of the normal component $P^* = \frac{\bar{p}_z^+}{n_2} - \frac{\bar{p}_z^-}{n_1}$ of the quasi-particle, generally not expressible in a simple form; if $P^*$ is lower than 0, it is a backward quasi-particle evolving in medium 1 and $\bar{p}_z^+ = 0$, while if $P^*$ is positive, it is a forward quasi-particle like in the isotropic case, evolving in medium 2 and $\bar{p}_z^- = 0$.

Obviously, if no transmission is possible above the cut line, all the incident rays above the cut-line are completely “absorbed”, similar to a total reflection in medium 1 and possible for $\bar{p}_z^+ = 2 n_2 N_{1i} \cos\alpha_1 \cos\xi_{1i}$.

In the case where $N_{2t}\cos\hat{\xi}_{2t} + N_{1i}\cos(\xi_{1i} - \alpha_1) = 0$, possible only for incident rays above the virtual interface of medium 1, the conservation law $r + t = 1$ leads to:

$$\frac{\left[N_{2t}\sin\hat{\xi}_{2t} - N_{1i}\sin(\xi_{1i} - \alpha_1)\right]\bar{P}_z}{N_{1i} n_2} = 2\left[N_{2t}\sin(\hat{\xi}_{2t} - \alpha_1) - N_{1i}\sin\xi_{1i}\right]\cos\xi_{1i}$$

note that if $N_{2t}\cos\hat{\xi}_{2t}+N_{1i}\cos(\xi_{1i}-\alpha_1)=0$ and $N_{2t}\sin\hat{\xi}_{2t}-N_{1i}\sin(\xi_{1i}-\alpha_1)=0$, this leads to $(N_{1i}-N_{2t})^2+4N_{1i}N_{2t}\cos^2\left(\frac{\xi_{1i}+\hat{\xi}_{2t}-\alpha_1}{2}\right)=0$, possible only for $N_{1i}=N_{2t}$ and $\xi_{1i}+\hat{\xi}_{2t}-\alpha_1=\pi[2\pi]$, obviously leading to $N_{2t}\sin(\hat{\xi}_{2t}-\alpha_1)-N_{1i}\sin\xi_{1i}=0$, i.e. the solution(s) of the following equation:

$$\begin{aligned}&\{\varepsilon'_{\perp 2}(1-\eta_2)\cos[2(\alpha_2-\alpha_1)]-\varepsilon'_{\perp 1}(1-\eta_1)\}\cos^2\xi_{1i}+\varepsilon'_{\perp 2}(1-\eta_2)\sin[2(\alpha_2-\alpha_1)]\sin\xi_{1i}\cos\xi_{1i}\\&=\eta_1\varepsilon'_{\perp 1}-[\sin^2(\alpha_2-\alpha_1)+\eta_2\cos^2(\alpha_2-\alpha_1)]\varepsilon'_{\perp 2}\end{aligned}$$

which can be reduced to a 4th algebraic equation in $\cos\xi_{1i}$; but the generalized Descartes' law binding $\xi_{1i}$ and $\xi_{2t}$ being transcendent, except when $\alpha_1,\alpha_2\in\left\{0,\frac{\pi}{2}\right\}$, one cannot have simultaneously the two later quantities equalling zero, so that in the general case one can express the normal component under $\overline{P}_z=2N_{1i}n_2\frac{[N_{2t}\sin(\hat{\xi}_{2t}-\alpha_1)-N_{1i}\sin\xi_{1i}]\cos\xi_{1i}}{N_{2t}\sin\hat{\xi}_{2t}-N_{1i}\sin(\xi_{1i}-\alpha_1)}$, from which Eqs.(55) can be rewritten after calculation:

$$\begin{aligned}r&=\frac{N_{2t}\sin\hat{\xi}_{2t}+N_{1i}\sin\hat{\xi}_{1i}}{N_{2t}\sin\hat{\xi}_{2t}-N_{1i}\sin(\xi_{1i}-\alpha_1)}-\frac{\overline{P}_x\cos\hat{\xi}_{2t}}{n_2N_{1i}\sin(\xi_{1i}+\hat{\xi}_{2t}-\alpha_1)}\\t&=\frac{2N_{1i}\sin\alpha_1\sin\xi_{1i}}{N_{2t}\sin\hat{\xi}_{2t}-N_{1i}\sin(\xi_{1i}-\alpha_1)}-\frac{\overline{P}_x\cos(\xi_{1i}-\alpha_1)}{n_2N_{1i}\sin(\xi_{1i}+\hat{\xi}_{2t}-\alpha_1)}\end{aligned},\qquad(57)$$

For this particular direction, the unknown value of $\overline{P}_x$ can be obtained when using the reflectivity rule $t_{1\to 2}(\xi_{1i})=t_{2\to 1}(\xi_{2t})$, where $t_{2\to 1}(\xi_{2t})=t_{1'\to 2'}(\xi'_{1i})$ is the transmission factor without normal component given by Eq. (56); for incident rays above the virtual interface, if the quantity $N_{2t}\cos\hat{\xi}_{2t}+N_{1i}\cos(\xi_{1i}-\alpha_1)=0$ changes its sign on the considered angular area, and more generally, one uses the non developed Eqs. (55) combined to the reflectivity rule, from which it comes after calculation:

$$\begin{aligned}\frac{\overline{P}_x}{n_2N_{1i}}&=\sin\hat{\xi}_{1i}-r_{2\to 1}\sin(\xi_{1i}-\alpha_1)-\frac{N_{2t}}{N_{1i}}t_{2\to 1}\sin\hat{\xi}_{2t}\\\frac{\overline{P}_z}{n_2N_{1i}}&=\cos\hat{\xi}_{1i}+r_{2\to 1}\cos(\xi_{1i}-\alpha_1)-\frac{N_{2t}}{N_{1i}}t_{2\to 1}\cos\hat{\xi}_{2t}\end{aligned},\qquad(58)$$

where $r_{2\to 1}$ and $t_{2\to 1}$ are the reverse reflection (absorption) and transmission factors without normal components given by Eqs. (56) setting $\overline{P}_z=0$ with the substitutions $\alpha_1\equiv\alpha_2$, $N_{1i}\equiv N_{2t}$, $\hat{\xi}_{1i}\equiv\hat{\xi}_{2t}$ and $\xi_{1i} \equiv \xi_{2t}$, and finally the expression of the reflection and transmission factors is simply $t_{1\to 2}(\xi_{1i}) = t_{2\to 1}(\xi_{2t})$ and $r_{1\to 2}(\xi_{1i}) = r_{2\to 1}(\xi_{2t})$, which was the expected result.

One concludes for optical axes not coinciding with the geometrical axis of the slab:

- for incident rays travelling in medium 1 towards medium 2 inside the transmission cone, the reflection (absorption if the incident ray is above the virtual interface in medium 1) and transmission factors are $r_{1\to 2}(\xi_{1i}) = r^*_{12} = \frac{N_{2t}\cos\hat{\xi}_{2t} - N_{1i}\cos\hat{\xi}_{1i}}{N_{1i}\cos(\xi_{1i} - \alpha_1) + N_{2t}\cos\hat{\xi}_{2t}}$ and $t_{1\to 2}(\xi_{1i}) = t^*_{12} = \frac{2N_{1i}\cos\alpha_1\cos\xi_{1i}}{N_{1i}\cos(\xi_{1i} - \alpha_1) + N_{2t}\cos\hat{\xi}_{2t}}$, associated to a phonon $\frac{\overline{P}_x}{n_2 N_{1i}} = \frac{2\left[N_{1i}\sin\xi_{1i} - N_{2t}\sin(\hat{\xi}_{2t} - \alpha_1)\right]\cos\xi_{1i}}{N_{1i}\cos(\xi_{1i} - \alpha_1) + N_{2t}\cos\hat{\xi}_{2t}}$ parallel to the interface, if $0 \le r^*_{12} \le 1$ and $0 \le t^*_{12} \le 1$

- for incident rays travelling in medium 1 towards medium 2 inside the transmission cone, the reflection (absorption) and transmission factors are deduced from the reflectivity rules, so that $r_{1\to 2}(\xi_{1i}) = r_{2\to 1}(\xi_{2t})$ and $t_{1\to 2}(\xi_{1i}) = t_{2\to 1}(\xi_{2t})$ if $r^*_{12} < 0$ or $r^*_{12} > 1$ (respectively *t*), associated to a phonon and a normal component given by Eqs. (58); note that if $\overline{P}_z < 0$ the quasi-particle is backwards so that $\frac{\overline{P}_z}{n_2 N_{1i}} \equiv \frac{\overline{p}^-_z}{n_1 N_{1i}}$ and $\frac{\overline{P}_x}{n_2 N_{1i}} \equiv \frac{\overline{p}^-_x}{n_1 N_{1i}}$, while if $\overline{P}_z > 0$, the quasi-particle is forwards with $\frac{\overline{P}_z}{n_2 N_{1i}} \equiv \frac{\overline{p}^+_z}{n_2 N_{1i}}$ and $\frac{\overline{P}_x}{n_2 N_{1i}} \equiv \frac{\overline{p}^+_x}{n_2 N_{1i}}$. Like for the isotropic case, one defines the photon-interface interaction energy by

$$\overline{E_{t0}}(\xi_{1i}) = \begin{cases} \sqrt{\overline{p}^{+2}_x + \overline{p}^{+2}_z} \\ \sqrt{\overline{p}^{-2}_x + \overline{p}^{-2}_z} \end{cases} - t_{1\to 2}\left(N^2_{1i} - N^2_{2t}\right),$$

- for incident rays outside the transmission cone, there is a total reflection (absorption) with $r_{1\to 2}(\xi_{1i}) = 1$, always associated to a forwards quasi-particle of normal component $\overline{p}^+_z = 2n_2 N_{1i}\cos\alpha_1\cos\xi_{1i}$ and phonon $\overline{p}^+_x = 2n_2 N_{1i}\sin\alpha_1\cos\xi_{1i}$, parallel to the optical axis and of energy $\overline{E}^+_{io} = 2n_2 N_{1i}\cos\xi_{1i}$ like for isotropic media.

The corrected directional transmission factor tau and photon-interface interaction energy are presented on the following figures 11a-b, for the same numerical case previously studied and reported on Figs. 9, 10: negative angles (relatively to the geometrical axis of the slab) correspond to incident angles from right to left, and positive angles to incident angles from left to right; the plain curves represent the transmission and interaction energy behaviour from crystal 1 to crystal 2, while the dashed lines illustrate the equivalent behaviours from crystal 2 to crystal 1: like for isotropic media, $t_{2\to 1}(\hat{\xi}_{1i}) \ne t_{1\to 2}(\hat{\xi}_{1i})$ but $t_{2\to 1}(\xi_{2t}) = t_{1\to 2}(\xi_{1i})$. For incident rays from medium 2 to medium 1 (dashed lines), the two discontinuous transmission areas $\hat{\xi}_{1i} \in [-37.41°, -31.64°] \cup [-19.80°, 48.75°]$, above and below the discontinuity line, are clearly distinguishable; when the incident ray is between the real interface and the first transmission cone, the quasi-particle's impulsion is always parallel to the optical axis of medium 2, the interaction energy being a simple cosine function; for rays inside the first transmission cone, the induced quasi-particle has only a longitudinal component parallel to the interface, and the interaction energy is discontinuous at $\hat{\xi}_{1i}=-37.41°$, first boundary of the cone, with $\overline{E_{t0}}\left(\hat{\xi}^{-}_{1i}\right)=1.09$ and $\overline{E_{t0}}\left(\hat{\xi}^{+}_{1i}\right)=0.19$, and at $\hat{\xi}_{1i}=-31.64°$, second boundary of the transmission area, with $\overline{E_{t0}}\left(\hat{\xi}^{-}_{1i}\right)=-0.03$ and $\overline{E_{t0}}\left(\hat{\xi}^{+}_{1i}\right)=0.24$; between the boundaries of the two transmission areas, the induced quasi-particles have an impulsion parallel to the optical axis and the interaction energy is a cosine function; when entering continuously the second transmission area, i.e. $\overline{E_{t0}}\left(\hat{\xi}^{-}_{1i}\right)=\overline{E_{t0}}\left(\hat{\xi}^{+}_{1i}\right)=1.49$, the induced quasi-particles have a normal and a longitudinal components inside the whole transmission cone, and when leaving the second transmission are, the induced quasi-particles remain parallel to the optical axis, with a cosine interaction energy function. For the opposite situation, i.e. from medium 1 to medium 2 (plain lines), all the incident rays can be transmitted in medium 2, except at the discontinuity line $\hat{\xi}_{1i}=-60.°$ where the transmission factors are $t\left(\hat{\xi}^{-}_{1i}\right)=0.057$ and $t\left(\hat{\xi}^{+}_{1i}\right)=0.$, and the interaction energies are $\overline{E_{t0}}\left(\hat{\xi}^{-}_{1i}\right)=0.32$ and $\overline{E_{t0}}\left(\hat{\xi}^{+}_{1i}\right)=0.$; on the angular area $\hat{\xi}_{1i}\in\left[-60.°,90.°\right]$ the induced quasi-particles have a single longitudinal component, while for $\hat{\xi}_{1i}\in\left[-90.°,-60.°\right]$ the induced quasi-particles also have a normal component, such that the impulsion is nearly parallel to the optical axis of medium 1.

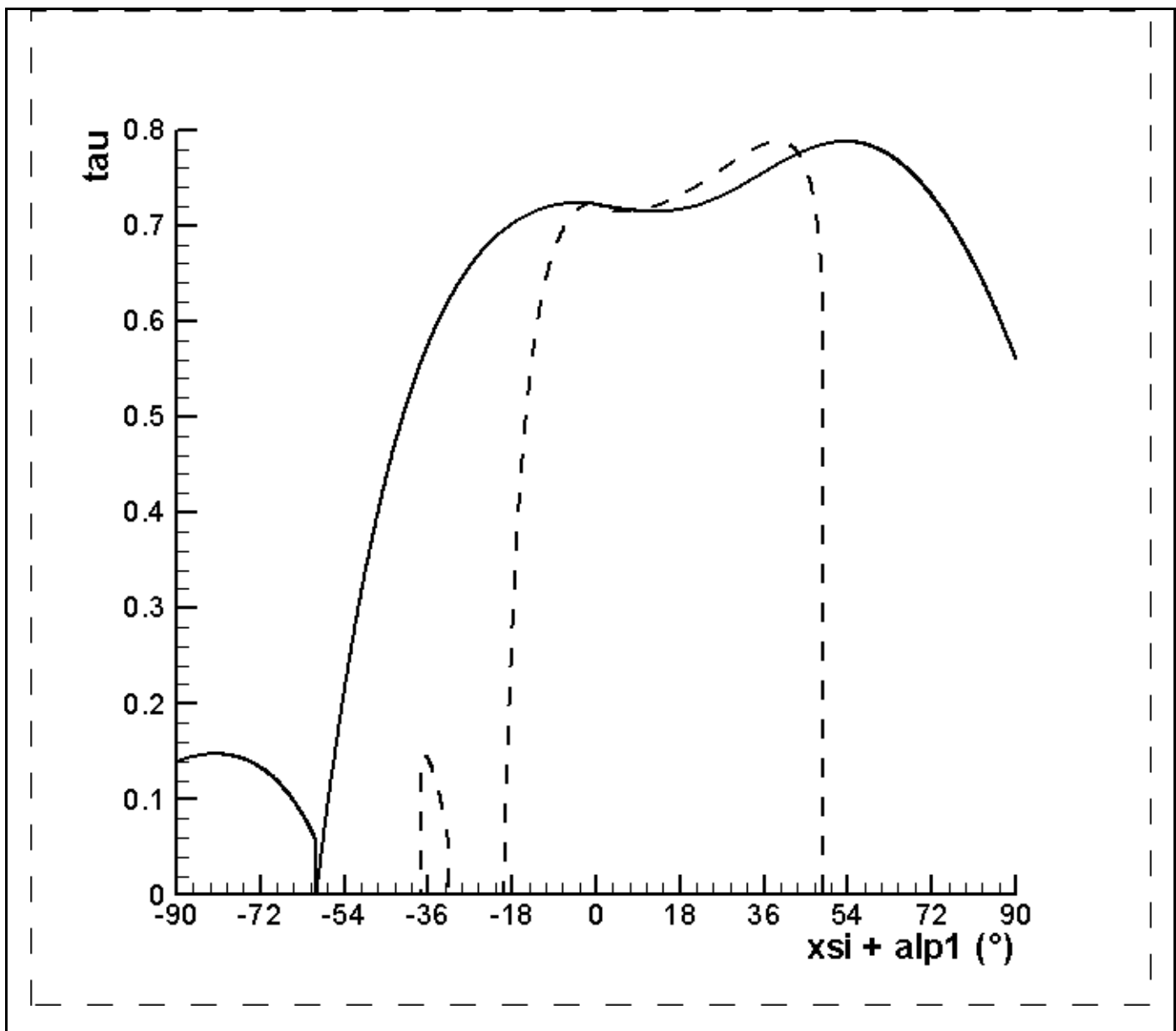

Fig. 11a: directional transmission factors

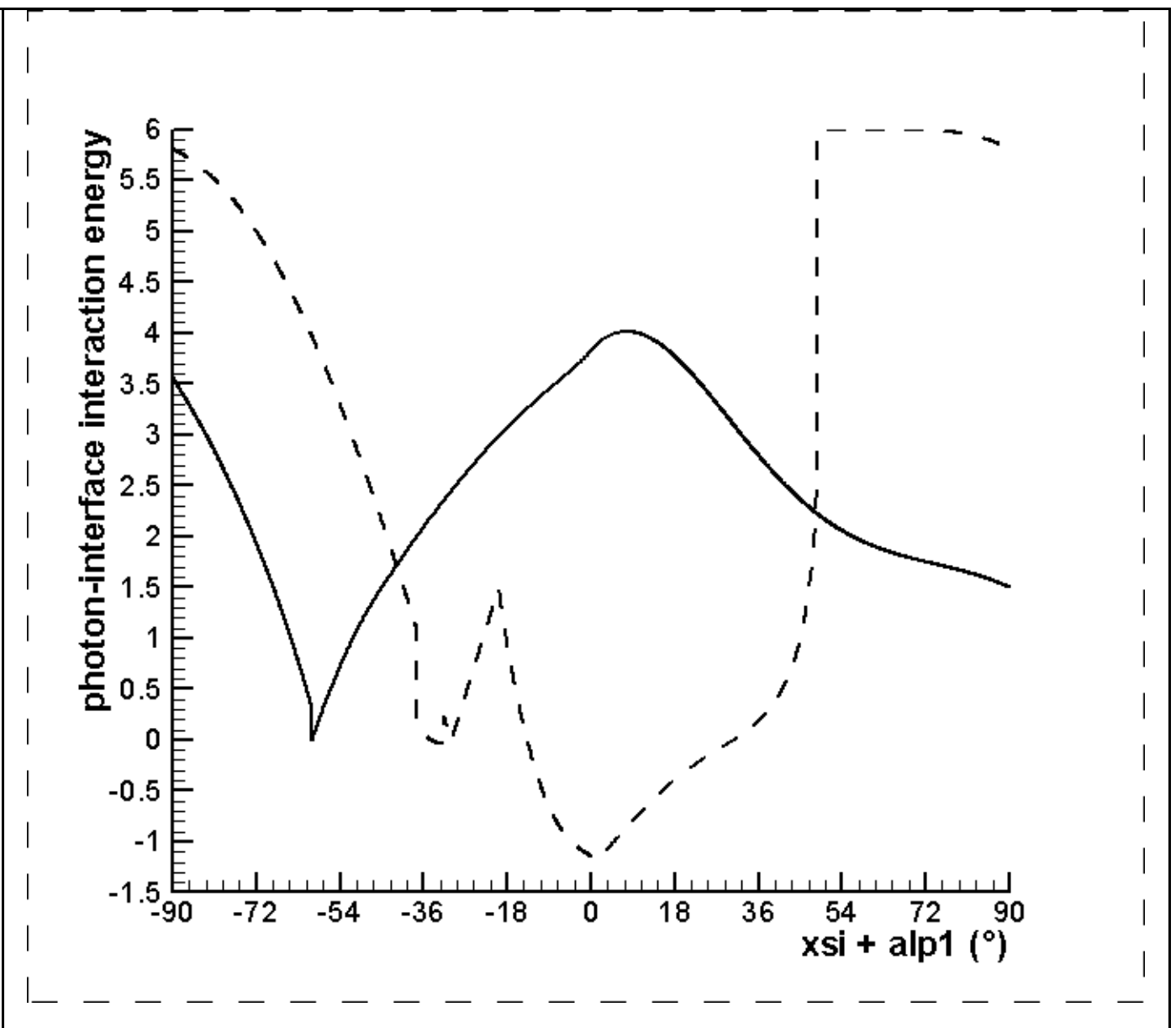

Fig. 11b: directional photon-interface interaction energy

***Remark:*** for isotropic media, the number of photons conservation equation takes the form:

$$\left(n_2 \sin\xi_{2t}-n_1 \sin\xi_{1i}\right)\overline{P}_z-\left(n_2 \cos\xi_{2t}+n_1 \cos\xi_{1i}\right)\overline{P}_x=2\, n_1 n_2 \cos\xi_{1i}\left(n_2 \sin\xi_{2t}-n_1 \sin\xi_{1i}\right)$$

and due to the isotropic Fermat's principle, $n_2 \sin\xi_{2t} = n_1 \sin\xi_{1i}$, from which it comes $\overline{P}_x = 0$: in isotropic media, the induced quasi-particles have a normal impulsion to the separating interface and cannot have a longitudinal component: the longitudinal components exist only for anisotropic media and should be associated to the phonons.

Similarly, for anisotropic media with optical axes coinciding with the geometrical one, there is no reason a priori for non normal induced impulsions, so that $\overline{P}_x = 0$, and $\overline{P}_z = 2 N_{1i} n_2 \cos\xi_{1i}$ since $N_{2t} \sin\xi_{2t} \neq N_{1i} \sin\xi_{1i}$ for a non normal incident direction; hence the reflection and transmission factors verify *r = 1* and *t = 0*, i.e. a total extraordinary reflection, possible only for $\eta_1 \varepsilon'_{\perp 1} \geq \eta_2 \varepsilon'_{\perp 2}$ and $\xi_{1i} > \xi_{1m}$: then there is always a tangential induced impulsion, i.e. a phonon along the interface, whatever the optical axis is. The transmission factor and the photon-interaction energy are depicted on the two following figures 12a-b for two crystals whose optical axes coincide with the geometrical one (i.e. $\alpha_1 = \alpha_2 = 0°$), their numerical constants being $\sqrt{\varepsilon'_{\perp 1}} = 1.5$, $\eta_1 = 2.0$, $\sqrt{\varepsilon'_{\perp 2}} = 2.0$ and $\eta_2 = 2.0$; no surprisingly these quantities are symmetric relatively to the geometrical axis and the transmission factor is extremely similar to the one of isotropic media (see Fig. 8b); the interaction energy is also very close to the one of isotropic media (see Fig. 8a), with however a main difference: for an incident angle parallel to the geometrical axis, the interaction energy is $\overline{E_{t01\rightarrow 2}} = \overline{E_{t02\rightarrow 1}} = 2 n_1 (n_2 - n_1)$ since for isotropic media there is no quasi-particle with an impulsion's longitudinal component; on the contrary, for uniaxial media the induced quasi-particles impulsion always has a longitudinal component so that the interaction energy is, for an incident direction parallel to the geometrical axis, $\overline{E_{t01\rightarrow 2}} = 2 N_{1i} (N_{2t} - N_{1i}) = 2 n_1 (n_2 - n_1)$, while from what precedes it is easy to obtain $\frac{\overline{P}_x}{n_2} = 0$ and $\frac{\overline{P}_z}{n_2} = 2 (N_{1i} - N_{2t})$ from which $\overline{E_{t02\rightarrow 1}} = 2 (N_{2t} - n_2)(N_{2t} - N_{1i}) = 0$.

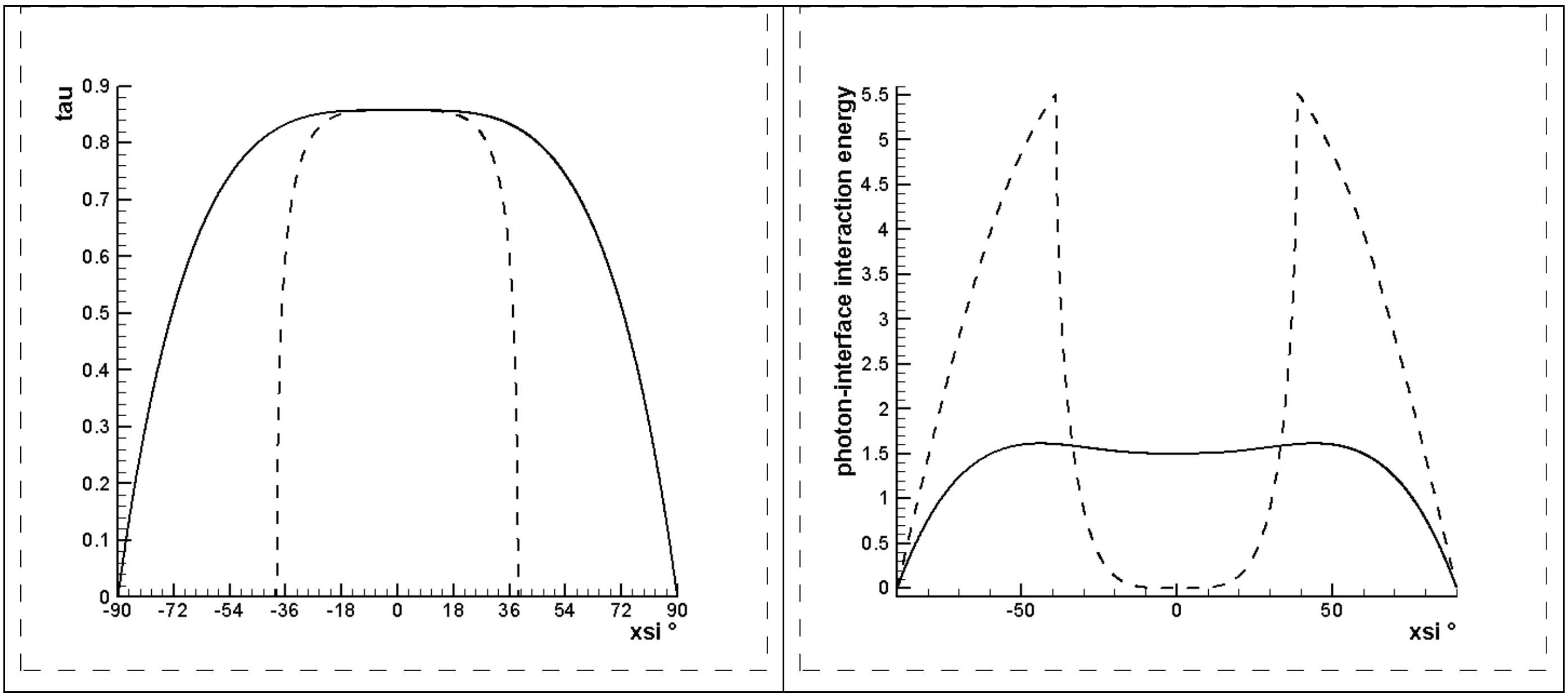

Fig. 12a: directional transmission factors Fig. 12b: directional photon-interface interaction energy

To look at the influence of the optical axis in medium 1 on the transmission factor and interaction energy, one chooses two crystals such that $\sqrt{\varepsilon'_{\perp 1}}=1.5$, $\eta_1=2.0$, $\sqrt{\varepsilon'_{\perp 2}}=2.0$, $\eta_2=2.0$ and $\alpha_2=60.0°$, $\alpha_1$ being a free parameter; note that for $\alpha_1<18.58594°$ there is no transmission between the discontinuity line and the interface, with $\hat{\xi}_{1i}\in[-71.414°,90.°]$ for $\alpha_1=18.58594°$; from $\alpha_1=18.58594°$ to $\frac{\pi}{2}$ however, there are two distinct transmission areas, and if $\alpha_1=30.0°$, the discontinuity line is one boundary of the two transmission areas; the transmission factors and interaction energy are presented on the two following figures 13a-b for several optical axes ($\alpha_1\in\{15°,30°,45°,60°,75°,90°\}$); for $\alpha_1=15°$ (solid line) there is only one transmission cone below the discontinuity line, while for $\alpha_1=90°$ (dash-dot-dot line), there is also only one small transmission cone, with small transmission factors and a cosine interaction energy function on a large angular area.

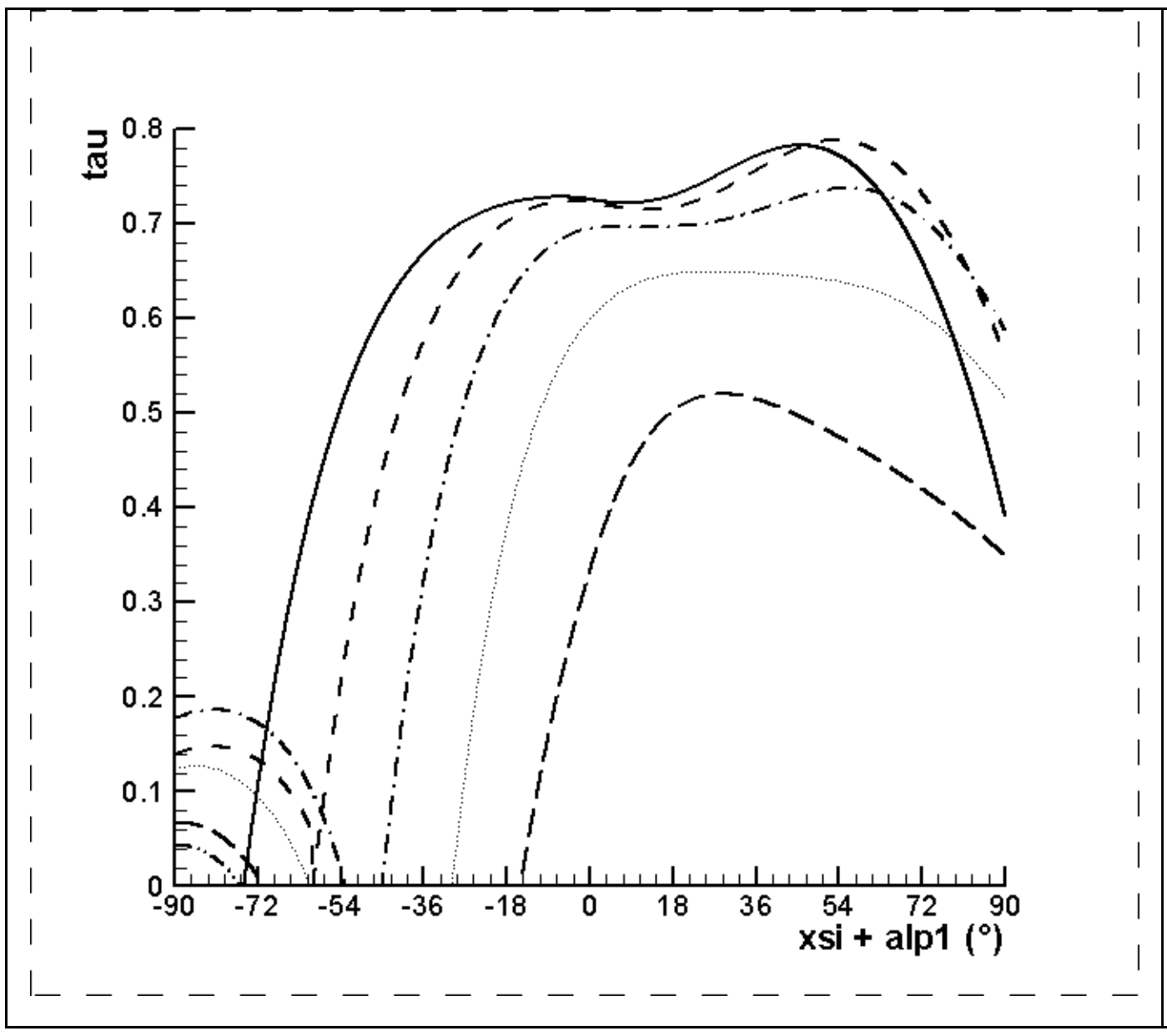

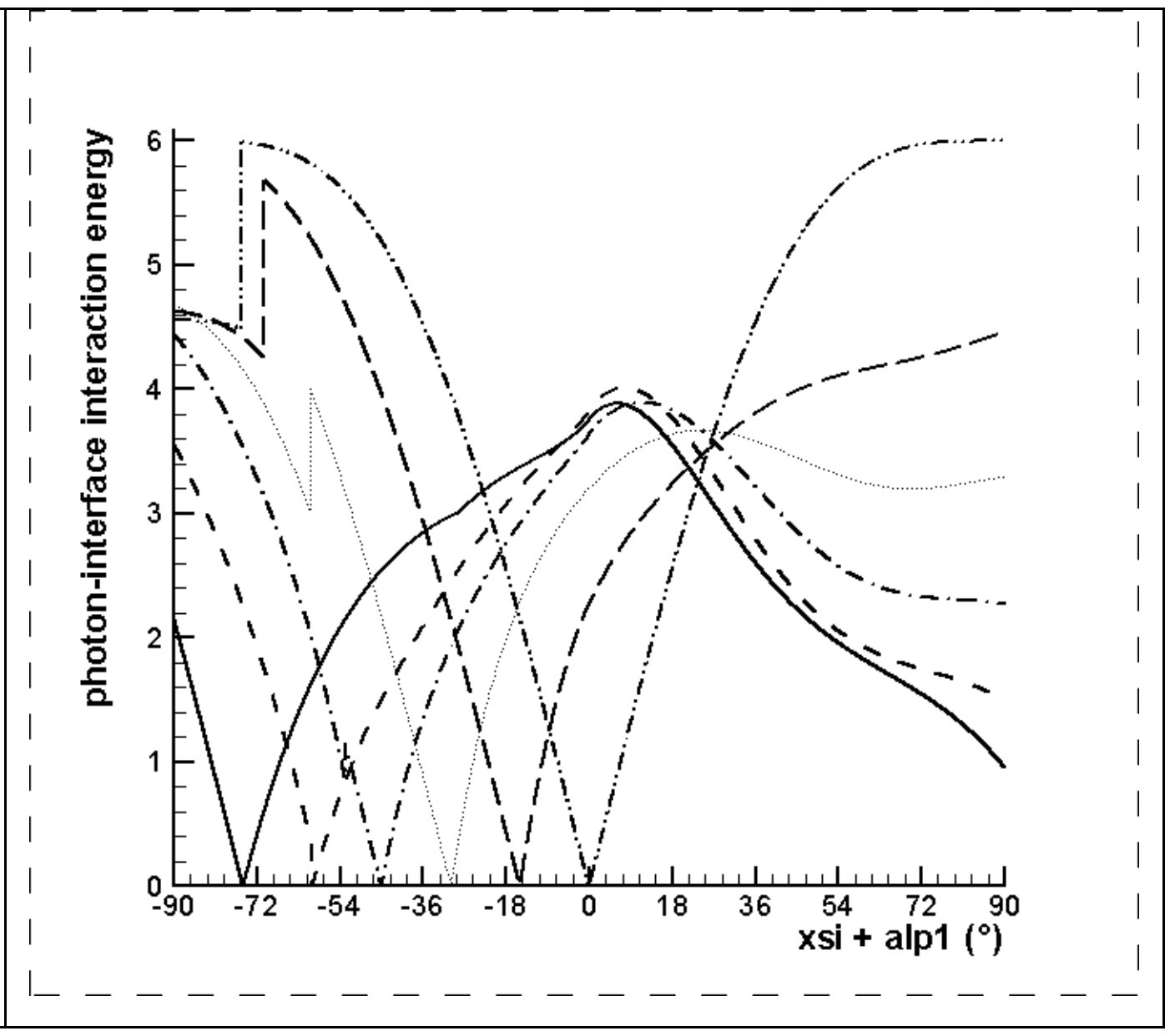

Fig. 13a: directional transmission factors Fig. 13b: directional photon-interface interaction energy

To examine now the influence of the anisotropy factor $\eta$ in medium 1, one chooses two crystals such that $\sqrt{\varepsilon'_{\perp 1}}=1.5$, $\alpha_1=30.0°$, $\sqrt{\varepsilon'_{\perp 2}}=2.0$, $\eta_2=2.0$ and $\alpha_2=60.0°$, $\eta_1>0.4444$ being a free parameter; note that for $1.8312\le\eta_1\le 2.6982$ there is a transmission between the discontinuity line and the interface, but if $\eta_1$ is outside this set, a transmission is possible only for incident rays below the discontinuity line; the transmission factors and interaction energy are presented on the two following figures 14a-b for several anisotropy factors ($\eta_1\in\{0.5,2.,2.5,3.,5.,10.\}$); for $\eta_1=2.$ (dashed line) and $\eta_1=2.5$ (dash-dot line), the two transmission cones are perfectly distinguishable, and the transmission factor can reach high values for incident angles above the cut-line, sometimes higher than for directions below the cut-line; for anisotropy factor values higher than 2.7 (in this particular example), the transmission area below the cut line becomes smaller and can even tend towards a single direction for very high anisotropy factor values, with a transmission factor close to 1; the interaction energies have complex forms, due to the presence of quasi-particles with normal and parallel impulsions.

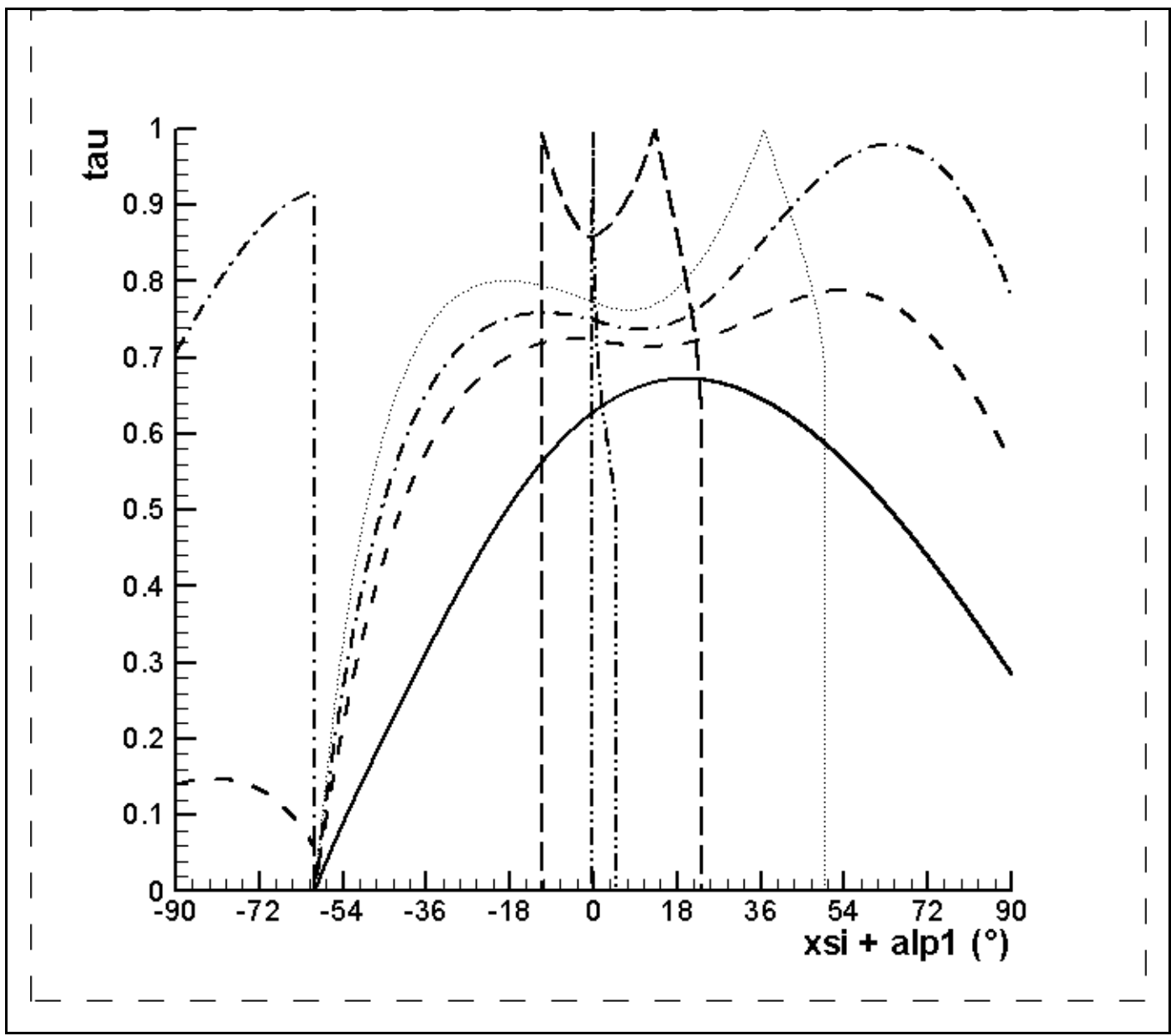

Fig. 14a: directional transmission factors

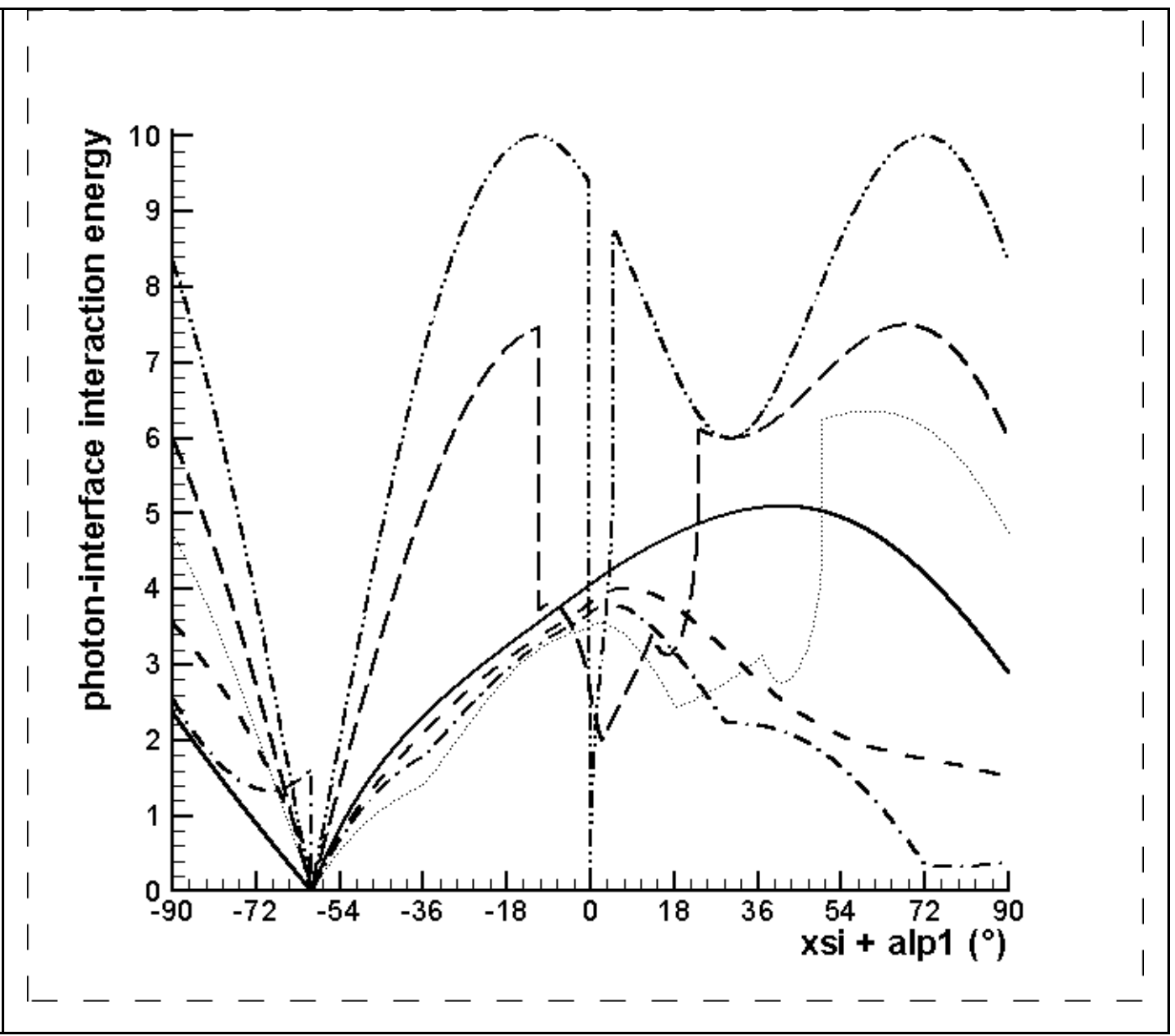

Fig. 14b: directional photon-interface interaction energy

## V – CONCLUDING REMARKS

In this paper we first derived a generalized “Descartes’ law” from the generalized Fermat’s principle valid in uniaxial crystals, basis of the geometric optics in such media, which gives the necessary conditions for a luminous ray evolving in an uniaxial crystal characterized by its dielectric permittivity diagonal tensor and its optical axis, to be transmitted inside another uniaxial crystal when the separating interface is a smooth specularly reflecting plane; a general study of these particular conditions of reflection and transmission at the interface is fully examined, which enhances the strong influence of the respective optical axis positions relatively to the geometrical axis of the whole system, and of the anisotropy factor, on the behaviour of the luminous rays in such a configuration: some spectacular results show transmission areas restricted to only very small angular domain tending to single directions in extreme conditions. From this complete geometric light description of the reflection/transmission phenomena at the interface between two uniaxial crystals, we secondly introduced a consistent definition of the energetic transmission/reflection factors from the photons impulsion-energy 4-vectors conservation, since up to now the geometric optics description of light in anisotropic media appears to be apparently non compatible with an electromagnetic one. This combined geometric/energetic way in the calculation of the reflection factors can be understood as an extension of the isotropic case: for this latter configuration, the Fermat's principle and the eikonal developments give identical results for the light trajectories seen as luminous rays, and a process based on the impulsion-energy conservation leads to reflection factors extremely similar to those obtained by a simple electromagnetic calculation, in the sense that the "photon" reflection factor is the square value of the electromagnetic perpendicular reflection factor: obviously the photons impulsion conservation process ignores the electromagnetic polarization effects, but it introduces normal non luminous quasi-particles strongly associated to the incident photons, in a way that could be summed up: the induced normal quasi-particles associated to the photons travel from the most refractive medium to the less refractive one, similarly as heat travels from the hottest medium to the coldest one. Such a result can be then extended to the uniaxial media, with the major difference that the induced quasi-particles associated to the photons must exhibit an impulsion component parallel to the interface separating the two crystals. Nevertheless the extension is of easy use and gives coherent results for the reflection/transmission factors between two uniaxial crystals for extremely various situations combining optical axes, anisotropy factors variations.

## REFERENCES

1] V. Le Dez and H. Sadat, "Derivation of the radiative transfer equation inside a moving semi-transparent medium of non unit refractive index", *Electronic journal of theoretical physics*, Vol. 4, n°14, pp. 113-150, 2007

2] B. van Tiggelen and H. Stark, "Nematic liquid crystals as a new challenge for radiative transfer", *Rev. Mod. Phys.* 72, pp. 1017-1039, 2000

3] J. F. Carinena and J. Nasarre, "Presymplectic geometry and Fermat's principle for anisotropic media" *J. Phys. A: Math. Gen.* 29, 1996

4] W. A. Newcomb, Generalized Fermat Principle", *American Journal of Physics*, Vol. 51 (4), pp 338-340, 1983

5] N. O. Naida, "Propagation of electromagnetic waves in an anisotropic inhomogeneous medium with spherical symmetry", *Journal of Radiophysics and Quantum Electronics, 12(4) Springer NY*, pp 450-452, april 1969

6] A. A Fuki, A. Kravtsovyu, N. O. Naida, Ann R. Webb, "Geometrical Optics of Weakly Anisotropic Media", Taylor & Francis, *Gordon and Breach Science Publishers*, 1998

7] V. Cerveny, "Fermat's variational principle for anisotropic inhomogeneous media", *Stud. Geophys. Geod.*, 46, pp 567-588, 2002

8] V. Le Dez and H. Sadat, "On the Fermat's principle in a semi-transparent sphere of uniaxial crystal", *European physical journal-Applied physics*, n°37, pp. 181-190, 2007

9] J. Lekner, “Reflection and refraction by uniaxial crystals”, *J. Phys Condens Matter* 3, pp 6121-6133, 1991

10] M. Sluijter, D. K. G. de Boer and J. J. M. Braat, “General polarized ray-tracing method for inhomogeneous uniaxially anisotropic media”, *JOSA A*, Vol. 25, Issue 6, pp. 1260-1273, 2008

11] S. Fumeron, P. Ben Abdallah and A. Charette, “Thermal shield effect with uniaxial crystals”, *J. Quant. Spectros. Radiat. Transfer*, Vol. 104 n°3, pp 474-481, 2007

12] L. Landau et E. Lifchitz, ”Physique théorique (tome 8), Electrodynamique des milieux continus”, *Ed. Librairie du globe, Editions MIR* (2nde édition), 1990

13] G. Stephenson, ”La géométrie de Finsler et les théories du champ unifié”, *Annales de l’IHP*, tome 15 n°3, pp 205-215, 1957